# The College Melting Pot:
# Peers, Culture and Women's Job Search

Federica Meluzzi

September 30, 2024

### Abstract

Gender norms are widely recognized as key determinants of persistent gender gaps in the labor market. However, our understanding of the drivers of gender norms, and their implications for preferences, remain lacking. This paper addresses this gap by examining how cultural assimilation from college peers influences women's early-career labor market decisions. For identification of causal effects, I exploit cross-cohort idiosyncratic variation in peers' geographical origins within Master's programs, combined with unique administrative and survey data covering the universe of college students in Italy. The main finding is that exposure to female classmates originating from areas with more egalitarian gender culture significantly increases women's labor supply, primarily through increased uptake of full-time jobs. A one standard deviation increase in peers' culture increases female earnings by 3.7%. The estimated peer effects are economically significant, representing more than a third of the gender earnings gap. Drawing on comprehensive data on students' job-search preferences and newly collected data on their beliefs, I shed novel light on two distinct mechanisms driving peer influence: (i) shifts in preferences for non-pecuniary job attributes, and (ii) social learning, particularly on the characteristics of the job offer distribution.

I am grateful to Pierre Cahuc and Arne Uhlendorff for their invaluable guidance, and to Alberto Bisin, Patrick Kline, Camille Landais, Yves Le Yaouanq, Roland Rathelot, Pauline Rossi and Michela Tincani for their support and suggestions at different stages. For very helpful comments, I would like to thank Yann Bramoullé, David Card, Pauline Carry, Patricia Cortes, Gordon Dahl, Stefano Della Vigna, Xavier D'Haultfoeuille, Matthias Doepke, Christina Felfe, Antoine Ferey, Raquel Fernandez, Albrecht Glitz, Paola Giuliano, Claudia Goldin, Libertad Gonzalez, Hilary Hoynes, Matthew Jackson, Xavier Jaravel, Andreas Kotsadam, Francis Kramarz, Alan Manning, Arnaud Maurel, Guy Michaels, Tatiana Mocanu, José Montalbán Castilla, Paula Onuchic, Ricardo Perez-Truglia, Barbara Petrongolo, Anna Raute, Alessandro Riboni, Nina Roussille, Nicolas Salamanca, Kjell G. Salvanes, Francesco Trebbi, Danny Yagan, Alessandra Voena, Christopher Walters, Basit Zafar and Yves Zenou. I thank audiences at UC Berkeley Labor Lunch Seminar, UC Berkeley IRLE Seminar, CREST Microeconomics Seminar, LSE Labour WiP Seminar, LSE Public WiP Seminar, Insead/Collège de France Brown Bag Seminar, AMSE PhD Seminar, CEPR Paris Symposium, Cesifo Labor Area Conference, SOLE Annual Meeting, EALE Annual Conference, IZA Summer School, CEPR-PSE Policy Forum, QMUL PhD Workshop, SEHO Annual Meeting and Search and Matching Workshop. I am very grateful to the AlmaLaurea consortium, and in particular to Silvia Ghiselli, for enabling data access. I gratefully acknowledge financial support from the CREST Economics Department, Fondazione Luigi Einaudi, Chaire sécurisation des parcours professionnels. I thank Alessandra Casarico, Pamela Campa and Paola Profeta for sharing with me data on Italian provinces from Indagine Excelsior and from the World Value Survey.

CREST, Institut Polytechnique de Paris. E-mail: federica.meluzzi@ensae.fr

# 1   Introduction

Cultural norms are ubiquitous and shape payoffs from many individual decisions. One critical area where their influence is particularly strong is in the economic decisions of men and women. By shaping the beliefs and preferences of both genders, gender norms are recognized as key determinants of the persistent gender gaps in the labor market, over and above traditional economic factors such as human capital accumulation, comparative advantage and discrimination (Cortes and Pan 2023, Bertrand 2020). Prior research has highlighted that their stickiness over time has prevented gender convergence in the labor market (Kleven 2024, Fernandez 2013, Fortin 2015).

Understanding the determinants of cultural change is therefore a significant yet insufficiently understood problem. One hypothesis is that culture evolves through social learning (Fernandez 2013 and Fogli and Veldkamp 2011). Despite the popularity of these theories, the empirical evidence is scarce. This is primarily due to the scarcity of natural experiments and of data sources that allow to study empirically how gender norms are formed and transmitted. In this paper, I address this gap by exploring cultural assimilation from college classmates. Using comprehensive administrative and survey data that covers 93% of all college students in Italy, I provide the first quasi-experimental evidence on the effects of peers' gender culture on women's early-career choices. By leveraging comprehensive data on students' valuations of job attributes and newly collected data on students' beliefs over time, I identify two specific mechanisms through which college peers shape women's outcomes.

Owing to features of this setting, university degrees can be thought to reproduce a *melting pot*, where students born in places with very different gender culture mix together within the same programs. Indeed, Italy is a salient example of a country with remarkable spatial differences in gender culture, as reflected in a variety of indicators, as self-reported gender attitudes, employment patterns and the magnitudes of child penalties (Campa, Casarico, and Profeta 2011, Casarico and Lattanzio 2023, Carlana 2019). The magnitude of these geographical differences is comparable to that of wide cross-country differences.

For instance, the share of women (15-64) participating in the labor force ranges from 29% to 67% across provinces (NUTS 3 classification), and the share of firms reporting hiring preferences for male workers ranges between 46% to 84%. A key feature of this setting is that a high share of students - around 57% - migrate outside their province of origin to attend university. This allows the cultural composition of degrees to be very heterogeneous: in the median degree, half of the students are coming from above-median FLFP areas and the other half from below-median FLFP areas. Such a setting is therefore ideal to study how culture evolves when individuals migrate, and whether peers shape its evolution.

I start by establishing that the childhood environment in which a woman is raised has a lasting influence on her labor supply decisions at the start of her career—a phenomenon I refer to as *cultural persistence*. I substantiate this claim relying on the epidemiological approach of Fernandez 2007 and re-adapting it to analyze granular within-country variations in cultural norms through the behavior of movers—individuals who work in a different province from where they were born[1]. The effect of early cultural exposure is estimated based on the relationship between the labor supply decisions of movers and several measures of gender culture in their place of birth. These measures include standard indicators, as female labor force participation (FLFP) relative to men across age groups, as well as indicators of firms' gender culture and local role models of previous cohorts of female graduates, drawn from the behavior of stayers in the sample. The findings reveal that women who move from more gender-egalitarian areas—e.g. reflected in higher FLFP—have significantly higher labor supply compared to similar peers from less egalitarian provinces. This pattern holds even when controlling for factors such as working in the same local labor market and graduating from the same Master's program with similar academic performance. The observed difference is both statistically significant and economically meaningful, translating into a 7.6% increase in weekly hours worked and a 2.2 percentage point higher likelihood of full-time employment. Leveraging detailed information on individual characteristics before entering the Master's program (e.g., academic records, family background), I show that this relationship is unlikely driven by differential selection of movers from different areas. Moreover, the role of local cultural exposure is both quantitatively larger and not confounded by maternal role models.

---

1. This approach is similar to recent studies by Kleven 2024, Kerwin, Guryan, and Pan 2024 and Boelmann, Raute, and Schönberg 2023.

What happens to women's labor market decisions when they get exposed to college classmates with different gender norms? This paper explores the transmission of gender norms during college years and their impact on women's labor market choices. For identification of causal effects, I exploit idiosyncratic within-Master, between-cohort variation in students' geographical origins. The analysis draws from data covering 1,572 two-year Master's programs across 71 universities and all fields of study, spanning cohorts enrolled between 2012 and 2016. This approach, first proposed by Hoxby 2000, rests on the assumption that there exists some variability in the composition of peer groups across adjacent cohorts within a degree program, which is beyond the control of individual students. I bolster the validity of this design through a large battery of randomization checks. Importantly, I provide evidence that cross-cohort changes in peers' cultural composition are unrelated to pre-determined covariates (ability, family background, socio-demographics) that explain changes in students' labor market outcomes.

My main finding is that exposure to peers from provinces with more egalitarian gender culture increases women's labor supply along the intensive margin, leading to greater take-up of full-time jobs and longer weekly working hours. The magnitude of this effect is large: a one standard deviation increase in the culture of female peers (8.33 pps.) leads to a 3.3% increase in their weekly hours and in a 1.9 percentage points increase in the likelihood of fulltime employment one year after graduation, translating to a 3.7% increase in monthly earnings. These estimated peer effects are economically significant, comparing to 33% – 41% of the size of the gender differences in the same outcomes. Notably, the effects are gender-specific, with no influence from male peers. As a placebo, I conduct the same analysis on male students and find that peers do not affect men's earnings, labor supply, or other job characteristics.

Identifying the precise mechanisms through which peer effects operate has long been a challenge, primarily due to data limitations (Sacerdote 2011). As clearly stated by Barrios Fernandez 2023, understanding these mechanisms is crucial, as this would allow to design policies that could replicate the benefits of peer exposure without altering peer group composition. With unique comprehensive data on students' valuations of job attributes and original survey data on students' beliefs, this paper advances our understanding by identifying two main mechanisms driving peer effects. First, I show that peer effects are not mediated by changes in academic performance or geographic mobility, allowing me to rule out local labor market networks as a key factor. Instead, I uncover two channels of peer influence: (i) shifts in preferences for non-pecuniary job attributes and (ii) social learning, likely related to the job offer distribution. Using data from a mandatory pre-graduation survey, I show that exposure to peers from more gender-egalitarian areas changes women's preferences, reducing the importance they place on non-pecuniary job factors such as flexible hours and social utility—consistent with cultural transmission. Moreover, I find strong asymmetry in peer effects: women from areas with below-median FLFP are positively influenced by exposure to egalitarian peers, while the reverse does not occur. This asymmetry, which starkly reject conformism as an underlying mechanism, aligns with social learning explanations.

To investigate social learning and identify which beliefs are affected, I have designed an original survey and collected students' beliefs at two points in time. This includes beliefs about gender roles, perceptions of employers' discrimination, beliefs regarding the distribution of job offers, and expectations of future fertility and child penalties. The survey also gathers information on the network structure and perceived peer influence. It was administered to a random sample of current female students across all disciplines at one large university. With in-person administration and lottery incentives, I achieved a 97% response rate among attending students. I use this data to (1) examine asymmetries in beliefs based on the gender culture in their province of birth, (2) test whether these beliefs predict job acceptance decisions, and (3) analyze how beliefs evolve within the social environment. Analysis of these data reveal that women from low-FLFP areas have more pessimistic expectations on the arrival rates of job offers, and expect a significantly share of part-time job offers relative to full-time ones compared to their peers from high-FLFP areas. Consistent with the predictions of a McCall type model (McCall 1970) incorporating heterogeneous workers' beliefs, my results show that students' beliefs are relevant in the decision to accept part-time jobs, explaining about one fifth of the part-time acceptance gap between women from high- and low-FLFP areas. Lastly, I provide evidence of belief convergence after peer exposure, driven primarily by strong beliefs' updating concentrated among women from low-FLFP areas.

**Related literature.** This article contributes to several strands of literature. First, it relates to a burgeoning literature that, since the seminal contribution of Fernandez 2007, has explored the interplay between cultural norms and women's economic decisions. Several works have highlighted the role of cultural factors in shaping women's labor supply (Fortin 2015, Blau, Kahn, and Papps 2011, Bertrand, Kamenica, and Pan 2015), the magnitude of *child penalties* (Kleven 2024, Cortés, Kosฺar, Pan, and Zafar 2022, Boelman, Raute, and Schönberg, Forthcoming), fertility choices (Alesina, Giuliano, and Nunn 2013, Fernandez and Fogli 2006), marriage prospects (Bertrand et al. 2021) and time allocation (Ichino et al. 2024). In many of these studies, the impact of culture is estimated by comparing outcomes among first- or second-generation immigrants within a host country, with cultural differences stemming from cross-country variations in norms. These works typically focus on the broader female population and how cultural influences shape (work) decisions around motherhood or marriage. My contribution to this literature is twofold. First, I focus on a narrower segment—young, educated women—and exploit granular within-country variations in gender norms, in a setting where I can rule out many confounders, such as ability and family background. My findings complement existing research by showing that childhood exposure to gender norms influences women's labor supply decisions at the very start of their careers, which may play a crucial role in explaining the higher child penalties observed later in life. Second, while existing research often remains agnostic about the mechanisms behind *cultural persistence*, I uncover a previously unexplored channel: information asymmetries. With this finding, I extend the conventional view of cultural persistence beyond its influence on preferences and attitudes, and show that childhood exposure to local role models also shapes women's beliefs about labor demand and employers' tastes. This finding carries important policy implications, suggesting that information provision, e.g. on the job offer distribution, can be an effective tool in reducing *cultural persistence* and gender disparities.

A second strand focuses on the transmission of gender norms. Theories developed by Fernandez 2013 and Fogli and Veldkamp 2011 emphasize social learning as a key driver of cultural evolution. They model a process in which beliefs are shaped both within the family—via mothers' behavior (vertical transmission, in the terminology of Cavalli-Sforza and Feldman 1981)—and through observing the labor force participation of previous generations of women (oblique transmission). Consequently, current shocks to female labor supply trigger the dynamics of beliefs' updating. Empirical evidence supporting the importance of childhood exposure has been provided by Olivetti, Patacchini, and Zenou 2020 and Mertz, Ronchi, and Salvestrini 2024, who show that women's career decisions, such as whether to work after motherhood or their occupational choices, are influenced not only by their own mothers' behavior but also by the behavior of other mothers in their social network, such as their peers from primary or high-school. An emerging and directly related literature examines the horizontal transmission of norms. Boelmann, Raute, and Schönberg 2023 and Jessen, Schmitz, and Weinhardt 2024 show that increased exposure to East Germans, either through coworkers or friendships, influences West Germans' behavior by reducing their parental leave or increasing labor supply at the intensive margin. Similarly, Maurin and Moschion 2009 find that the labor supply decisions of female neighbors positively affect other women's labor participation at the extensive margin in France. My contributions to this literature is twofold. To the best of my knowledge, my paper provides the first large-scale empirical evidence on the role of college peers in the transmission of gender norms. Specifically, I exploit quasi-random variation in the gender norms to which young, educated women are exposed during college, allowing me to directly compare the influence of horizontal peer effects with childhood exposure. My findings reveal that the social environment in college plays a critical role in shaping women's early-career labor supply decisions, demonstrating that peers can nearly offset the impact of limited female role models during childhood. This challenges the view that gender norms are primarily formed in early childhood and remain sticky over time. Second, I offer direct evidence on the mechanisms behind peer influence. I show that peers shape women's preferences for job attributes, using data from a compulsory survey of all graduates. Newly collected survey data on students' beliefs reveal that initial information gaps about job opportunities between women from different cultural backgrounds narrow substantially, supporting social learning explanations. With the latter finding, I closely relate to a contemporaneous literature on the role of biases in beliefs and the stickiness of gender norms (Cortés, Koşar, Pan, and Basit 2022, Bursztyn et al. 2023).

This paper also contributes to a broad body of work on gender gaps in the labor market. Recent evidence has shown that, in the skilled population, gender differences in the valuation of temporal flexibility, coupled with increasing returns to the provision of long hours, largely contribute to earnings inequalities (Cortes and Pan 2019, Wiswall and Zafar 2018, Blau and Kahn 2017, Azmat and Ferrer 2017, Flabbi and Moro 2012, Bertrand, Goldin, and Katz 2010). Aligning with previous work, I document large differences in hours worked and earnings emerge between female and male graduates at labor market entry. My findings suggest that preferences for job attributes are endogenous to the social environment and can explain part of early-career gaps. Specifically, I show that 30% of the initial gap can be closed through peer influence. Finally, my paper contributes to a rising literature that has focused on the interplay between biased beliefs, e.g. on job finding probabilities or on the wage distribution, and job-search behavior (Jäger et al. 2024, Cortes et al. 2023, Mueller and Spinnewijn 2023, Conlon et al. 2018, Mueller, Spinnewijn, and Topa 2021, Alfonsi, Namubiru, and Spaziani 2024, Bandiera et al. 2023, Caliendo, Cobb-Clark, and Uhlendorff 2015). Relative to previous studies, usually focussed on long-term unemployed individuals, I examine the effects of pessimism on arrival rates of job offers and workers' acceptances of part-time jobs.

The rest of the article is organized as follows. Section 2 describes the institutional setting and the data sources and provides a description of the sample. Section 3 describes the college melting pot. Section 4 describes the early-career gender earnings gap and the phenomenon of cultural persistence. Section 5 explains the identification strategy and discusses its validity. Section 6 presents baseline estimates of peer effects. Section 7 discusses estimates from a battery of robustness exercises. Section 8 explores non-linearities in peer effects. Section 9 presents evidence on the mechanisms of peer influence. Section 10 analyzes original data on students' beliefs and presents a job search model to illustrate their implications for job search.

## 2    Institutional Background and Data

### 2.1    Admission and Structure of Graduate Education in Italy

This subsection provides an overview of the key institutional features governing tertiary education in Italy. Section 3 will then zoom in on the specific aspects that make this context particularly well-suited for studying cultural assimilation.

**Admission to Master's degrees**. Since the early 2000s, degrees are organized as bachelor's (three years) and master's (two years)[2]. Most students pursue a two-year master's program after completing their bachelor's degree. Admission criteria for master's programs are determined autonomously by each academic institution. However, certain fields—such as medicine, health sciences, architecture, psychology, and primary education— are subject to selective national entry exams under Law 264/1999. A common requirement for master's admission is the completion of a bachelor's degree, along with specific curricular prerequisites, typically fulfilled through credits in required courses. This system allows students some flexibility to switch fields between their bachelor's and master's studies, provided they meet the relevant eligibility criteria. Therefore, students have the flexibility to change fields between the bachelor's and master's levels, subject to meeting the eligibility requirements[3]. Beyond these prerequisites, many programs also include selective entry exams, bachelor's grade requirements, and interviews in their admission process. Admission is often competitive, with applicants ranked based on entry exam scores or bachelor's GPA, and additional requirements like English proficiency, motivation letters, or reference letters may apply.

**Tuition fees**. Ninety percent of Italian students attend public universities (ISTAT 2016), with tuition fees varying based on the degree, institution, and family income. Regional governments set income thresholds for need-based grants, which cover tuition, housing, and meal vouchers for low-income students (Rattini 2022). On average, 23% of the students in my sample receive such grants. For those not eligible, the average annual tuition fee is €1262 (Commission/EACEA/Eurydice 2016).

## 2.2 Data Sources and Sample

The empirical analysis relies on two main sources of data. The primary source comprises comprehensive administrative and survey data, which collectively cover 93% of the uni-verse of college students in Italy, obtained from the AlmaLaurea consortium. Specifically, the dataset encompasses all students enrolled in 1,572 2-year Master's degree programs across 71 universities, spanning enrollment cohorts between 2012 and 2016. This database consists of administrative data from university records, institutional survey data, and post-graduation follow-up surveys.

---

2. Italy adheres to the Bologna process (1999) that ensures comparability in higher education standards across the European Higher Education Area (EHEA), which comprises 48 European and Central Asian countries.

3. On average, students must complete 77 constrained credits to qualify for a master's program, though requirements vary across fields. For instance, a student entering a master's program in economics must have earned at least 53 credits in economics, statistics, or other social sciences (Brandimarti 2023).

1. **Administrative student-level information**, from university records, for all students. This source provides information on academic performance during the Master (number of exams, GPA, final grade), and demographic information (age, immigration status, municipality of birth and residence), as well as unique identifiers of Master's programs within universities and enrollment and graduation dates. Importantly, I use this source to identify college classmates and construct measures of their gender culture based on their birth province. Due to the administrative nature of the data, all information is available for the entire student population, ensuring that I observe the characteristics of all peers.

2. **Institutional pre-graduation survey.** Universities administer this survey to all students as part of the graduation process. At the end of their final year, students are required to complete a compulsory survey, with a response rate close to 100%. This survey collects detailed information on students' job search intentions and preferences, including their valuation of several job attributes. Additionally, the survey gathers data on students' socio-economic background, including parents' occupations and education levels. Furthermore, it collects detailed information on students' educational histories, such as previous education in high school and Bachelor's degree programs, their grades in previous education, and the working activities they participate in during their studies.

3. **Follow-up surveys.** Students are contacted by the AlmaLaurea consortium for follow-up surveys one, three, and five years after graduation. These surveys gather comprehensive information on realized job characteristics, such as net monthly earnings, usual weekly hours worked, contract type (part-time vs. full-time), job security, occupation, industry, sector and location. The responses to the earnings questions, and usual weekly hours worked, are collected in discrete bins that I transform into real-valued variables (at the mid-point of each bin)[4]. Additionally, they include retrospective information on the job-search process and current job search activities. While participation in these surveys is voluntary and does not involve monetary incentives, the response rate remains high (e.g., 74% after one year).

**Original survey on students' beliefs**. To investigate the mechanisms of peer influence, I have designed an original survey to elicit students' beliefs regarding gender attitudes and various future outcomes. This includes perceptions of employers' discrimination, beliefs regarding the distribution of job offers, and expectations of future fertility and child penalties. The survey also gathers information on the network structure and perceived peer influence. It has been administered to a random sample of students from several fields at a large university (enrollment cohorts 2022-2023). With in-person administration and lottery incentives, I achieved a nearly 100% response rate among attending students. Detailed information on the survey and elicitation methods is provided in Section 10.

## 2.3 Sample description

In this paper, I use data on students enrolled between 2012 and 2016. Since data are collected from graduating cohorts, I recostruct enrollment cohorts using students' enrollment and graduation dates from university records. A student's classmates, or peers, are defined as all students who enroll in the same university major, or degree, in the same cohort, and who remain enrolled for the entire duration of the Master[5]. The sample is composed of students from a panel of Master degrees that (i) count at least one man and one woman in the same cohort (excludes 3.55% of students) and (ii) exist and respect (i) for at least 2 consecutive years (excludes 6% of students). The final sample is composed of 316,470 students from 1,572 degrees and 71 universities. Table A.1 and Table A.2 provide descriptive statistics of the analysis sample. The last column of the tables reports the p-value of the test of equality of the means across gender.

---

4. Possible answers to the earnings question were < €250, €250–€500, €500–€750, €750–€1000, €1000–€1250, €1250–€1500, €1500–€1750, €1750–€2000, €2000–€2250, €2250–€2500, €2500–€3000, and > €3000. I converted the answers into a real-valued earnings variable at the mid-point of each earnings bin; I assigned earnings of €187,5 to those that responded earning less than €250 and earnings of €3750 to those who indicated earning more than €3000. The response bins for usual weekly hours worked were < 5 hours, 5-9 hours, 10-14 hours, 15-19 hours, 20-24 hours, 25-29 hours, 30-34 hours, … , 55-59 hours, and > 60 hours.

5. A drawback is that I lose track of dropouts, which account for 6% of the enrolled students between 2012 and 2016 (ANVUR 2023).

**Background characteristics and academic records**. Table A.1 presents summary statistics on the main background characteristics and academic variables in the sample, disaggregated by student gender. These data are drawn from both administrative records and the institutional survey. The sample includes 182,792 women and 133,678 men. On average, women outperform men academically, as evidenced by higher GPAs and final grades during their master's studies, as well as stronger prior academic records, such as bachelor's and high school grades. Additionally, the variation in these outcomes is smaller among women compared to men. Women are also more likely to have attended a general high school (*liceo*) than men—84.1% versus 71.3%. In terms of field specialization, women are underrepresented in scientific tracks and more concentrated in humanities, both in high school and at university. The largest disparities are in engineering and humanities: 27% of men study engineering, compared to just 8.2% of women, while 24.7% of women and only 10.4% of men pursue humanities. Regarding family background, around one-fifth of students have parents with tertiary education, and roughly one-third come from families where the father is in a high-SES profession. Additionally, 71% to 73% of students have mothers in the labor force. Women, however, are less likely to come from wealthier families compared to men, as indicated by a lower share of parents with tertiary education and a higher share of parents in low-SES occupations.

**Labor market outcomes**. Table A.2 presents summary statistics on students' labor market outcomes, based on responses from the follow-up survey conducted one year after graduation. Response rates to this survey are 73.7% and 73.2% in the samples of women and men, respectively, very high compared to traditional survey data. At the time of the survey, about 12% of both female and male students are pursuing further education, either at the master's or PhD level, while the majority are participating in the labor market. 67.3% of women and 71.8% of men are employed, either with a standard labor contract or via an internship. A slightly higher share of women (20.5%) than men (15.1%) are unemployed and actively searching for a job, while less than 1% of both genders are unemployed and not looking for work. The observed gender differences in employment patterns can largely be attributed to differences in field specialization, as discussed in Section 4. Among employed graduates, women experience less favorable outcomes than men. On average, women earn €1,077.8 per month, compared to €1,324 for men, and are significantly less likely to hold full-time positions (69.3% vs. 86.2%). Women are also underrepresented in occupations and industries with above-median earnings or above-median shares of full-time jobs. A higher proportion of women work in the public sector (16.5% vs. 11% for men). Regarding the job-search process, around 80% of both men and women are in their first job after graduation. While men and women begin their job search at similar times, women tend to search longer before accepting their first job offer.[6]

# 3   The College Melting Pot

This section outlines three key features of the Italian context that make it uniquely suited for studying cultural assimilation from peers on a large scale.

## 3.1   Spatial Differences in Gender Culture

Studying cultural assimilation requires significant variation in gender norms among students. Italy provides an ideal setting, offering granular yet wide geographical variations in gender culture. Spatial differences in gender culture are striking, comparable in magnitude to wide cross-country differences. These differences are reflected in a variety of indicators, from self-reported gender attitudes to labor market attachment, as documented in previous studies (Campa, Casarico, and Profeta 2011, Carlana 2019, Casarico and Lattanzio 2023, Carrer and Masi 2024). A particularly important feature is the substantial geographical variation in traditional labor market outcomes for women. For example, female labor force participation rates among those aged 15-64 range from 29% to 67% across provinces, and for women aged 25-34, the range is even wider, from 38% to 86%. In contrast, male labor force participation varies much less, ranging from 64% to 82% for the 15-64 age group. These differences in female labor market outcomes are accompanied by substantial heterogeneity in gender attitudes. For instance, the proportion of individuals who disagree with statements such as "*Being a housewife is just as fulfilling as working for pay*" or "*Men should be given priority when jobs are scarce*" varies from 16% to 67% across Italian regions (NUTS 2), according to recent waves of the European Values Survey (EVS 1990-2008). Throughout this paper, provinces (the NUTS-3 classification) are used as the main geographical unit. Italy is partitioned into 103 provinces, which are administrative divisions of intermediate level between a municipalities and regions[7].

---

6. This variable takes the value -1 if a student started searching for a job before graduation, 0 if they began searching at the time of graduation, and 1, 2, 3, etc., if they started their search 1, 2, 3, etc., months after graduation.

I define the gender culture of a province using several alternative measures:

1. Female labor force participation (FLFP hereafter) of different age groups at the province level (NUTS 3 level);

2. Ratio of female to male labor force participation (FLFP/MLFP hereafter) of different age groups at the province level (NUTS 3 level);

3. Indicator of firms' gender culture, e.g. the share of firms in the private sector without hiring preferences for male workers, at the province level (NUTS 3 level);

4. Share of female college graduates that are employed full-time at the outset of the career at the region level (NUTS 2);

5. Share of female vs. male college graduates that are employed full-time at the outset of the career at the region level (NUTS 2);

Students are assigned to provinces based on their province of residence at the enrollment date, as recorded in university registers. Such province should be interpreted as the place where the student grew up. Since the objective is to capture the gender culture and female role models that students were exposed to while growing up, most of these measures are based on data prior to their university enrollment, particularly during their adolescence. The first two measures refer to averages of FLFP and FLFP/MLFP from 2004 to 2007. Indicators of firms' gender culture in (c) are constructed based on answers to a survey of a nationally representative sample of 100,000 Italian firms in 2003 (*Indagine*

---

7. The average population of a province was 551,000 as of 2010, but there is large heterogeneity. The largest province, Rome, has over 4 million residents and contains 121 different municipalities. The smallest province, Ogliastra (Sardinia), has less than 60,000 residents and only includes 23 municipalities.

*Excelsior, Unioncamere*[8]). Each year, firms report their hiring intentions, specifying whether they prefer to hire male or female workers, or if they have no preference. At the province level, I define firms' gender culture as the proportion of firms that either prefer female workers or are indifferent between male and female hires. Additionally, I introduce novel measures related to the employment patterns of previous cohorts of female graduates, who serve as a more direct reference group for the students in the sample. To construct measures (d) and (e), I map local labor market opportunities for female and male college graduates, focusing on graduates who remain employed in their province of birth. Using labor market data from cohorts preceding those analyzed, I calculate the share of female college graduates employed full-time one year after graduation, both in absolute terms and relative to male graduates. Due to limited data in smaller provinces, these measures are aggregated at the regional level (NUTS 2). Table 1 shows descriptive statistics for these different indicators in the sample, obtained by assigning to each student the value of the indicator corresponding to the student's province (or region) of origin. Table A.3 disaggregates these statistics by gender, showing that there are no significant differences in the geographical origins of female and male graduates. A visual illustration of the spatial distribution of the FLFP and of the ratio of FLFP/MLFP, for all ages, is presented in Figure 1. Similar spatial patterns emerge for other indicators of gender culture, such as participation rates for young women, firms' gender preferences, and individual gender attitudes, as illustrated in Figure A.1. All of these measures reveal substantial variation across provinces. For example, in some areas, the FLFP is as low as 29%, similar to that of low-income countries, while in other areas, significantly higher shares of women are in the labor force, aligning with or exceeding the OECD average. These disparities are similarly pronounced among young women (aged 25-34), where FLFP ranges from 38% to 86%. Importantly, these differences in labor market opportunities for women are not merely a reflection of spatial heterogeneity in economic activity, as reflected in the nearly identical variation in the FLFP/MLFP ratio, whose geographical distribution closely follows that of FLFP. The ratio shows a wide range: between 43% and 86% of women of all ages, and

---

8. Starting from 1997, the Excelsior Survey represents one of the main sources of information on the Italian labor market. The survey is administered by Unioncamere, in partnership with the Ministry of Labor, ANPAL and the European Union. The survey is conducted on firms operating in the manufacturing and service sectors with at least 0.5 employees/year. It excludes employers in the agricultural and public sector, as well as those that are not registered with the Chambers of Commerce. The sample of firms represents roughly one third of the total in the respective population.

| Variable | Mean | SD | Min | Max | Obs |
|---|---|---|---|---|---|
| Female labor force participation (age: 15-64) | 49.7 | 11.2 | 27.3 | 66.7 | 316470 |
| Female/Male labor force participation (age: 15-64) | 66.7 | 11.8 | 43.0 | 85.7 | 316470 |
| Female labor force participation (age: 25-34) | 65.1 | 15.2 | 38.0 | 86.0 | 316470 |
| Female/Male labor force participation (age: 25-34) | 74.3 | 13.0 | 47.0 | 95.0 | 316470 |
| Male labor force participation (age: 15-64) | 73.8 | 4.5 | 62.9 | 81.9 | 316470 |
| Male labor force participation (age: 25-34) | 86.7 | 6.5 | 71.4 | 97.1 | 316470 |
| % of female graduates in full-time job | 56.4 | 9.5 | 40.1 | 68.9 | 316470 |
| % of female/male graduates in full-time job | 71.8 | 6.7 | 55.3 | 83.5 | 316470 |
| % of firms without hiring pref. for male workers | 34.8 | 7.8 | 16.0 | 54.0 | 316470 |

Notes: The Table presents summary statistics f the measures of gender culture 1-5 presented in Section 3 The unit of observation is a student. Students are assigned to provinces based on their residence province prior to enrollment in the Master.

between 47% and 95% of young women, participate in the labor force relative to men in the same age group. Additionally, the ratio of female to male college graduates in full-time employment also varies considerably by region, from 55% to 83%. Looking at firms' stated preferences, between 16% and 54% of firms report having no strict preference for male workers.

## 3.2 Students' Mobility

Within this context, a second key aspect is the high mobility of students who study outside their home province. Due to historical and institutional factors, the majority of students in Italy relocate to attend university (ANVUR 2023). This phenomenon has longstanding origins and has been relatively stable over time. During the period covered by this analysis, more than 57% of students moved to another province, and about 31% moved to a different region to pursue higher education. To better understand the strong migration of students away from their province of origin, it is useful to describe the institutional landscape in Italy. In 2016, public universities, which account for over 90% of students, offered Master's degrees at 89 institutions. However, these universities were concentrated in only 52 of Italy's 103 provinces. Furthermore, not all fields of study are available at every institution, as some universities specialize in specific disciplines. For instance, degrees in information

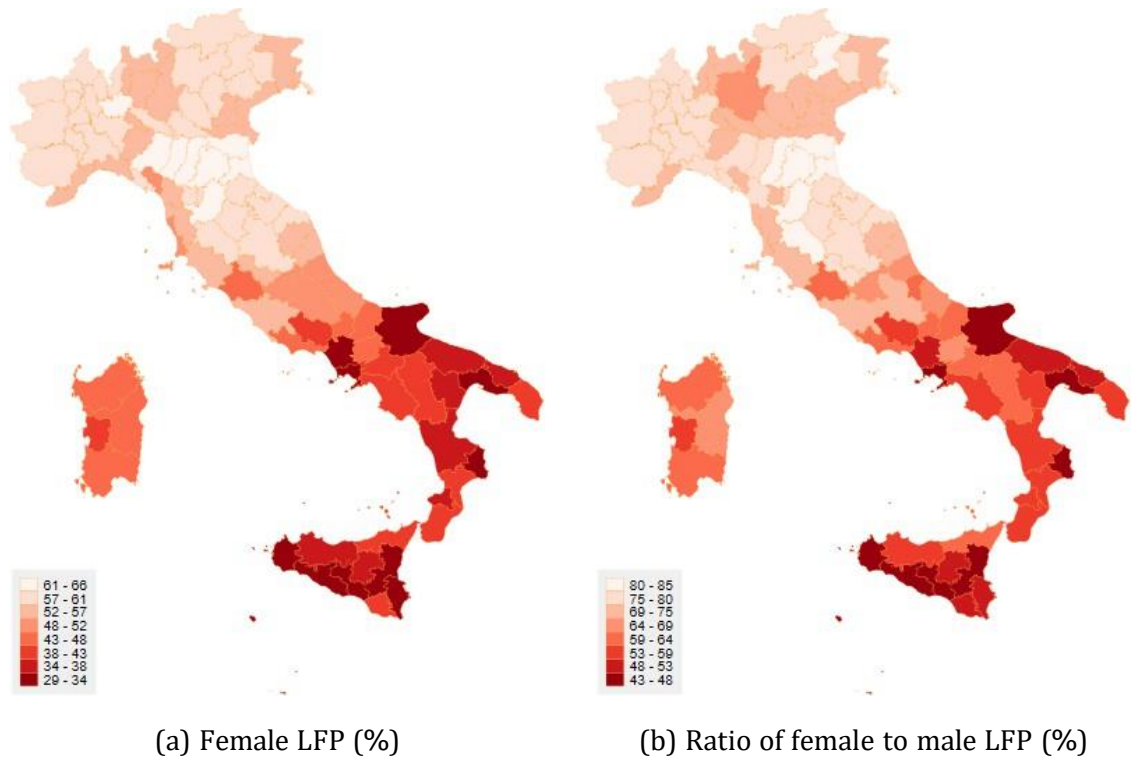

(a) Female LFP (%)      (b) Ratio of female to male LFP (%)

Notes. The maps present the FLFP (Panel a) and the FLFP/MLFP (Panel b) across provinces in Italy. Individuals between 15 and 64 years of age are considered. Each geographical partition is a province (NUTS 3 classification) and there are 103 provinces in total. Both measures are constructed as averages of years 2004-2007. Source: Labor Force Survey (Istat).

technology were offered at only 29 universities, and agriculture and veterinary sciences at just 24 institutions. This uneven distribution of universities results in limited local options for many students. About 20% of Italians aged 18-19 live in provinces without any higher education institutions, and only about 77% have access to both STEM and non-STEM universities in their home province (Braccioli et al. 2023). Moreover, students are free to apply to any university in the country, regardless of their place of residence. This flexibility, combined with the uneven availability of university programs and variation in the quality of these programs, fuels the high levels of student mobility observed across Italy.

**Mobility patterns in the sample**. Table A.4 present summary statistics of students' mobility by gender. 58.9% of women 55.4% of men have moved outside their birth provice for their studies. The table also describes the local gender culture in the provinces where these students are studying. Overall, the mobility destinations for both genders do not exhibit significant differences, with both groups studying in areas characterized by more egalitarian gender cultures compared to their provinces of origin. Additionally, Table A.5 provides summary statistics on mobility patterns to local labor markets for students employed at the time of the follow-up survey. A large proportion of students work outside their province of origin, with this share being higher for male students (51.4%, compared to 44% for women), indicating that women are more likely to return to their home provinces after completing their studies. The majority of students—68.4% of women and 65.2% of men—secure their first job in the same region where they studied, and approximately 5% of both men and women move abroad for work. The locations where men and women work show no significant differences in terms of local gender culture.

Since the focus of the paper will be on gender norms and women's outcomes, Table A.6 zooms in on the mobility patterns of women, by contrasting the behavior of those originating from provinces in the first or the fourth quartile of FLFP. Both groups exhibit high levels of migration outside their provinces of origin, with women from more egalitarian regions migrating at a higher rate. Women from less gender-egalitarian areas, however, are significantly more likely to migrate longer distances, such as outside their home region. Among those who relocate, women from high-FLFP areas tend to move to regions with higher FLFP levels compared to women from low-FLFP areas. This results in unequal exposure to peers from high-FLFP regions: on average, women from high-FLFP provinces attend programs where 67% of their peers are from above-median FLFP areas, compared to only 27% for women from low-FLFP provinces. In terms of academic specialization, women from low-FLFP areas are slightly more likely to pursue degrees in scientific disciplines, engineering, and psychology, while being less likely to choose fields such as humanities, economics, statistics, and architecture. However, these differences in specialization are relatively modest in magnitude.

**Peers' composition**. A direct consequence of these high mobility rates is that they make degrees very heterogeneous in terms of students' geographical origins. Panel (A) of Figure 2 plots the distribution of degrees by the share of students that have moved away from another province. Less than 1% of programs have no movers. On the contrary, 50% of degrees have more than 56% of students who are movers, as represented by the red line. Panel (B) offers a visual snapshot of the "melting pot" within degree programs, mapping each program's cultural composition through a scatter plot. The y-axis represents the share of students from provinces in the highest FLFP quartile, while the x-axis captures those from below-median FLFP areas. Notably, about one-third of degree programs are entirely filled with students from below-median FLFP provinces, clustered in the right tail. However, a large portion of programs display significant diversity, with representation from all geographical groups. A small minority—less than 5% of students—attend programs comprised solely of peers from above-median FLFP areas, as illustrated in Figure A.2.

FIGURE 2. Geographical composition of students within degrees

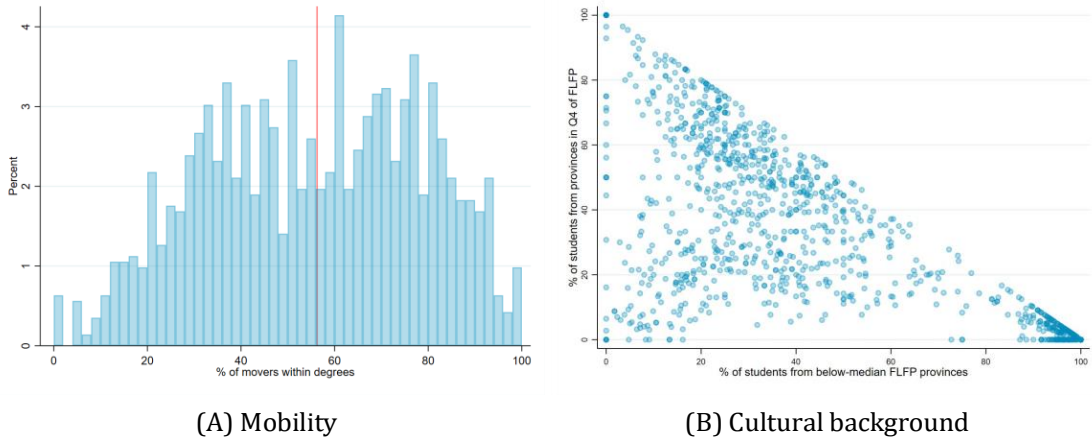

(A) Mobility                                        (B) Cultural background

Notes. Panel (A) represents degrees by the % of movers in 2016. One unit corresponds to a degree (N=1,572). The red line corresponds to the median % of movers across degrees. In Panel (B), each dot corresponds to a degree program (N=1,572). For each degree, the figure plots the share of students from provinces in the highest quartile of FLFP (y-axis), alongside the share of students from provinces with FLFP in the first or second quartiles (x-axis). Data refer to the enrollment cohort of 2016.

## 3.3   Size and Relevance of Peer Groups

A third crucial feature of this context makes it particularly well-suited for studying peer effects. Unlike prior studies that define peer groups broadly as all students within the same school cohort, I focus on a much narrower and possibly more relevant set: students enrolled in the same Master's degree cohort. The typical program is small, with a median (mean) size of 34 (46) students, as shown in Figure 3 (Panel A), which plots the distribution of programs' size. This is in a range between 4 and 410. Smaller peer

FIGURE 3. Degree size and gender composition

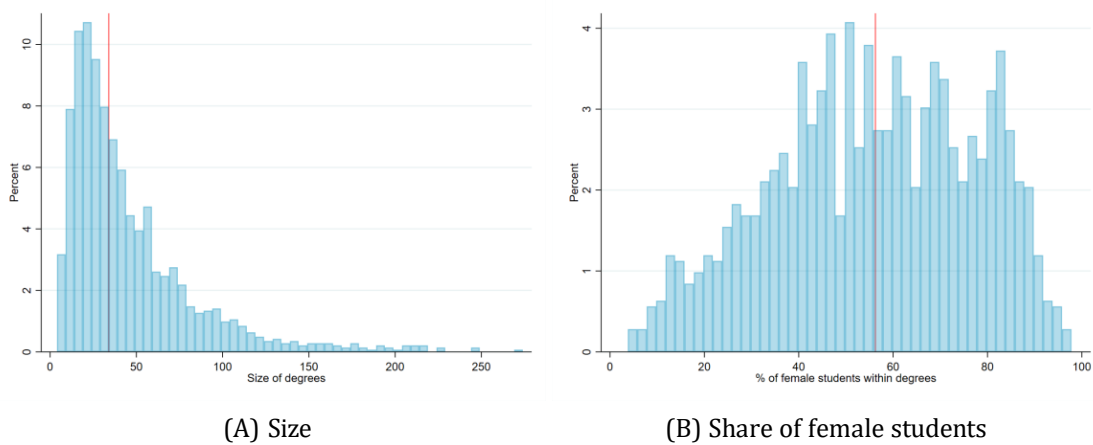

(A) Size                                  (B) Share of female students

Notes. Panel (A) represents the distribution of degrees by their size in 2016. The red line corresponds to the median size. Panel (B) represents the distribution of degrees by the share of female students. The red corresponds to the median share of female students. In both panels, one unit corresponds to a degree (N=1,572).

groups offer several advantages when studying peer effects. First, in smaller peer groups, it is very likely that students regularly interact with most of their peers and closely observe their behaviors. Moreover, the long duration of exposure—at least two years spent with the same cohort—intensifies these peer interactions. Section 10 provides direct evidence on the nature and frequence of such interactions, through a survey to a sample of prospective students. Second, smaller groups reduce the likelihood of endogenous sorting or self-selection into subgroups, a mechanism often present in larger settings (Carrell, Sacerdote, and West 2013). Third, the reduced size of degrees is crucial for identifying peer effects through cross-cohort variations in characteristics. In smaller programs, even minor idiosyncrasies within a cohort can lead to significant shifts in its composition. Conversely, as program sizes increase, the law of large numbers indicates that cohort compositions will tend to the average. I will discuss this point in greater detail in Section 5. Furthermore, note that 50% of a degree's course content and credit distribution is fixed, and that students are free to allocate only about 10% of their total credits (Ministerial Decree 270/2004). This structure ensures that students are consistently exposed to the same peers throughout their studies. Additionally, Panel (B) plots the distribution of degrees according to the share of female students, and descriptive statistics of degrees are presented in Table A.7.

# 4 Two Novel Facts About Early-Career Gender Gaps

## 4.1 Fact 1: the early-career gender earnings gap

Despite women achieving higher levels of human capital accumulation, as reflected in their higher college attendance and GPA, they fare significantly worse than men at the start of their careers. One year after graduation, equally productive women earn 11% less than their male counterparts from the same Master's program (Table 2)[9]. This gap is both statistically significant and economically meaningful: it represents €1,795 every year, on average. The earnings gap is primarily driven by differences in the intensive margin of labor supply: women are 5 percentage points less likely to be employed in full-time jobs and work 8% fewer hours per week compared to male students with similar academic performance[10]. The residual gap in hourly wages, however, is much smaller at 2.9%. These differences in labor supply are not attributable to geographic mobility and are only mildly related to differences in occupational and industry sorting by gender (Table A.9). Furthermore, while the results are presented at the mean, gender gaps are pervasive across types of programs and fields. This is illustrated in Figure 4, which plots binned degree-specific effects on monthly earnings and full-time employment for women against the corresponding effects for men. OLS estimates of the slope are presented, after re-weighting each degree for the share of students it represents across all years. Although there is a strong correlation between the average premiums for male and female students within each degree, women across all degrees are systematically employed in jobs with lower earnings (Panel A) and fewer weekly hours (Panel B). Moreover, these gaps remain stable and do not shrink within the first five years of entering the labor market. Overall, these findings provide new and comprehensive evidence of systematic gender differences in earnings and labor supply at the start of the career among highly skilled individuals in Italy. Previous research on gender gaps at labor market entry among highly skilled individuals had mainly focused on the U.S. and narrower groups (such as graduates from elite universities or specific fields). These studies found little or no gender differences in

---

9. This is consistent with estimates by Bovini, De Philippis, and Rizzica 2023, who use individual-level administrative data for all university graduates in Italy, linking education records from the Ministry of Education with social security data from 2011 to 2018.

10. Note, instead, that there are no differences in the extensive margin of labor supply (Table A.8).

TABLE 2. The gender earnings gap at labor market entry

| | (1) Log(monthly earnings) | (2) Log(weekly hours) | (3) Pr(fulltime) | (4) Log(wage) |
|---|---|---|---|---|
| Female | -0.113*** | -0.083*** | -0.051*** | -0.029*** |
| | (0.004) | (0.003) | (0.003) | (0.003) |
| GPA | ✓ | ✓ | ✓ | ✓ |
| Degree FEs | ✓ | ✓ | ✓ | ✓ |
| Cohort FEs | ✓ | ✓ | ✓ | ✓ |
| Observations | 127,153 | 127,153 | 127,153 | 127,153 |
| R-squared | 0.294 | 0.259 | 0.293 | 0.089 |

Notes: The table reports coefficients from regressions of graduates' labor market outcomes on a female dummy, after including degree and cohort fixed effects and controlling for GPA. The sample consists of female and male students who are employed one year post graduation. Standard errors are clustered at the degree level.

working hours at labor market entry (Cortes et al. 2023) and documented that such differences typically become significant a few years after entering the labor market (Bertrand, Goldin, and Katz 2010, Azmat and Ferrer 2017).

**Fertility and couple decisions**. Importantly, while the focus of a prominent body of literature has been on the gendered role of parenthood in explaining gender differences in labor supply[11], realized fertility can be ruled out as a major factor in this setting. In the sample, the average age of women is 24, with only a small fraction having children or being married or cohabiting (3.7% and 16.1%, respectively). Excluding these groups from the analysis does not alter estimates of the gender earnings gap, as shown in Table A.10. Moreover, analysis of newly collected expectations data from the the students' survey suggests that anticipated fertility is unlikely to be a major factor in women's overrepresentation in part-time jobs. On average, women in this sample expect to have their first child at age 31—well after they have entered the labor market—and only a minority foresee working part-time or exiting the workforce due to motherhood. A more comprehensive examination of this potential channel will be provided in Section 10.

---

11. Examples include: Bertrand, Goldin, and Katz 2010, Kleven, Landais, and Leite-Mariante, Forthcoming, Angelov, Johansson, and Lindahl 2016, Cortes and Pan 2023, Kleven 2024, Kleven, Landais, and Søgaard 2019, Bertrand, Goldin, and Katz 2010, Azmat and Ferrer 2017, Casarico and Lattanzio 2023, Altonji and Blank 1999)

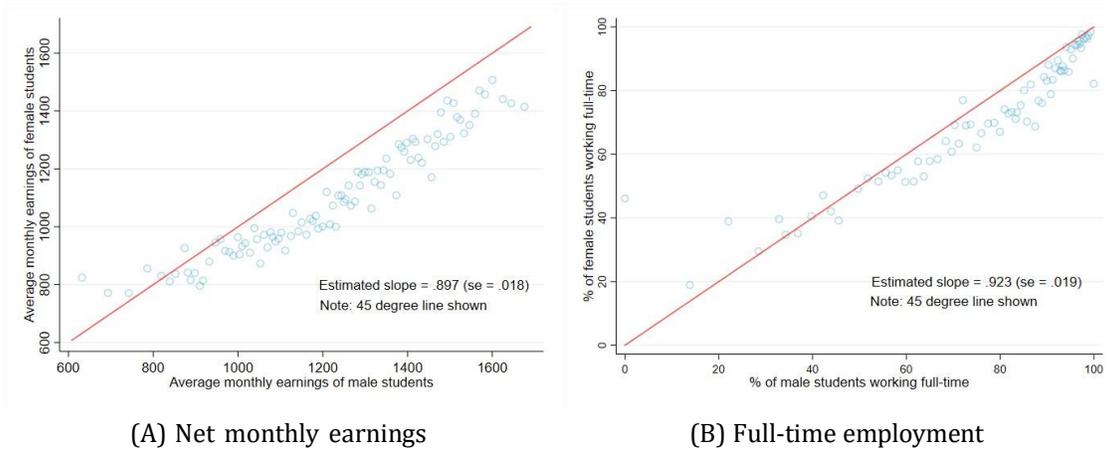

(A) Net monthly earnings        (B) Full-time employment

Notes: The figure shows binned scatter plots of estimated degree effects for female students against estimated firm effects for male students. Each degree is characterized by the average earnings (or full-time employment) of male students (x-axis) and the average earnings (or fulltime employment) of female students (y-axis), both computed across all years. The slope is estimated across degrees by OLS, after re-weighting each degree for the share of students it represents. In the regression, a unit corresponds to a degree. A 45-degree line is shown in red.

**Timing of job offer acceptances**. Previous research has highlighted that gender differences in job search behavior, such as women accepting job offers earlier due to higher risk aversion, play a significant role in explaining early-career earnings disparities (Cortes et al. 2023). However, these data do not support this explanation. In this study, both women and men start their job search at similar times, yet women actually accept job offers later than their male counterparts (Table A.2). Furthermore, the gender earnings gap does not appear to shrink over the course of the job search period.

## 4.2 Fact 2: Cultural Persistence

A second motivating fact is that women's labor market opportunities are shaped by the societal role models they were exposed to during childhood—specifically, the working behavior of other women. I document that the gender culture in a woman's province of origin has a lasting impact on her early career labor supply decisions—a phenomenon I refer to as *cultural persistence*. I substantiate this claim, relying on the epidemiological approach of Fernandez 2007 and re-adapting it to analyze granular within-country variations in cultural norms through the behavior of movers—individuals who work in a different province from where they were born[12]. To isolate the influence of culture from other local factors such markets and institutions, I examine the working behavior of female movers using information on their province of residence prior to entrance in college and their province of work. In this context, *movers* are defined as Italian-born individuals working in a province different from their birthplace. The role of early cultural exposure is captured by examining the labor supply of movers and linking it to various measures of gender culture in their birthplace, as outlined in subsection 3.1. The fundamental idea underlying this approach is straightforward: movers to the same local labor market share the same market conditions and institutional settings, but they may not necessarily share the same information, beliefs or preferences. According to theories of cultural transmission, one key mechanism through which cultural traits evolve is through learning from societal role models[13]. In the models of Fernandez 2013 and Fogli and Veldkamp 2011, one prominent channel leading to the formation of gender culture, besides maternal influences, is the observation of other women's work behavior, a process referred to as *oblique transmission* in the terminology of Cavalli-Sforza and Feldman 1981. A key prediction of these theories is that women's beliefs, broadly defined, will differ depending on the working behavior of women in their place of origin. I test this hypothesis in Table 3, which compares the labor supply of female movers depending on the quartile of FLFP in their province of origin, controlling for the province where they work, as well as degree and cohort fixed effects (Columns 2 and 5). The corresponding specification is:

$$(1) \qquad Y_{idcp} = \beta_0 + \alpha \times Q4FLFP_{idcp} + \theta_d + \alpha_c + \gamma_p + \left( \sum_{k=1}^{X} \beta_k x^k_{idcp} \right) + \epsilon_{idcp}$$

where i refers to a female graduate in the subsample of movers, in a given cohort c who has completed a Master's degree d and is employed in province p. Q4FLFP is an indicator variable that equals 1 if a graduate was born in a province with FLFP in the top quartile, and 0 if from the lowest quartile. $\theta_d$, $\alpha_c$, $\gamma_p$ are a set of degree, cohort and province of employment fixed effects. In some specifications, I add to the baseline model a set of covariates (GPA and fixed effects for parents' occupations). The results indicate that fe-male movers originating from high-FLFP provinces have significantly higher labor supply compared to those from low-FLFP provinces, even when they work in the same local labor market and graduate from the same Master's program. This difference is statistically significant and substantial, translating to a 7.6% increase in weekly hours worked and a 2.2 percentage point higher likelihood of full-time employment. This results in a 6.2% increase in earnings, equivalent to €802 annually. Moreover, estimates vary little when accounting for student's performance and parental background. Robustness checks confirm that these findings remain consistent across alternative measures of gender culture at the provincial level.

---

12. This approach is closely related to studies by Kleven 2024, Kerwin, Guryan, and Pan 2024 and Boelmann, Raute, and Schönberg 2023.

13. See, for instance, Bisin and Verdier 2000 and Cavalli-Sforza and Feldman 1981.

TABLE 3. Estimates of gender culture on women's labor supply at labor market entry

| | (1) | (2) | (3) | (4) | (5) | (6) |
|---|---|---|---|---|---|---|
| | Log(weekly hours) | | | Pr(fulltime) | | |
| Q4 vs. Q1 FLFP | 0.072*** | 0.076*** | 0.074*** | 0.025*** | 0.022*** | 0.021*** |
| | (0.011) | (0.011) | (0.011) | (0.008) | (0.008) | (0.008) |
| Province of job FEs | | ✓ | ✓ | | ✓ | ✓ |
| GPA | | | ✓ | | | ✓ |
| Parental background | | | ✓ | | | ✓ |
| Degree FEs | ✓ | ✓ | ✓ | ✓ | ✓ | ✓ |
| Cohort FEs | ✓ | ✓ | ✓ | ✓ | ✓ | ✓ |
| Observations | 15,838 | 15,835 | 15,835 | 15,838 | 15,835 | 15,835 |
| Nb. of degrees | 1,218 | 1,218 | 1,218 | 1,218 | 1,218 | 1,218 |
| R-squared | 0.293 | 0.302 | 0.304 | 0.331 | 0.348 | 0.350 |

Notes: The table reports coefficients from separate regressions of women's labor market outcomes on a dummy variable indicating whether the student originates from a province with FLFP in the highest vs. lowest quartile. All regressions include controls for degree and cohort fixed effects. Controls for parental background include: FEs for education titles of mother and father (10 classes), FEs for occupations of mother and father (12 classes). The sample consists of female *movers*, defined as women working in a different province from their birth province, who are employed one year post-graduation. Standard errors are clustered at the degree level.

While these results are striking, one concern in attributing these differences to gender culture is that movers born in low-FLFP or high-FLFP provinces might differ in other dimensions that impact their labor market outcomes. To investigate the importance of such concerns, Table A.11 compares the ability, educational histories, and socio-economic backgrounds of female movers based on the FLFP of their birthplace (top versus bottom quartile). Specifically, it provides predictions from the empirical model 1 for these two groups of women. A key advantage of the data is its rich set of pre-Master's degree characteristics, which allows to assess potential confounders. The table shows that, conditional on attending the same degree program, movers from high-FLFP provinces are similar to those from low-FLFP provinces in terms of ability and and socio-economic background. In particular, the two groups are equivalent in parental education, while there are some differences in parental occupations: mothers from low-FLFP areas are less likely to hold low-SES jobs and more likely to hold medium-SES jobs than those from high-FLFP areas, though there are no significant differences in their representation in high-SES jobs. A notable exception is the labor market participation of mothers[14], which strongly correlates with local gender culture. Despite these differences in maternal role models, they do not mediate the relationship between local gender culture and women's labor supply. As shown in Table, women's labor supply has little to no correlation with their mother's employment status. Furthermore, women from low-FLFP areas appear to be better selected in terms of their educational histories: they are more likely to have attended general high-school tracks, particularly in science and humanities, compared to their peers from high-FLFP areas. Therefore, evidence presented here does not suggest that female students from low-FLFP areas have characteristics that explain why they fare significantly worse in the labor market than women from high-FLFP areas. These findings thus reduce concerns about differential selection of movers based on birth province and support the validity of the epidemiological approach. A second concern arises from the the fact that other local factors, such as economic activity, labor market conditions and measures of social capital, correlate with spatial disparities in FLFP, as documented in Acciari, Polo, and Violante 2021. Should these factors influence the beliefs and preferences of individuals, the estimated relationship would not purely capture the effects of local gender culture. To explore this possibility, I present an epidemiological analysis focusing on the subsample of male students in Table A.12. While there is some positive relationship between the gender culture in males' province of origin and their working hours, the magnitude of the coefficients is less than half of those observed for women.

---

14. This refers to the student's response about whether their mother is currently employed at the time of the survey.

This main goal of this subsection has been to provide comprehensive evidence that, within degrees, labor supply differs systematically among women with similar academic credentials based on the gender culture of their birthplaces. By focusing on a specific population—young female graduates entering the labor market—and leveraging a setting where admission rules are typically based on academic performance, this approach minimizes potential confounders and differential selection biases related to birth province. Several factors may explain these observed differences, including differences in preferences, information disparities, as well as employers' taste. While this section remains agnostic about the precise mechanisms behind *cultural persistence*, Section 10 will provide direct evidence of significant asymmetries in specific beliefs shaped by early cultural exposure.

How are gender norms transmitted? What happens to women's labor supply decisions when they get exposed to peers raised in a different cultural environment? This setting, together with unique characteristics of the data, allow to reproduce this experiment at a large scale and explore the mechanisms of cultural change. Section 5 presents the empirical framework used to identify the causal impact of peers' background characteristics on women's labor market outcomes.

## 5  Identification Strategy and Empirical Model

**Challenges and intuition**. The identification of peer effects is notoriously challenging. The main threat to their identification relates to *selection*, or endogenous peer formation. In this setting, peer groups are not formed at random, since individuals choose their majors and universities. As a result, the characteristics of the peers they are exposed to are likely correlated with their unobserved characteristics that drive their choice of a specific peer group in the first place and that plausibly affect their success in the labor market, leading to *correlated effects* in the Manski terminology (Manski 1993). In the absence of randomization of students into peer groups, which is difficult to implement on a large scale, my identification strategy overcomes the selection issue by leveraging within-degree variation in the geographical origins of the peers to which consecutive cohorts of students are exposed. This approach has been first proposed by Hoxby 2000 to assess the impact of classmates gender and race on students' educational outcomes, and has become widely used for studying peer effects in education[15]. A key assumption in this strategy is that changes in the geographical origins of students enrolling in a given program over consecutive years are *as good as random*, i.e. no variables simultaneously influence the characteristics of the students and the outcomes of interest. Several tests, detailed in the next section, are conducted to verify whether this crucial identifying assumption holds in practice. Implementing this approach requires access to longitudinal administrative data, which is essential for the analysis. My dataset is particularly valuable because it includes a comprehensive panel of Master's programs - covering nearly all students in the country - across multiple enrollment cohorts from 2012 to 2016. This level of coverage is unique in the context of university education, as previous datasets were limited to small, selected samples, often from a single university.

## 5.1   The empirical model

The associated empirical model is:

$$(2) \quad Y_{idc} = \theta_d + \alpha_c + \gamma FLFP_{idc} + \delta^{FP} \overline{FLFP}^{FP}_{-i,dc} + \delta^{MP} \overline{FLFP}^{MP}_{-i,dc} + \left( \sum_{k=1}^{K} \beta_k x_{idc}^k \right) + \epsilon_{idc}$$

where $i$ refers to a student who attended degree $d$ in cohort $c$. The main outcomes of interest are monthly earnings, weekly hours of work, an indicator variable for full-time employment and hourly wages. Additionally, the empirical model will also be used to assess the effects of peers' characteristics on a number of other job characteristics, such as the type of occupation and industry, location, or contract type. The empirical model is estimated separately in the two subsamples of female and male students, allowing for gender-specific peer effects. The main parameters of interest are $\delta_{FP}$ and $\delta_{MP}$, i.e. the treatment effects of exposing a student to a set of peers that are coming from places where the FLFP is, on average, one percentage point higher. These estimates are causal under the assumption that within degree-across cohort variation in the geographical origins of students is as good as random. I provide evidence in favor of this assumption. The associated variables are the sample moments of the leave-one-out distribution of the FLFP in the province of origin of students who belong to a specific gender, degree and cohort:

$$\overline{FLFP}^{FP}_{-i,dc} = \frac{\sum_{j \neq i} FLFP_{jdc}}{n^F_{dc}-1} \text{ if female=1;} \qquad \overline{FLFP}^{MP}_{-i,dc} = \frac{\sum_{j} FLFP_{jdc}}{n^M_{dc}} \text{ if female=1;}$$

$$\overline{FLFP}^{FP}_{-i,dc} = \frac{\sum_{j} FLFP_{jdc}}{n^F_{dc}} \text{ if female=0;} \qquad \overline{FLFP}^{MP}_{-i,dc} = \frac{\sum_{j \neq i} FLFP_{jdc}}{n^M_{dc}-1} \text{ if female=0;}$$

Since the leave-one-out strategy introduces a mechanical negative correlation between the FLFP in the own province of origin of a student and the average FLFP in the provinces of her same-sex peers (Angrist 2014), I also condition for the FLFP in a student's own province of origin, $FLFP_{idc}$. Alternatively, in a sensitivity analysis, I condition for province of origin FEs instead of the FLFP in the province of origin. Note that, the linear-in-means model assumes that what matters in a peer group is the average peers' characteristic, regardless of which peers' allocation is leading to it. However, as pointed out by Boucher et al. 2024, in many real-life situations peer effects are not linear in means. Additionally, the most relevant policy implications of peer effects rely on non-linear models in which not all individuals affect and are affected by their peers in the same way. In Section, I will provide an analysis of non-linearities and heterogeneity.

Taking advantage of the panel nature of the data source (repeated observations on Master's programs), the inclusion of degree fixed effects $\theta_d$ allows to account for degree characteristics that are constant across cohorts, for example whether the program tends to be attended by students with specific set of background characteristics. More specifically, it accounts for time-invariant unobserved determinants of earnings of students of a given gender who graduate from a given program. The baseline specification also include enrollment cohort fixed effects $\alpha_c$, to account for confounding factors affecting all individuals within the same cohort. Finally, $\epsilon_{idc}$ is the error term, which is composed of a degree-specific random element and an individual random element. Standard errors are clustered at the degree level to account for potential correlation in students' outcomes within degrees.

**Discussion**. OLS estimates of $\delta^{FP}$ and $\delta^{MP}$ are unbiased if $\overline{FLFP}^{FP}_{-i,mc}$ and $\overline{FLFP}^{MP}_{-i,mc}$ are uncorrelated with time-varying unobserved determinants of students' earnings, conditional on degree and cohort FEs. Therefore, for equations (1) to yield valid causal estimates of these paramaters, the key identifying assumption is that cross-cohort changes in students' geographical origins are random within degrees. Conceptually, this assumption is likely to hold given the rules governing university admission in Italy. In fact, a majority of Master's degrees are selective (more than 55% according to estimates of the Ministry of Education), meaning that admission is limited to a fixed number of students. Typically, admission rules are decided by universities and involve an entrance exam, a standardized test, or consideration of average grades from the Bachelor's degree. Hence, in selective programs, variation in $\overline{FLFP}^{FP}_{-i,mc}$ and $\overline{FLFP}^{MP}_{-i,mc}$ reflect year-to-year variations in the geographic origins of students whose admission scores are high enough to be admitted into a program. This design assumes that such variations are idiosyncratic, conditional on the student's bachelor's GPA. In contrast, in non-selective programs, year-to-year changes in the geographical origins of students stem from shifts in the applicant pool's composition. Other sources of cross-cohort variations in students' composition can include changes in admission policies, such as adjustments in the size of programs or in the admission requirements. In the absence of granular data on admission policies, in some sensitivity analyses I will use a data-driven approach to detect possibly non-random changes in size and/or composition and exclude this subset of degrees from the analysis. The objective of the next section is to describe the sources of the identifying variation and to provide credible evidence that this variation is idiosyncratic.

## 5.2   Validity of the empirical strategy

**Threat to identification.** One critical concern in identifying peer effects is the presence of *correlated effects*, in the Manski terminology (Manski 1993). In essence, similarities in economic outcomes among individuals within a peer group are likely to stem from shared individual characteristics or common shocks, rather than from social influence alone. In this context, this translates into the possibility that cross-cohort changes in students' geographical origins within master's programs can correlate with time-varying unobserved determinants of students' labor market outcomes. For example, this could occur if labor market trends in a specific region influence the applicant pool for programs in that region, or if shifts in the student composition within a program impact the selection of new students. If this happens, then $\overline{FLFP}^{FP}_{-i,mc}$ and $\overline{FLFP}^{MP}_{-i,mc}$ are correlated with time-varying determinants of outcomes $\epsilon_{idc}$, leading to biases in $\delta^{FP}$ and $\delta^{MP}$. For this identification strategy to effectively capture social influence, it is crucial that these cross-cohort fluctuations are effectively random. The objective of this section is to examine the validity of this identifying assumption through a series of checks.

**Balancing tests for cohort composition.** One empirical test of this assumption is to verify that there is no selection, based on observables, into peer groups. Precisely, while sorting of students into programs based on time-invariant characteristics - such as the average peers' composition - is admitted, I need to rule out that students systematically sort into programs based on the specific composition of their cohort. To assess the plausibility of the key identifying assumption that time-variant and unobservable factors are not driving the results, I test whether there is systematic selection based on a wide range of observable student characteristics. Specifically, I perform a battery of balancing checks in which I test whether the peer composition in a Master's cohort is systematically related to a large vector of high-quality measures of student background characteristics observable in the institutional data. For these placebo tests, I pick as characteristics pre-determined covariates, that cannot be causally affected by peers but that might be correlated with unobserved characteristics of other students enrolling in the same programs. These characteristics include academic performance in previous education (e.g., Bachelor's degree or high school grades), and indicators of family socio-economic status, derived from detailed information about the occupations and educational backgrounds of both parents. Tables B and B present the results of these placebo checks on the subsamples of female and male students, respectively. They report OLS estimates of $\delta^{FP}$ and $\delta^{MP}$ from equation 2. Each column corresponds to a different regression, where the dependent variable is a different pre-determined covariate. Results indicate that none of the estimated correlations appear to be significantly different from zero in the model, indicating that the exposure to peers from egalitarian provinces in the Master is unrelated to outcomes measured before entry in the Master.

I also use a second approach to test whether students in a cohort with a different peers' geographical composition have characteristics that are associated with different labor market outcomes. I proceed in two steps. I first run separate regressions for female and male students, where I predict their labor market outcomes - e.g. their labor supply at the extensive and intensive margin as well as their earnings - based on all observable pre-determined covariates (age, type of high-school (10 classes), high-school grade, BSc grade, dummy for mother's/father's citizenship (Italian vs. not), FEs for education titles of mother and father (10 classes), FEs for occupations of mother and father (12 classes)). Second, I use the predicted outcomes as dependent variables in 2 to check balancedness with respect peers' composition. The advantage of this approach is that it quantifies any non-balancedness of student characteristics. Table 4 summarizes these results, separately for female and male students. The predicted labor market outcomes for the two sub-samples appear balanced, providing strong evidence that cross-cohort changes in peers' geographical composition are not related to covariates that explain changes in students' labor market outcomes.

Finally, I perform a third test. I use all observable pre-determined covariates to predict the two treatment variables - $\overline{FLFP}^{FP}_{-i,mc}$ and $\overline{FLFP}^{MP}_{-i,mc}$ - in two regressions that include cohort and degree fixed effects (FEs). I then perform an F-test for the joint significance of all regressors. The two F-test results are 0.381 and 0.221, respectively, indicating that we cannot reject the hypothesis that the regressors jointly do not explain the treatment variables. Overall, I take the results from these two sets of tests as encouraging indication that the treatment variable is unlikely to be correlated with other time-varying unobservable individual determinants of labor market outcomes. In fact, drawing from Altonji and Blank 1999, we can reasonably infer that the degree of selection on observable characteristics serves as a reliable indicator of the degree of selection on unobservables.

**Identifying variation**. Implementing this empirical strategy requires that there is sufficient variation in the geographical composition of peers across cohorts within a master's program. This variation is necessary to obtain precise estimates of the peer effects I aim to identify. Table 5 reports descriptive statistics for the average FLFP in the province of origin of peers within a degree, before and after removing degree and cohort fixed effects. In the cross-section, the standard deviation of this measure is 8.50 percentage points among female peers and 8.59 percentage points among male peers, and is reduced to 1.86 and 2.13 after netting out degree and cohort fixed effects. The same descriptive statistics for the other measures of gender culture are shown in A.15. Most of the variation in students'

**Panel A. Female sample**

|  | (1) Pr(employed) | (2) Log(monthly earnings) | (3) Log(weekly hours) | (4) Pr(fulltime) |
|---|---|---|---|---|
| $\hat{\delta}^{FP}$ | -0.000 (0.001) | -0.000 (0.002) | -0.001 (0.001) | -0.000 (0.001) |
| $\hat{\delta}^{MP}$ | -0.000 (0.001) | -0.000 (0.002) | 0.002 (0.001) | 0.001 (0.001) |
| Degree FE | ✓ | ✓ | ✓ | ✓ |
| Cohort FE | ✓ | ✓ | ✓ | ✓ |
| Observations | 146,476 | 146,476 | 146,476 | 146,476 |
| R-squared | 0.119 | 0.246 | 0.239 | 0.273 |

**Panel B. Male sample**

|  | (1) Pr(employed) | (2) Log(monthly earnings) | (3) Log(weekly hours) | (4) Pr(fulltime) |
|---|---|---|---|---|
| $\hat{\delta}^{FP}$ | -0.000 (0.001) | -0.001 (0.001) | -0.001 (0.001) | -0.001 (0.001) |
| $\hat{\delta}^{MP}$ | 0.001 (0.001) | 0.001 (0.002) | 0.002 (0.001) | 0.001 (0.001) |
| Degree FE | ✓ | ✓ | ✓ | ✓ |
| Cohort FE | ✓ | ✓ | ✓ | ✓ |
| Observations | 106,448 | 106,448 | 106,448 | 106,448 |
| R-squared | 0.204 | 0.259 | 0.316 | 0.324 |

Notes: OLS estimates of a regression of predicted labor market outcomes on: the average FLFP in the provinces of origin of female and male peers and the FLFP in the own province of origin. Regressions include cohort and degree fixed effects. The estimating equation is 2 in the main text. All regressors are standardised. Standard errors are clustered at degree level. Labor market outcomes are predicted separately for female and male students based on regressions of labor market outcomes on a set of pre-determined invididual covariates: age, type of high-school (10 classes), high-school grade, BSc grade, dummy for mother's/father's citizenship (Italian vs. not), FEs for education titles of mother and father (10 classes), FEs for occupations of mother and father (12 classes).

characteristics occurs across different degrees, indicating that there is sorting of students, as expected. However, there is some variation within a degree over time. Between one-fourth and one-fifth of the total variation in peers' gender culture remains unexplained after accounting for fixed effects: I rely on this variation to estimate peer effects, and all estimates are precisely estimated. Interestingly, the magnitude of this residual variation aligns with what we would expect from random fluctuations. To explore this further, I conducted 500 simulations in which students were randomly assigned to degrees and cohorts - think about it, using a distribution of FLFP that matches the mean and standard deviation of the actual sample distribution. Across these simulations, the residual standard deviation of the average FLFP in the province of origin of peers is centered around 1.57 percentage points, with a range of 1.53 to 1.62 percentage points, consistent with the values observed in the actual sample. As a further randomization check, I inspect whether

TABLE 5. Raw and Residual Variation of Peers' Gender Culture

|  | Mean | SD | Min | Max |
|---|---|---|---|---|
| **A: Avg FLFP in province of origin of female peers** | | | | |
| Raw cohort variable | 49.65 | 8.50 | 29.87 | 66.66 |
| Residuals: net of degree and cohort FEs | 0.00 | 1.86 | -13.80 | 10.79 |
| | | | | |
| **B: Avg FLFP in province of origin of male peers** | | | | |
| Raw cohort variable | 49.72 | 8.59 | 27.33 | 66.66 |
| Residuals: net of degree and cohort FEs | 0.00 | 2.13 | -16.80 | 14.51 |

Notes: The table reports descriptive statistics for the average FLFP in the province of origin of female (Panel A) and male (Panel B) students within degrees, before and after removing degree and cohort fixed effects. The unit of observation is a degree-cohort pair, leading to a total of 7,160 observations.

the variation in students' geographical composition is consistent with variation that we would expect with natural random fluctuations. Figure 5 plots the average FLFP in peers' provinces, with separate panels for female peers (Panel a) and male peers (Panel b), after residualizing on degree and cohort fixed effects. Figure A.3 provide the same graphical evidence using all other measures of gender culture used in the paper. Deviations from the average students' composition closely follow the normal distribution, which is plotted for comparison. The shape of the distribution further supports the idea that student's geographical composition is as good as random, conditional on the included controls. Another graphical illustration of the identifying variation is shown in Figure A.4, which plots the time series of the average FLFP of female peers across the years, for a randomly picked program within each decile of program's size.

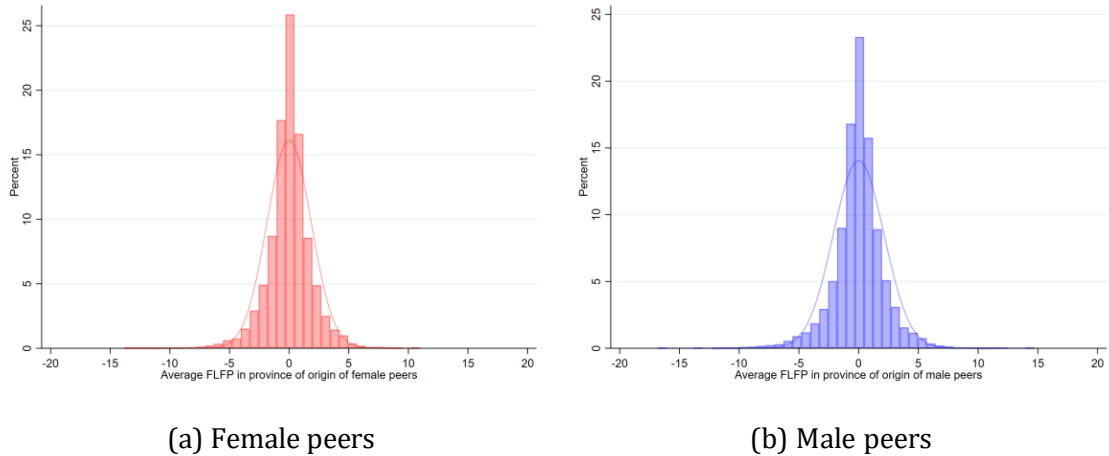

(a) Female peers                                (b) Male peers

Notes: The figure plots the distribution of residuals from a OLS regression of the average FLFP in the province of origin of female (Panel a) or male students (Panel b) on cohort and degree fixed effects. One observation corresponds to a degree-cohort pair. Histograms are presented by bins of 0.75. The normal distribution is plotted for comparison.

**Identifying variation and program's size**. If the variations in students' characteristics across cohorts within a degree program were random, the law of large numbers would predict that the average characteristics of peers in a cohort would converge to the true value of the program as the program size increases. This leads to a testable prediction: the magnitude of cross-cohort changes in students' geographical origins within a Master's program should become smaller as the program size grows. This is shown in Figure 6, which plots year-to-year changes in students' origins for degrees in the lowest and highest quintile of size. The lowest quintile corresponds to degrees with an average size below 22 students, while the highest quintile includes degrees with an average size between 70 and 413 students. It is also further tested in Table A.16, which provides descriptive statistics for the main variable defining peers' gender culture, in different groups of degrees categorized by the quintiles of their size. It is clear that while the raw standard deviation of the variable shows little change across quintiles, the residual variation decreases significantly as program size increases, consistent with random fluctuations. These findings underscore a key advantage of this dataset compared to those used in previous studies on peer effects, that examined entire schools as peer groups. Specifically, the granular data on program composition allows for the analysis of peer groups with small cohort sizes (median=34, mean=47), where cross-cohort changes in peers' characteristics are likely driven by ran-dom fluctuations.

FIGURE 6. Year-to-Year Variation in Students' Geographical Origins by Program Size

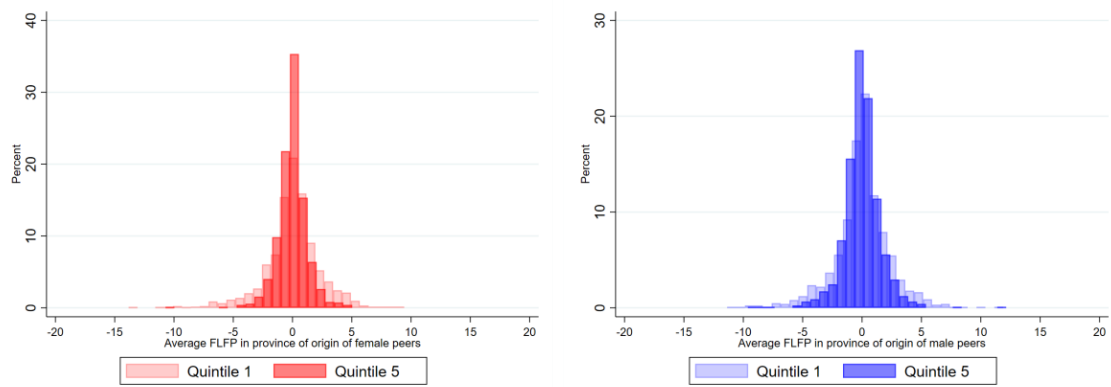

(a) Female peers                    (b) Male peers

Notes: The figure plots the distribution of residuals from a OLS regression of the average FLFP in the province of origin of female (Panel a) or male students (Panel b) on cohort and degree fixed effects. The distributions are shown separately for degree programs in the first and highest quintiles of size. Degree programs are divided into quintiles based on their average size across five cohorts: the first quintile includes degrees with fewer than 21 students, while the fifth quintile includes degrees with 70 to 413 students. Each observation represents a degree-cohort pair. Histograms are presented by bins of 0.75.

**Other checks.** The evidence provided in this section supports the hypothesis that year-to-year changes in the geographical composition of students are random within degrees. However, it does not entirely rule out the influence of simultaneous institutional factors, such as fluctuations in the local labor market, that might affect outcomes across different student cohorts.To address these potential concerns, I undertake several robustness checks. First, I extend the baseline model by incorporating either (i) degree-specific linear time trends or (ii) region-specific linear time trends. Second, I perform sensitivity analyses excluding degree programs that might exhibit non-random changes in size and/or composition. This includes programs with observable trends in size or geographical composition, as well as those with changes in the average or standard deviation of students' academic performance. These results are included as part of the robustness checks.

# 6  Baseline Estimates of Peer Effects

## 6.1  Estimation Results

This section presents the baseline estimates of gender-specific peer effects $\delta^{FP}$ and $\delta^{MP}$, derived from the empirical model of equation 2. These estimates come from specifications that condition on degree and cohort fixed effects, as well as the FLFP in the own province of origin of a student, and correspond most closely to the hypothetical experiment described in Section . The analysis is first conducted on the sample of female students, and is then extended to the sample of male students as a placebo.

**Estimates of peer effects on female earnings and labor supply**. Estimates of the empirical model on the sample of female students are presented in Table 6. The outcome variables are monthly earnings, weekly hours worked and hourly wages, all in logarithmic forms, and an indicator variable of fulltime employment. Regressors are standardised. Results indicate that women who study in cohorts where female classmates are born in places with higher FLFP increase their labor supply along the intensive margin, both through higher take-up of full-time jobs and increases in weekly hours worked (Columns 2 and 3). The magnitude of this effect is large: a one standard deviation increase in $\overline{FLFP}^{FP}_{-i,mc}$ (8.50 percentage points) is associated with a 3.3% increase in weekly hours and in a 1.9 percentage points increase in the likelihood of fulltime employment one year after graduation, a 2.5% increase relative to the mean. This translates into a 3.7% increase in their monthly earnings. These estimated peer effects are economically significant, comparing to $33\% - 40\%$ of the size of the gender differences in the same outcomes and accounting for $45\% - 76\%$ of the gap between women born in low-FLFP and high-FLFP provinces. These findings also highlight the presence of gender-specific peer effects for women. Specifically, the positive effects on earnings and hours worked are entirely attributable to variations in the geographical origins of female peers. In contrast, the estimates for $\delta^{MP}$ are zero across all outcomes, suggesting that exposure to male classmates from areas with higher FLFP does not affect women's earnings or labor supply decisions. There are several reasons why peers' gender might matter in this context. One key reason involves the mechanisms generating peer effects. Earlier, I documented a much stronger relationship between female outcomes in the labor market and gender norms in their province of origin, whereas men's outcomes showed little to no relationship with these cultural measures. A natural implication is that increasing the share of women from high-FLFP areas within a cohort systematically raises the representation of women who hold beliefs and preferences associated with acceptances of full-time jobs. This association is much weaker among men. Consequently, if peer effects operate through mechanisms such as (i) conformism or (ii) social learning, it is reasonable to expect little or no influence from male peers in this setting. However, opposite-sex peers might influence women's labor market decisions through other channels, especially in contexts where couples are likely to form (Bursztyn, Fujiwara, and Pallais 2017). For instance, men raised in more egalitarian environments might have different expectations regarding their partners' work behavior, as shown in Fernández, Fogli, and Olivetti 2004. If students pair up with classmates, this could be a significant channel through which the gender culture of men might impact women's choices. Yet, this potential explanation is not supported by these findings or by evidence from the original survey. As I will discuss in Section 10, the share of couples formed among students in the same program is low. Another possible explanation for why peer gender matters relates to the structure of social networks. Previous studies in sociology have documented that the quality and quantity of interactions differ between same-sex and opposite-sex peers (see, for instance, **cook**). Homophily—the tendency for peer groups to be segregated by gender—can be critical in determining how information spreads (Currarini, Jackson, and Pin 2009). To explore the relevance of this channel, I collected data on network structure through the original survey. The results will be presented in Section 10.

**Estimates of peer effects on female occupational choices**. Table 7 presents estimates of peer effects on the types of occupations and industries women enter one year after graduation (Columns 1-4). Occupations and industries are categorized based on their average monthly earnings and the share of full-time employment. The dataset includes 20 occupations and 21 industries. An occupation or industry is classified as "high-earnings" or "high-fulltime" if it ranks above the median in these distributions. These four indicators are used as outcome variables. The results indicate that exposure to female classmates from high-FLFP areas also affects women's occupational choices (Columns 1-2). Specifically, a one standard deviation increase in $\overline{\overline{FLFP}}^{FP}_{-i,mc}$ is associated with a 1.7 percentage point

TABLE 6. Estimates of peer effects on earnings and labor supply - Female sample

| | (1) Log(monthly earnings) | (2) Log(weekly hours) | (3) Pr(fulltime) | (4) Log(hourly wage) |
|---|---|---|---|---|
| $\hat{\delta}^{FP}$ | 0.037*** | 0.033*** | 0.019** | 0.003 |
| | (0.013) | (0.012) | (0.009) | (0.012) |
| $\hat{\delta}^{MP}$ | -0.000 | 0.001 | -0.002 | -0.002 |
| | (0.010) | (0.009) | (0.007) | (0.010) |
| Degree FE | ✓ | ✓ | ✓ | ✓ |
| Cohort FE | ✓ | ✓ | ✓ | ✓ |
| Observations | 69,645 | 69,645 | 69,645 | 69,645 |
| R-squared | 0.287 | 0.246 | 0.280 | 0.100 |

Notes: OLS estimates of a regression of women's earnings and labor supply one year after graduation on: the average FLFP in the provinces of origin of female and male peers and the FLFP in the own province of origin. Regressions include cohort and degree fixed effects. All the estimates are done on the sample of women who are employed one year after graduation and with non-missing information on these variables. Standard errors clustered at degree level. All regressors are standardised.

increase in the likelihood of choosing a high-earnings occupation and a 1.5 percentage point increase in the likelihood of selecting an occupation with a high share of full-time jobs, representing a 4.6% and 2.9% increase relative to the mean, respectively. As highlighted in the previous paragraph, these peer effects are gender-specific. To assess the role of occupational changes in the rise of women's labor supply, I re-estimated the model for weekly hours worked, including occupation and industry fixed effects. Although the estimated coefficient for $\delta_{FP}$ is reduced by about one-third, the coefficient on weekly hours remains large and statistically significant. This suggests that changes in occupations account for only part of the increase in women's labor supply. This finding aligns with earlier evidence showing that occupational differences between men and women explain less than a third of the total gap in labor supply.

**Estimates of peer effects on other job characteristics.** While exposure to female peers from high-FLFP provinces affects women's labor supply, earnings, and occupational choices, it does not impact sorting along other observable dimensions, such as employers' characteristics. For example, hourly wages are not affected (Table 6, Column 4), and there

TABLE 7. Estimates of peer effects on occupations and industries - Female sample

| | (1) | (2) | (3) | (4) | (5) |
|---|---|---|---|---|---|
| | Occupation | | Industry | | |
| | High-earn | High-fulltime | High-earn | High-fulltime | Log(weekly hours) |
| $\hat{\delta}^{FP}$ | 0.018** | 0.016* | 0.014 | 0.008 | 0.023** |
| | (0.009) | (0.009) | (0.010) | (0.009) | (0.011) |
| $\hat{\delta}^{MP}$ | -0.004 | -0.005 | -0.003 | -0.009 | -0.000 |
| | (0.006) | (0.007) | (0.007) | (0.006) | (0.009) |
| Occ. & ind. FE | | | | | ✓ |
| Degree FE | ✓ | ✓ | ✓ | ✓ | ✓ |
| Cohort FE | ✓ | ✓ | ✓ | ✓ | ✓ |
| Observations | 68,216 | 68,216 | 68,419 | 68,419 | 69,645 |
| R-squared | 0.361 | 0.466 | 0.272 | 0.398 | 0.349 |

Notes: OLS estimates of regressions of types of occupations and industries one year after graduation on: the average FLFP in the provinces of origin of female and male peers and the FLFP in the own province of origin. The dependent variables in Columns (1) and (3) are constructed from the distribution of earnings across occupations and industries, respectively. Specifically, indicators of high-earning occupations (industries) are based on whether an occupation (industry) pays above-median earnings. The dependent variables in Columns (2) and (4) are constructed from the distribution of fulltime jobs across occupations and industries, respectively. Specifically, indicators of high-fulltime occupations (industries) are based on whether an occupation (industry) has above-median shares of fulltime jobs. Regressions include cohort and degree fixed effects. In Column 5, I add occupation FEs (20 classes) and industry FEs (21 classes). All the estimates are done on the sample of women who are employed one year after graduation and with non-missing information on these variables. Standard errors clustered at degree level. All regressors are standardised.

is no significant effect on the industry in which women are employed. Table A.17 presents estimates of the empirical model for other job characteristics observed in the data, such as whether the employer is in the public or private sector or the type of employment contract (permanent, no contract and self-employment). None of these variables show any influence from peer exposure.

**Effect of peers on male outcomes**. In Table 8, I repeat the analysis on the male sample as a placebo test. The rationale here is that if the FLFP (and the other measures used in this paper) in a student's province of origin accurately reflects a set of beliefs and preferences that are gender-specific, we should not expect the gender culture of peers to have a direct impact on men's labor supply and earnings—just as men's decisions are not influenced by these measures in their own province of origin. However, indirect effects, e.g. in the form of spillovers, could still occurr, for instance, if some men feel pressure due to women's rising aspirations. Results on male students indicate that exposure to female or male peers from high-FLFP provinces has no impact on men's weekly hours and the likelihood of fulltime employment. Peers of both genders have only small positive effects on men's hourly wages.

TABLE 8. Estimates of peer effects on earnings and labor supply - Male sample

| | (1) Log(monthly earnings) | (2) Log(weekly hours) | (3) Pr(fulltime) | (4) Log(hourly wage) |
|---|---|---|---|---|
| $\hat{\delta}^{FP}$ | 0.013 | -0.000 | -0.001 | 0.014* |
| | (0.008) | (0.008) | (0.006) | (0.008) |
| $\hat{\delta}^{MP}$ | 0.013 | -0.005 | 0.004 | 0.018* |
| | (0.011) | (0.010) | (0.008) | (0.010) |
| Degree FE | ✓ | ✓ | ✓ | ✓ |
| Cohort FE | ✓ | ✓ | ✓ | ✓ |
| Observations | 57,476 | 57,476 | 57,476 | 57,476 |
| R-squared | 0.246 | 0.233 | 0.270 | 0.107 |

Notes: OLS estimates of a regression of men's earnings and labor supply one year after graduation on: the average FLFP in the provinces of origin of female and male peers and the FLFP in the own province of origin. Regressions include cohort and degree fixed effects. All the estimates are done on the sample of men who are employed one year after graduation and with non-missing information on these variables. Standard errors clustered at degree level. All regressors are standardised.

# 7 Robustness and Validation of the Design

This section has two primary goals. The first is to assess the validity of the key identifying assumption - that cross-cohort changes in students' geographical origins are random within degrees - using a series of tests that address different forms of selection and the presence of *correlated effects*. Once these concerns about the identification of peer effects are mitigated, the second objective is to investigate whether the estimated effects can be attributed to peers' gender culture or other characteristics. Additionally, I re-estimate the model using alternative specifications to ensure that the results are robust and not overly dependent on specific assumptions regarding the empirical model and the clustering of standard errors.

## 7.1 Validation of the design

To organize the analysis, consider that the unobserved determinant of students' earnings, represented by $\epsilon_{idc}$ in equation 2, is composed of two elements: a degree-specific random element, $v_{dc}$, which reflects time-varying inputs at the degree level, and an individual random element, $u_{idc}$. For the identification strategy to be valid, changes in $\overline{FLFP}_{-i,mc}^{FP}$ and $\overline{FLFP}_{-i,mc}^{MP}$ must be uncorrelated with both $u_{idc}$ and $v_{dc}$. In Section, results from a series of balancing tests demonstrated that changes in $\overline{FLFP}{-i,mc}^{FP}$ and $\overline{FLFP}{-i,mc}^{MP}$ are uncorrelated with several observed individual characteristics, such as students' demographics, average ability, education history, and socio-economic status (all measured prior to entry into the program). A key finding is that improvements in the labor market outcomes of different cohorts of women are not driven by differences in their observed characteristics, as shown by the balancedeness of their predicted outcomes. Given the high number of observed characteristics, these tests provide credible evidence that changes in $\overline{FLFP}{-i,mc}^{FP}$ and $\overline{FLFP}{-i,mc}^{MP}$ are likely uncorrelated with changes in $u_{idc}$. Therefore, the main objective of this section is to test for potential biases due to *correlated effects*. There are two primary sources of time-varying changes in the degree-specific random component: (i) changes in program characteristics, such as those driven by admission policies affecting program size and the dispersion of students' abilities, and (ii) regional shocks that impact the labor market outcomes of all students working within a given local labor market.

**Trends.** In the baseline specification, the inclusion of degree fixed effects accounts for time-invariant factors that affect the outcomes of students graduating from a given program. However, concerns might arise that the estimates of the main parameters could be confounded by degree-specific trends. This could happen if, for example, trends in students' outcomes of a specific program, e.g. related to changes in its quality - systematically change the selection of students into the program. In this case, changes in $\overline{FLFP}_{-i,dc}^{FP}$ and $\overline{FLFP}_{-i,dc}^{MP}$ would be correlated with time-varying unobserved factors driving the trends in outcomes, which would bias the estimates. To address these concerns, I extend the baseline specification by incorporating degree-specific linear time trends to account for potential changes in degree effects over time. The empirical model now becomes:

$$(3) \qquad Y_{idc} = \theta_d + \alpha_c + \varphi_d \times \tilde{t} + \gamma FLFP_{idc} + \delta^{FP}\overline{FLFP}^{FP}_{-i,dc} + \delta^{MP}\overline{FLFP}^{MP}_{-i,dc} + \epsilon_{idc}$$

where $\varphi_d \times \tilde{t}$ is a degree-specific linear time trend, and $\tilde{t}$ indicates the distance between a students' cohort and a reference cohort (i.e. 2012). Table A.18 reports the estimates of $\hat{\delta}^{FP}$ and $\hat{\delta}^{MP}$ from the augmented model. The results are similar to those of the baseline model, suggesting that it is unlikely that students select into degrees based on trends in outcomes.

As an additional robustness check, I re-estimate the main specification augmented with region-specific linear time trends, following the specification below:

$$(4) \qquad Y_{idc} = \theta_d + \alpha_c + \psi_r \times \tilde{t} + \gamma FLFP_{idc} + \delta^{FP}\overline{FLFP}^{FP}_{-i,dc} + \delta^{MP}\overline{FLFP}^{MP}_{-i,dc} + \epsilon_{idc}$$

where $\psi_r \times \tilde{t}$ is a region-specific linear time trend, and $\tilde{t}$ indicates the distance between a students' cohort and a reference cohort (i.e. 2012). Including time trends at the region of study level helps control for potential trends in student outcomes within a particular region that could influence students' selection. The results of this specification are reported in Table A.19. The estimates remain unchanged, which reduces concerns about potential confounding factors related to regional trends.

**Non-random changes in peers' composition**. This empirical strategy assumes that that cross-cohort variations in the geographical origins of students within degrees are due to random fluctuations. In selective programs, this assumption is plausible because changes in $\overline{FLFP}^{FP}_{-i,dc}$ and $\overline{FLFP}^{MP}_{-i,dc}$ result from shifts in the geographical origins of students with sufficiently high admission scores. In non-selective programs, however, these changes are driven by shifts in the applicant pool—are these random, or influenced by selection? Previous checks suggested that it is unlikely that students systematically select based on trends in outcomes within specific degrees or regions. Ideally, I would like to conduct an additional test, by restricting the analysis to selective degrees that don't experience changes in admission requirements. However, due to data limitations, I cannot perform this test, instead I use a data-driven approach to identify degrees likely to experience non-random changes in student composition. First, I re-estimate the baseline model on a subset of degrees that show no trends in size. To conduct this exercise, I run separate regressions for each degree in the dataset, treating the size of a program in a given cohort as the dependent variable and regressing it on a constant and a linear time trend. A degree is considered to have a trend in size if the p-value from the statistical test on the time variable does not exceed 0.10, which applies to about one-fourth of the degrees. I then re-estimate the baseline model on the degrees that were not flagged. The results, shown in Table A.20, are consistent with those from the main sample, with coefficients being generally higher in this cleaner subset. I conduct an additional set of sensitivity analyses to examine whether the estimates vary across different samples of degrees by excluding those with significant changes in student composition over time. Specifically, I investigate whether degrees experience *too large* shocks in the following student characteristics: (i) average student ability, measured by the average final grade in the Bachelor's degree (prior to entering the Master's program), (ii) the dispersion of student ability, assessed through the standard deviation of Bachelor's grades within a cohort, and (iii) cohort size. Such changes might indicate shifts in admission policies or in students' selection. To gauge the severity of these shocks within programs over time, I analyze the residual variation in characteristics (i)-(iii) after accounting for degree and cohort fixed effects[16]. Specifically, for each degree d in the dataset and each characteristic Y in (i)-(iii), I construct a measure summarizing the cross-cohort variation in peers' characteristics within a degree:

$$(5) \qquad Z_d^Y = \frac{1}{T_{max}} \sum_{t=1}^{T_{max}} |r_{dt}^Y|$$

where $r_{dt}^Y$ represents the residual obtained from regressing the average characteristic Y in a cohort of a degree on degree and cohort fixed effects. $T_{max}$ denotes the maximum number of cohorts a degree is observed in the data (for 92% of degrees, $T_{max} = 5$). I transform $Z_d^Y$ into a relative measure by dividing it by the mean characteristic of the program across years, $\frac{1}{T_{max}} \sum_{t=1}^{T_{max}} Y_{dt}$. Next, I categorize degrees based on their position in the distribution of this relative measure, separately for each characteristic Y. I then conduct a set of sensitivity analyses, as shown in Table A.21, with the dependent variable being log(monthly earnings). The estimates across different samples are very similar, and in some cases higher, to the benchmark specification, which is presented in Column 1.

---

16. Note that the unit of observation is a degree-cohort pair.

## 7.2   Sensitivity to Sample Restrictions

This sub-section assesses the sensitivity of peer effect estimates to alternative sample restrictions. Specifically, I investigate whether the results vary among samples defined by degree size and the proportion of students who completed their Bachelor's degree at the same university. The results on log(monthly earnings) are presented in Table A.23. I then discuss how these restrictions impact the baseline specification, which is provided in Column (1) for comparison.

**Heterogeneity by degree size**. One concern is that my identification strategy might not be valid in very small programs, where students have more chances to anticipate the peers' composition they will face. However, Column (2) shows that the benchmark estimates remain almost unchanged when excluding degrees in the bottom decile of the size distribution (where the average number of students per degree is fewer than 13). This suggests that the benchmark estimates are not influenced by noise or possible endogenous peer formation stemming from very small programs. Second, estimated effects are similar to those of the benchmark specification when degrees in the highest decile of the size distribution are dropped (where, on average, there are more than 87 students), as indicated in Column (3). In this sample of large programs, the effects vanish and standard errors become very large, as shown in Column (4). The latter finding is consistent with evidence provided in Section that the magnitude of cross-cohort variations in peers' geographical origins become significantly smaller as the degree size increases. Lastly, estimated peer effects are significantly higher in degrees with sizes below the mean (43 students), while they become smaller and less precisely estimated in degrees above the mean. This aligns with the idea that, in large programs, the network structure is likely to adjust to changes in the initial cohort composition. For example, students can segregate into distinct social networks, possibly related to their background characteristics, resulting in decreased beneficial social interactions among out-group members, as in Carrell, Sacerdote, and West 2013.

**Heterogeneity by proportion of students with Bachelor at the same institution**. Another concern arises in degrees where a significant portion of students completed their Bachelor's degree at the same institution. In such cases, students may have moved together from a shared Bachelor's program to a shared Master's program. This could result in cross-cohort variation in peers' origins being driven by selection rather than being idiosyncratic. To address this concern, I exclude degrees where the vast majority of students completed their Bachelor's at the same institution (Column 6). The estimates remain robust after this exclusion. Additionally, I provide results from the empirical model on a sample of degrees where the proportion of students who completed their Bachelor's at the same institution falls within the bottom 25% (Column 7). Despite the significant reduction in sample size, the estimated effects are larger in magnitude than the benchmark estimates and precisely estimated.

**Heterogeneity by student attendance to classes**. As a placebo, I conduct the analysis focusing on the subset of students who do not attend classes in the Master's program. This subset is identified through the pre-graduation survey, where some students report working full-time for the entire duration of the Master's program, constituting 8.7% of the sample. Since these students likely have limited interactions with their peers, peer effects within this subgroup should be minimal. The results of the analysis regarding heterogeneity by students' attendance are detailed in Table **??**. As expected, the findings reveal significant heterogeneity in the effects of peers: while the peer effects are substantial for students with high attendance to classes, there is no evidence of an effect for students with low attendance.

## 7.3    Sensitivity to Measures of Gender Culture

This sub-section assesses the sensitivity of peer effect estimates to different measures of local gender culture. Specifically, I explore whether the results change when using six additional proxies for local gender culture, as described in Section. The results for log(monthly earnings) are presented in Table A.24. Each column corresponds to a separate regression using a different proxy for peers' gender culture, as indicated by the column labels. Column (1) replicates the baseline specification from Column (1) of Table 6 for reference. Across the different measures of students' gender culture, the estimates remain consistent, with slightly larger effects observed when gender culture is defined using the labor market behavior of younger women and female graduates, suggesting that younger women might be a more relevant reference group. When gender culture is based on women's labor market behavior, the peer effect estimates range from 0.035 to 0.041. Using the proxy based on firms' gender culture, the estimates are somewhat smaller (0.016) but remain statistically significant.

## 7.4 Which peers' characteristics matter?

The main finding of this paper is that exposure to female classmates from provinces with more egalitarian gender cultures positively impacts women's labor market outcomes. But which peer characteristics drive this effect? This section investigates whether the observed effects are specifically due to peers' gender culture or other factors. In Section, I documented that movers from high- and low-FLFP areas do not differ significantly in key aspects. To further ensure the robustness of these findings, I incorporate controls for alternative peer characteristics into the analysis. The results of these tests are presented in Table A.25. Each column represents a separate regression on log(monthly earnings), following the specification in 2 and adding controls for different peer characteristics. For instance, Column 1 includes controls for the share of female and male peers with working mothers, while Column 6 accounts for the share of peers with above-median grades before entering the Master's program. Other regressions include controls for peers' socio-economic background and educational history. The peer effect estimates remain stable or even slightly increase with these additional controls. This suggests that the estimated effects are unlikely to be confounded by other peer characteristics, such as ability or socio-economic status. In Columns 8 and 9, I further control for characteristics of degree programs, such as program size and the share of female peers. Adding these controls does not change the estimates, reinforcing the conclusion that the peer effects are not driven by shocks occurring within specific programs.

**Other checks: fixed effects**. Finally, I modify specification 2 to include for province of origin FEs instead of controlling for the FLFP in the own province of origin of the student.

The adjusted empirical model now becomes:

$$(6) \qquad Y_{idcp} = \theta_d + \alpha_c + \omega_p + \delta^{FP}\overline{FLFP}^{FP}_{-i,dcp} + \delta^{MP}\overline{FLFP}^{MP}_{-i,dcp} + \epsilon_{idcp}$$

where $\omega_p$ are province of origin fixed effects. The results of this specification are equivalent to those of the main specification, as shown in Table A.22.

# 8 Beyond the Linear-in-Means Model

The Linear-in-Means (LiM) model, as presented in equation 2, is the most frequently estimated framework in the peer effects literature. This model posits that a student's outcome is a linear function of the average background characteristics of their peers. However, this approach imposes strict assumptions on the nature of peer effects. First, it constrains the size of peer effects ($\delta^{FP}$ and $\delta^{MP}$) to be the same regardless of where the student falls within the distribution of student background. Second, it assumes that peer effects operate exclusively through the mean, disregarding any influences from other aspects of the peers' background distribution. For instance, mean-preserving increases in the variance of peers' background characteristics are assumed to have no effect. The linear-in-means model has the virtue of simplicity but does suffer from the disadvantage of not being a convincing description of the world (Sacerdote 2014, 2011). Peer effects are often non-linear in various contexts (Boucher et al. 2024)[17]. The accuracy of the LiM as a model carries substantial implications for social welfare. Nonlinearities in peer effects open up the possibility that the outcomes of some students could be improved by a change in peers without detriment to others. Conversely, if peer effects are strictly linear in means, then regardless of how peers are arranged, society would have the same average level of outcomes. This section aims to explore the potential for such nonlinearities in peer effects.

First, I turn to specifications in which I allow the effects of peers to vary with a student's own gender culture. Specifically, I associate each student with the quartile of FLFP

---

17. Beginning with Hoxby and Weingarth 2005, a number of empirical studies have investigated these non-linearities, including Tincani 2024, Booij and Leuven 2017, Feld and Zölitz 2017, Lavy, Paserman, and Schlosser 2012, Imberman, Kugler, and Sacerdote 2012, Burke and Sass 2013, Carrell, Fullerton, and West 2009, Hanushek and Rivkin 2009.

in her province origin, relative to the distribution of FLFP in this sample.

# 9 Mechanisms

The previous section presents evidence that peers' gender culture significantly influences women's early career labor market decisions, with these effects being notably large. To gain further insight into the driving forces of peer effects, this section examines the relevance of possible mechanisms. In the first subsection, I examine whether endogenous variations in effort or networks can explain the results. In the following subsection, I draw on data from both the institutional survey and the original survey I conducted to explore the roles of cultural transmission and social learning.

## 9.1 What peers don't do

**1. Human capital**. As previously noted, the FLFP in a student's province of origin does not serve as a proxy for her ability (Table A.11). Women from provinces in the highest and lowest quartiles of FLFP show no significant differences in observed measures of ability, such as their grades in prior education (e.g., Bachelor's degree) or their GPA and graduation grades in the Master's program. However, there is a concern that local characteristics in a student's province of origin could influence unobserved factors, such as motivation. If women from high-FLFP areas possess higher levels of these unobserved traits, which could explain their better labor market outcomes, then exposure to a higher proportion of such individuals might affect other students' efforts. To investigate this possibility, I replicate the empirical analysis using indicators of women's academic performance as the outcomes, as shown in Table A.26. The outcome variables include GPA in the Master's program (on a 0–30 scale), graduation grade (on a 66–110 scale), and an indicator of delayed graduation (*fuoricorso*). The table provides the sample averages of these variables. The results indicate that peers' gender culture has no significant effect on women's academic performance. None of the estimated coefficients is statistically different from zero, and the magnitudes of the estimates are small. For example, a one standard deviation increase in the gender culture of female peers is associated with only a 0.071-point increase in graduation grade, representing a minimal 0.06% increase relative to the mean. Based on this evidence, I conclude that human capital is not a mediating factor in enhancing women's labor market outcomes.

**2. Geographic mobility**. Another plausible explanation is that exposure to peers from high-FLFP provinces directly influences women's decisions on where to search for a job. For example, women from high-FLFP areas might share information about their local labor markets, or the friendships formed during their studies could encourage women to follow their peers into these markets. If either of these channels operates, we would expect the geographic origins of a woman's peers to impact her mobility decisions. To test this hypothesis, I use characteristics of the local labor market where a woman works as outcome variables in the empirical model described by equation 2. The results, presented in Table A.27), indicate that the geographic origins of peers, as measured by the FLFP in their province of origin, do not affect women's mobility decisions. The outcome variables examined include the FLFP in the province of work, whether a woman is employed in the same region where she studied or elsewhere, and whether she is working in a province different from her birth province. Sample averages of these variables are provided in the table. // A few descriptive facts regarding women's mobility are worth noting: 68.4% of students find their first job in the same region where they studied, 5% work abroad, and the remainder work in a different region than where they studied, possibly their province of origin. Among those who migrated to another province for their studies, 39% return to their home province, with significant variation across quartiles of gender culture. Women from the highest quartile of FLFP are nearly twice as likely to return to their province of origin compared to those from the lowest quartiles. . Results from Table A.27 indicate that exposure to peers from high-FLFP provinces does not significantly impact women's mobility decisions, such as choosing to work in the region where they studied or opting to work outside their birth province. None of the estimated coefficients is statistically different from zero, and the magnitudes of the estimates are small. For instance, a one standard deviation increase in peers' gender culture corresponds to a mere 0.155 increase in the FLFP in the woman's province of employment, amounting to a marginal 0.28% increase relative to the mean. Furthermore, the main findings remain stable when accounting for fixed effects for the province of employment, as shown in Table. Overall, these results suggest that the observed peer effects on female earnings and labor supply are not driven by changes in geographic mobility.

**Networks**. Economically beneficial labor market connections, such as networks to better firms, offer an alternative explanation for the observed earnings gains among female students. Unfortunately, the absence of firm identifiers in the data precludes a formal test of the extent to which networks contribute to the estimated peer effects. Nonetheless, I provide suggestive evidence that these networks are unlikely to be a major driver of increases in women's labor supply. The underlying logic of this test is straightforward: cross-cohort changes in the average female labor force participation (FLFP) in peers' provinces within a degree program could correlate with shifts in the relative proportions of students who are "local" versus "movers." In this context, "local" refers to students who study in their province of birth. If locals have better connections to local firms, e.g. through their parents, they could share new information about job opportunities or provide job referrals to peers with different information sets. Given the documented importance of networks at labor market entry (Zimmerman 2019, Kramarz and Skans 2014, Hampole and Wong 2024, Fischer, Sandoy, and Walldorf 2023, Einiö 2023), this could be a significant channel influencing earnings. In practice, if locals indeed have better connections to local firms, two conditions would lead to an upward bias in the estimated effects of peers' culture: (i) cross-cohort variation in peers' gender culture correlates with increases in the share of locals within a program, and (ii) peers serve as networks facilitating access to better firms. To test this hypothesis, I include controls for the shares of female and male peers who are locals in the main specification. The results, presented in Table A.28, show that the estimates of the effects of peers' gender culture remain robust. Additionally, the magnitude of these estimates slightly increases, likely due to the small but significant negative effect of a higher share of local students on women's earnings and labor supply.

## 9.2 Cultural transmission and learning

Given the lack of empirical support for channels related to students' networks or human capital, I explore two alternative explanations that could rationalize my findings and align with existing theories of cultural change. One view is that peers could affect the beliefs and preferences of women in a way that influence their labor market behavior—a process known as *cultural transmission*, as outlined by Bisin and Verdier 2001. The second, not mutually exclusive, mechanism involves *social learning*, where students share new information with peers with different information sets, following the framework of Fernandez 2013. In this section, I present evidence supporting both mechanisms. First, I show that students' social environment affect their self-reported preferences for various job attributes, particularly those that have been found to be relevant to gender wage disparities. The following section will then explore the social learning mechanism, drawing on novel data on students' beliefs collected through the original survey.

**Endogeneous preferences**. This analysis leverages data from the institutional survey administered to all graduating students, with a near 100% response rate due to the survey's compulsory nature. This comprehensive dataset offers a significant advantage over previous studies, which relied on smaller, less representative samples from select fields and universities. In the survey, students' preferences for job attributes are directly elicited by asking them to rank various attributes on a scale from 1 to 5. Building on prior research into gender differences in job-search preferences (Wiswall and Zafar 2018, Mas and Pallais 2017, Eriksson and Kristensen 2014), I construct indexes to measure preferences for pecuniary aspects (i.e. salary and career progression) and flexibility (i.e. leisure time and hours flexibility). These indices are calculated as unweighted averages of the scores assigned to each job characteristic and are standardized for ease of interpretation. Additionally, I create a binary indicator to identify whether a student assigns the maximum value (5/5) to the job's social utility—a relevant attribute that has been overlooked in previous studies but where I observe significant gender differences. While prior research has typically treated these preferences as exogenous, I examine whether, and to what extent, these preferences are endogenous to the social environment students encounter in college. Accordingly, I use these measures as outcome variables in the empirical model, with the results presented in Table A.29. The findings suggest that women socialized in cohorts with more female peers from high-FLFP provinces tend to place less importance on non-pecuniary job factors, particularly those related to flexibility and social utility. Specifically, a one standard deviation increase in peers' gender culture decreases women's preferences for hours flexibility by 2.7% of a standard deviation. In contrast, there is no statistically significant impact on women's preferences for pecuniary aspects. These estimated effects on students' self-reported preferences align with cultural transmission explanations, indicating potential shifts in women's job preferences.

# 10    Original survey on students' beliefs

The primary objective of this survey is to complement existing data sources and gain a deeper understanding of the mechanisms of peer influence, with a focus on social learning. Specifically, the survey has been designed in a way to (1) shed light on possible asymmetries in women's beliefs depending on the gender culture in their province of birth, (2) test whether beliefs are relevant in their job acceptance decisions, and (3) explore beliefs' updating.

**Gender culture and women's beliefs.** Since the seminal contribution of Fernandez 2007, numerous papers have argued that the gender culture in a woman's country/region of origin or ancestry shapes her beliefs, thereby influencing her labor market choices (Boelmann, Raute, and Schönberg 2023, Kleven 2024, Ichino et al. 2024, Fernandez 2013, Fogli and Veldkamp 2011, Fernandez and Fogli 2009). While previous studies propose hypotheses about which beliefs are influenced by the environment—such as beliefs regarding gender identity or the perceived costs of working - all of these studies are essentially agnostic regarding the precise sources of gender norms. Conceptually, in this context, we can think of the gender culture in a woman's province of birth as affecting a wide array of beliefs: for example, beliefs about the role of women in society, perceptions regarding employers' discrimination, beliefs on the job offer distribution, as well as expectations about long-run outcomes, such as fertility and expectations of future labor supply at the time of motherhood. All of them, in turn, could influence the labour supply decisions of young women and, particularly, the acceptance of part-time job offers. I have designed this survey to elicit women's beliefs regarding these various aspects. In this section, I will first present evidence of disparities in beliefs stemming from local gender culture and show how these differences translate into acceptances of part-time jobs. I will offer a theoretical illustration of the mechanisms by which these beliefs affect women's decisions to accept part-time jobs. Finally, leveraging data from two survey waves, I will present evidence of beliefs' updating.

## 10.1   Survey Design and Administration

I have conducted the survey among graduate students currently enrolled at a large public university in Italy (the University of Bologna). This represents the largest university in Italy, contributing to approximately 7% of all graduates. Importantly, it offers a multitude of cultural backgrounds, as it attracts a significant number of students from various provinces and regions across the country (88.8% and 69.6%, respectively, in the sample, as shown in Table A.30).

To construct a sample of analysis, I have randomly selected a sample of Master's degree programs and, within each program, I have randomly chosen one course from the first semester in the first year and one from the first semester of the second year. Students attending these courses have been invited to take part in the survey. The administration was done in person. Specifically, upon agreement with lecturers, I went in person to one class - usually in the first/last 15 minutes - and I encouraged students to voluntarily complete a 10-minute questionnaire on their mobile phones through the SurveyMonkey platform. Before, I took some minutes to provide general information on the study[18]. To incentivize participation, students had the chance to enter three lotteries with gift cards worth €100 [19]. The response rate reached 97% among attending students. To have a sense of the representativeness of this sample, consider that around 77% of students attend classes regularly, based on self-reported attendance in the AlmaLaurea questionnaire. Students were not informed in advance about my intervention to ensure that their attendance in class would be orthogonal to the survey administration. These two features attenuate concerns related to selection. The survey was conducted between November 2023 and February 2024. I chose to run the survey 3-4 months after the start of the academic year to strike a balance between students being able to give informed responses to the questions—especially about the program's social environment— and learning of students in the first year being not yet complete. A total of 899 students from 34 Master's programs participated in the survey. Among them, 535 identified as women, 348 as men, and 13 as non-binary. The sample included 571 students in their first year and 322 in their second year. This disparity is attributed to the curriculum structure, with mandatory courses mainly offered in the first year. Consequently, the second-year cohort tends to be smaller due to the greater flexibility in choosing optional courses.

---

18. Specifically, I informed students that the questionnaire was about their beliefs and labor market expectations and was needed for a study on students' career decisions after college. To avoid priming, I did not disclose that the study focused on peer influence or its connection to gender inequalities.

19. The gift cards were generic and could be used across multiple brands or providers, to avoid any potential selection bias from the choice of a specific provider.

## 10.2 Sample Selection and Description

I exclude from the sample students who are on Erasmus or attending a bachelor's program (less than 1%), as well as international students or students with missing information on the country/province of origin (5.8%). The number of female students in the resulting sample is 490, 319 of whom are in the first year of the Master's program and 171 in the second year. The characteristics of this sample are summarized in Table A.30, which provides summary statistics for the entire sample and for sub-samples based on whether students come from provinces with below-median or above-median FLFP. This sample differs from the main one in a few dimensions: it over-represents students from high-SES backgrounds, indicated by a higher proportion of parents with university degrees, and includes more students from provinces with higher FLFP. There is also a higher percentage of students who migrated from other provinces and regions. While the distribution of fields of study is generally similar to the main sample—particularly in the representation of students in economics and the social sciences—this sample features a slightly higher representation of students in the humanities and sciences, with fields such as architecture and healthcare not represented. When comparing the characteristics of students by FLFP in their province of origin, it is immediate to observe that these groups differ in the role models they are exposed to, both in the family and in society. Students from low-FLFP provinces are more likely to come from families where fathers have higher education levels than mothers, while the opposite is true for students from high-FLFP areas. Additionally, mothers from low-FLFP areas are more likely to have experienced significant *child penalties*, such as pausing their careers during early childhood, compared to mothers from high-FLFP regions. Apart from these background characteristics, the two groups are similar in two key areas: women's fertility expectations and their intentions to search for a job after graduation.

## 10.3    Learning about The Job Offer Distribution

In this sub-section, I investigate the role of asymmetries in beliefs regarding the job offer distribution and the process of learning. The underlying idea is that women from areas with less favorable labor market conditions for women may hold different expectations about job offer arrival rates compared to those from more egalitarian areas. To investigate this, I gather students' expectations regarding key parameters of a job-search model, including the overall job offer arrival rate and the relative arrival rates of part-time versus full-time job offers.

**Elicitation.** Students' beliefs about the job offer distribution have been elicited through hypothetical scenarios that aim at reproducing a realistic setting of job search:

---

1. *Consider the following scenario: you have graduated from the Master's program in which you are currently enrolled and you start searching for a job. You submit 10 applications to positions aligned with your field of study. When applying, you don't know the specific working conditions—such as the monthly salary or whether the contract is part-time or full-time[a].*

- Out of these 10 applications, how many job offers do you expect to receive? ($\alpha$) Provide your answer on a scale from 0 to 10.

- You receive your first job offer. What do you believe is the probability that the employer will propose a part-time contract (less than 28 hours/week)? ($\gamma$) - Provide your answer on a scale from 0 to 100.

- On a scale from 1 (my answer is very close to the true probability) to 5 (I answered at random), how certain are you about your previous answer? ($\sigma$)

  2. *While waiting for responses to your applications, an employer contacts you and offers a part-time position (28 hours/week) with a net monthly salary in line with your expectations. You must decide whether to accept the offer or turn it down and wait for responses from the other applications.*

- What is the probability that you will accept this part-time job offer? Provide your answer on a scale from 0 to 100.

  ---
  *a.* Note that in Italy, 91% of online job postings do not include salary or salary ranges, and precise information about working hours is often limited (Burning glass data)

---

Note that, similar to Wiswall and Zafar 2021, students' intentions to accept a part-time job offer are elicited using a stated probabilities approach rather than a discrete choice approach. This method accounts for students' uncertainty when reporting their choices in the survey. A discrete choice model is, in fact, a special case of the stated probabilities approach, representing a scenario where there is no resolvable uncertainty. In this case, individuals would assign a probability of exactly 1 or 0 to the decision of accepting a part-time job. However, the data clearly reject this: only 4.83% of reported probabilities are 1, and less than 1% of probabilities are 0.

**Place of birth and asymmetries in beliefs**. The analysis of these beliefs is presented in Table 9. To explore potential asymmetries based on local gender culture, Panel (a) focuses on baseline beliefs collected during the students' first year, aiming to capture the student's initial perceptions before any influence from peers. Specifically, the table shows predictions from a linear regression model where each dependent variable—listed in the rows—is regressed on an indicator denoting whether a woman originates from a province with below-median or above-median FLFP, controlling for field fixed effects. The table provides the predicted values with their standard errors, as well as the p-values testing the significance of the differences between the two groups. The results indicate that women tend to have different expectations of these parameters depending on the FLFP in their province of origin. First, women from low-FLFP areas expect a slightly lower arrival rate of job offers ($\alpha$), though this difference is negligible and not statistically significant—out of ten applications, they expect to receive, on average, 3.21 offers compared to 3.52 for women from high-FLFP provinces. More notably, there is a substantial gap in their expectations regarding the proportion of part-time job offers. Women from low-FLFP areas expect a 6.45 percentage point higher likelihood of receiving a part-time versus full-time job offer, which represents a significant 12.6% increase compared to their high-FLFP counterparts. Additionally, these women are 7 percentage points more likely to indicate they would accept a part-time offer, marking a 12% increase relative to peers from high-FLFP provinces. Importantly, these differences cannot be attributed to differences in observed characteristics between the two groups, as they are unchanged after controlling for students' background characteristics (age, family background), job search intentions, and expected job location, as shown in Table A.31.

**Beliefs' updating.** Panel (b) of Table 9 investigates how these believes evolve over time,

| | Below-med FLFP | | Above-med FLFP | | |
|---|---|---|---|---|---|
| | Pred | SE | Pred | SE | P-value |
| **a. Baseline Beliefs (T=0)** | | | | | |
| α: Expected arrival rate of job offers (%) | 32.06 | 1.77 | 35.24 | 1.24 | 0.15 |
| γ: Expected % of part-time job offers | 57.64 | 2.29 | 51.19 | 1.61 | 0.02 |
| Perceived uncertainty (1-5) | 2.70 | 0.12 | 3.01 | 0.08 | 0.04 |
| Prob. to accept part-time job offer | 67.43 | 2.04 | 60.39 | 1.44 | 0.01 |
| | | | | | |
| **b. Updated Beliefs (T=1)** | | | | | |
| α: Expected arrival rate of job offers (%) | 32.20 | 2.17 | 32.19 | 1.73 | 1.00 |
| γ: Expected % of part-time job offers | 52.47 | 2.90 | 50.70 | 2.31 | 0.64 |
| Perceived uncertainty (1-5) | 2.73 | 0.14 | 2.75 | 0.11 | 0.90 |
| Prob. to accept part-time job offer | 62.48 | 2.81 | 62.37 | 2.23 | 0.98 |

Notes: This table presents predictions from a linear regression model, where the dependent variable is regressed on an indicator for whether the FLFP in the birth province is above or below the median, along with fixed effects for the field of study. Each row represents a different regression, with the dependent variable specified in Column 1. For each regression, the table reports the predicted dependent variable for women from provinces with low versus high FLFP, along with the standard errors. The last column provides the p-value for the difference between these two groups. In Panel (a), the sample consists of all first-year female Master's students (319), and in Panel (b), it includes all second-year female Master's students (164). Between 60% and 65% of the students are from provinces with above-median FLFP.

by focusing on answers from students in the second year. Overall, we observe convergence in these beliefs between the two groups, consistent with learning. More precisely, the results show that students significantly update their beliefs about the likelihood of receiving a part-time job offer, in a way that the gap originally observed between the two groups has narrowed considerably (by more than 70%). An analysis of the variance in beliefs within fields further supports evidence of learning. On average across degrees, the standard deviation of students' baseline beliefs γ in the first year is 24.28, which decreases by more than a third in the second year—a statistically significant reduction at the 1% level. What is particularly interesting is the asymmetry in this learning process. Women born in low-FLFP provinces experience strong beliefs' updating and revise their beliefs downwards regarding the probability of receiving a part-time offer relative to a full-time one by more than 5 percentage points (a 9% decrease), converging to the values expressed by their peers from high-FLFP areas who experience only little updating. Why do the two groups update their beliefs differently? One possible explanation is the initial asymmetry in the information available to them. Women from low-FLFP areas might have started with more biased beliefs about job offer arrival rates in their destination labor markets, leading to a more significant adjustment over time. This is a plausible channel, as women from low-FLFP areas are typically exposed to labor markets that differ substantially, in terms of women's outcomes, than those in their home regions, as reflected in their intentions to work predominantly in the North of Italy (Table A.30). These results provide evidence of asymmetric beliefs' updating, which is consistent with the asymmetry in the estimated peer effects. While I cannot precisely quantify the contribution of peers versus other social influences in the process of beliefs' updating, the results suggest social learning as a plausible mechanism driving peer influence in this context.

## 10.4   A Model of Job Search

To study the relation between beliefs and job search, I propose a McCall type model (McCall 1970) where risk-neutral female graduates search for their first post-graduation job. For the time being, I abstract from students' gender norms when I lay out the model, and I later introduce parameter heterogeneity when I discuss the model's prediction for differences in part-time acceptances between women from high- and low-FLFP areas.

## 10.5   Model Setup

My modeling framework is based on a standard model of labor market search à la McCall, augmented to allow for heterogeneous worker beliefs.

Consider an economy where three states exists: an individual can be unemployed, employed in a part-time job or employed in a full-time job. For simplicity, consider that part-time (P) and full-time jobs (F) are characterized by a fixed number of weekly hours. I express the per-period number of hours in a part-time job as $h^P = \theta h^F$, with $\theta < 1$. The model makes a number of key assumptions. Time t is discrete. All individuals discount the future at rate $\beta \in (0, 1)$. Students are risk-neutral: they have preferences over consumption represented by the instantaneous utility function u(c) = y, i.e. they maximize expected lifetime labor income. As is typical in job-search models, this model abstracts from other sources of income, so that instantaneous income is given by the following specification:

$$y = \begin{cases} y^F & \text{if employed in full-time job} \\ y^P & \text{if employed in part-time job} \\ b & \text{if unemployed} \end{cases}$$

b represents any income associated with not working, such as the pecuniary value of leisure and public unemployment insurance (UI) benefits. $y_P$ and $y_F$ are total per period income associated with either a part-time or a full-time job. For simplicity, I fix the number of hours in a full-time job to unity, so that $y_F = w$ and $y_P = \theta w$.

Unemployed jobseekers search for jobs and, in each period, job offers arrive with probability $\alpha^*$. A share $\gamma^*$ of job offers are part-time. A job offer is a random draw from a wage distribution F(w), which has support and non-zero density on $[w_{\min}, \bar{w}]$. Note that, for simplicity, I assume that the wage distribution is the same for part-time and full-time jobs. In each period, if an unemployed worker receives a job offer, she decides whether to accept the offer and leave unemployment or remain unemployed and enjoy the value of leisure b. I do not allow for on-the-job search or job destruction, meaning that employment - both part-time and full-time - is an absorbing state. I further assume that the environment is stable, i.e. that the arrival rates of full-time and part-time job offers do not change over the course of the search spell. Individuals are infinitely lived and, therefore, the model is stationary. Throughout, I use $*$ to indicate "true" or "actual" probabilities of receiving job offers, to distinguish these from the workers' beliefs.

## 10.6 Workers' Beliefs

I start with the notion that workers make decisions with possibly limited knowledge about job offer arrival rates. Specifically, I assume that workers do not necessarily know the per-period probability of receiving a job offer $\alpha^*$ and the relative share of part-time job offers $\gamma^*$. Define $\alpha_t$ and $\gamma_t$ as the worker's current beliefs about $\alpha^*$ and $\gamma^*$. I refer to biased beliefs if $\alpha_t \neq \alpha^*$ or $\gamma_t \neq \gamma^*$. While beliefs potentially evolve over time due to learning, in this version of the model I abstract from this possibility and assume that $\alpha$ and $\gamma$ are not time-varying. Workers take their decisions on whether to accept job offers based on their subjective beliefs $\alpha$ and $\gamma$. I abstract away from other potential biases in beliefs, for example on the wage offer distribution F(w), that I assume to be commonly known to all individuals.

## 10.7   Perceived Values of Employment and Unemployment

I characterize the perceived flow values of unemployment and employment[20]. For a worker with beliefs $\alpha$ and $\gamma$, the perceived value of unemployment equals

(7)

$$U(\alpha, \gamma) = b + \beta\alpha[\gamma \int_{y_P} \max\{V(y_P), U(\alpha, \gamma)\}\, dG(y_P) + (1 - \gamma) \int_{y_F} \max\{V(y_F), U(\alpha, \gamma)\}\, dG(y_F)]$$

$$+ \beta(1 - \alpha)U(\alpha, \gamma)$$

where b is the flow value of unemployment, and $\alpha$ and $\gamma$ are the worker's beliefs regarding the per-period probability of receiving a job offer and the relative share of part-time job offers. The values of part-time and full-time employment at wage w are respectively

(8)
$$V(y_P) = y_P + \beta V(y_P) \rightarrow (1 - \beta)V(y_P) = \theta w$$

(9)
$$V(y_F) = y_F + \beta V(y_F) \rightarrow (1 - \beta)V(y_F) = w$$

 Note that here, absent job destruction and on-the-job search, it is assumed that, once a worker has accepted a job offer, she remains at her current job at all future periods.

## 10.8   Reservation Wages

A job-seeker's decision to accept a job offer is determined by the reservation wage property: each job offering wages above the reservation value are accepted. The job seeker determines their reservation wage in order to maximize their perceived continuation value at any point during the search spell. I define the reservation earnings, $R(\alpha, \gamma)$, as the total per-period income at which a job seeker is indifferent between accepting a job

---

20. I refer to them as perceived as they are based on workers' beliefs about the arrival rates of part-time and full-time job offers instead of the actual ones.

and remaining unemployed. The resulting expression for the reservation earnings equals

(10) $$V(R(\alpha, \gamma)) - U(\alpha, \gamma) = 0 \rightarrow R(\alpha, \gamma) = (1 - \beta)U(\alpha, \gamma)$$

Note that, because part-time and full-time jobs differ in the number of working hours, the reservation earnings condition expressed above is verified for two different reservation wages, separately for the two job types. I define the reservation wages for part-time and full-time jobs as $w_{R,P}(\alpha, \gamma)$ and $w_{R,F}(\alpha, \gamma)$. The expressions are

(11) $$w_{R,P}(\alpha, \gamma) = \frac{R(\alpha, \gamma)}{\theta} \quad \text{and} \quad w_{R,F}(\alpha, \gamma) = R(\alpha, \gamma)$$

Any job offering wages above these values is accepted. Note that reservation wages are determined based on workers' beliefs. In the following propositions, I outline how biases in beliefs regarding $\alpha^*$ and $\gamma^*$ theoretically impact reservation wages.

**Proposition 1.** *Ceteris paribus, reservation wages are increasing in beliefs* $\alpha$.

**Proposition 2.** *Ceteris paribus, reservation wages are decreasing in beliefs* $\gamma$.

The proofs are contained in Appendix Section A.

## 10.9   Heterogeneous Beliefs and Model's Predictions

I now introduce heterogeneity in workers' beliefs into the model, considering that workers hold beliefs $\alpha_i$ and $\gamma_i$ about the arrival rates of job offers, where $i \in (L, H)$. However, throughout the model, I assume that all workers actually face the same true arrival rates, denoted as $(\alpha^*, \gamma^*)$. I define the beliefs of women from low-FLFP and high-FLFP provinces as $(\alpha_L, \gamma_L)$ and $(\alpha_H, \gamma_H)$, respectively. Drawing on the empirical analysis of students' beliefs from the previous section, I assume:

(12) $$\alpha_L < \alpha_H \quad \text{and} \quad \gamma_L > \gamma_H$$

i.e. women from low-FLFP areas expect a lower probability of receiving any job offer and a higher likelihood of receiving a part-time offer relative to a full-time one.

A direct corollary of Propositions 1 and 2 is that, all else equal, if women from low-FLFP provinces hold more pessimistic beliefs about the arrival rates of job offers and the likelihood of receiving part-time versus full-time offers compared to women from high-FLFP areas, they will have lower reservation earnings. Consequently, they will have a higher likelihood of accepting a part-time job offer. I test this prediction using data on students' expectations about job offer arrival rates and their intentions to accept a part-time offer, elicited on a probabilistic scale through a realistic hypothetical scenario. These intentions are indicative of how students might behave when faced with part-time job offers during their actual job search. Figure 7 illustrates the relationship between these beliefs and students' job search behavior, showing a strong correlation between their expectations of part-time versus full-time job offers and the likelihood of accepting a part-time offer. Estimates from a simple linear regression indicate that a one standard deviation increase (23 percentage points) in the expected probability of receiving a part-time offer results in an approximately 8 percentage point increase in the acceptance rate of part-time jobs, representing more than a third of the standard deviation in the sample. These beliefs alone explain 13% of the variation in students' acceptances of part-time jobs in the sample. Estimates are robust to the inclusion of field (or degree) fixed effects ($\beta$ = 0.27). Table A.32 shows the estimated coefficients from a regression of students' part-time job acceptances on their beliefs about $\gamma$ (Columns 1-2) and $\alpha$ (Columns 3-4), both with and without controls for field fixed effects. These results additionally show a negative relationship between expected job offer arrival rates and the likelihood of accepting a part-time offer, consistent with the model's predictions.

**Job finding probabilities**. The individual job-finding probability is defined as:

$$(13) \qquad \lambda_i = \alpha^*[\gamma^* P(y_P \geq R_i(\alpha_i, \gamma_i)) + (1 - \gamma^*) P(y_F \geq R_i(\alpha_i, \gamma_i))]$$

that I rewrite as

$$(14) \qquad \lambda_i = \alpha^*[\gamma^*(1 - F(\frac{R_i(\alpha_i, \gamma_i)}{\theta})) + (1 - \gamma^*)(1 - F(R_i(\alpha_i, \gamma_i)))]$$

where $\lambda_i$ represents the per-period probability of exiting unemployment. This probability

FIGURE 7. Acceptance of part-time jobs and expected share of part-time job offers

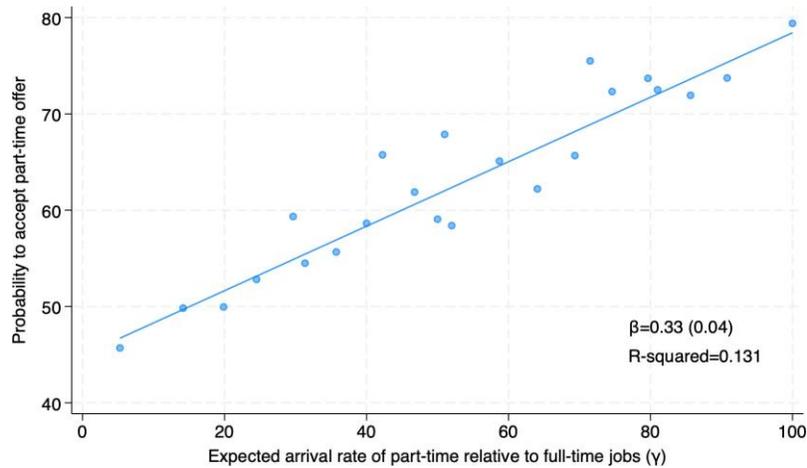

Notes. This figure presents binned scatter plots of the probability of accepting a part-time job against the expected arrival rate of part-time relative to full-time job offers (γ). One observation represents a student in the sample. The figure also displays the estimated coefficient β and its standard error from a simple linear regression of the intended probability of accepting a part-time offer on the expected percentage of part-time job offers.

depends on the true arrival rates of job offers, $\alpha^*$ and $\gamma^*$, as well as on women's beliefs about these parameters through their reservation earnings, $R_i(\alpha_i, \gamma_i)$, which are indexed by $i$ to reflect heterogeneity in workers' beliefs. A second implication of equation (12) in the model is that $\lambda_L > \lambda_H$ at any point in time, implying that women from low-FLFP areas have higher job-finding rates due to behavioral differences driven by their beliefs. Specifically, since women with more pessimistic beliefs are less selective and have lower reservation earnings, they are less likely to reject job offers and more likely to exit unemployment earlier. It is important to note that this result relies on a simplifying assumption of the model—that job search effort is exogenously determined. An extended version of the model, which includes endogenous job search effort (available in the appendix), can rationalize the dynamics of job-search behavior observed in the data.

## 10.10    Expectations of Fertility and Future Child Penalties

This sub-section explores how local gender culture shapes expectations about fertility and anticipated child penalties. Previous studies, such as Boelmann, Raute, and Schönberg 2023 and Kleven 2024, have shown that the gender norms women are exposed to during childhood have long-lasting effects on their labor supply decisions after becoming mothers. Even in the Italian context, the magnitude of child penalties varies significantly across geographical areas (Casarico and Lattanzio 2023). If women accurately anticipate these differences in future behavior, their expectations about fertility and future labor supply may vary based on the gender norms they were exposed to when growing up. For instance, women from less egalitarian backgrounds might expect higher employment costs associated with motherhood. As implied by a dynamic labor supply model in the vein of Adda, Dustmann, and Stevens 2017, these differences in expectations likely influence women's career choices even before they have children, particularly in selecting jobs or occupations with different opportunity costs of child-rearing. Whether women anticipate these differences in behavior is an empirical question, that I examine in this sub-section. To achieve this goal, I gathered data on women's expectations regarding fertility and future labor supply.

1. Would you like to have children in the future? Yes/No/Don't know/Already have

2. At what age do you expect to have your first child?

3. Expected labor supply at motherhood:

- *Scenario 1. Suppose that your partner is earning enough to support your family. What do you think you will do when your child is young (0-2 years)?* Answer: No work/Work part-time/Work full-time

- *Scenario 2. Suppose that your partner is earning enough to support your family and that in the area you live a full-day place in childcare is available to you. What do you think you will do when your child is young (0-2 years)?* Answer: No work/Work part-time/Work full-time

The analysis of their baseline and updated expectations is presented in Table A.33. At the start of the first year, there are no differences in fertility expectations that can explain the observed labor supply disparities between women from high and low-FLFP provinces. In fact, women from low-FLFP provinces are less likely to expect to have children and anticipate having their first child at a later age compared to their high-FLFP peers. Crucially, when considering future labor supply, those from low-FLFP areas are less likely to foresee reductions in working hours due to motherhood as they are more likely to expect to continue working full-time during the early years of parenthood. This pattern holds across both their unconditional expectations (elicited under scenario 1) and their expectations when full-day childcare is available near their residence (elicited under scenario 2). Additionally, both groups of women anticipate that childcare availability will impact their future labor supply, as the share expecting to work full-time increases by 43%-58%, depending on the group, when access to full-day childcare is available. One year later, both groups have revised their expectations upwards, particularly when full-day childcare is available. Around 80% of women in the two groups expect to work full-time in the early years of motherhood. Contrary to expectations, these results show that women from less egalitarian backgrounds do not anticipate higher employment costs of motherhood when entering the labor market, compared to women from high-FLFP areas. One plausible explanation is that young women may underestimate the career costs of motherhood— a phenomenon particularly pronounced among the college-educated (Kuziemko et al. 2018). Other sets of explanations relate to intergenerational shifts in the magnitude of child penalties or positive selection bias in the sample of women from low-FLFP areas. Importantly, unlike the asymmetric updating of beliefs regarding job offers, these results indicate that preferences and beliefs about maternal employment evolve symmetrically across both groups.

# 11   Conclusions

# Appendix

## A Appendix: Proofs

### Proof of Proposition 1

PROOF. The perceived value of unemployment can be rewritten as

$$(1-\beta)U(\alpha,\gamma) = b + \beta\alpha[\gamma \int_R^{\bar{y}_P}(V(y_P) - U(\alpha,\gamma))\,dG(y_P) + (1-\gamma)\int_R^{\bar{y}_F}(V(y_F) - U(\alpha,\gamma))\,dG(y_F)]$$

Using the reservation earnings rule and plugging the values of employment, this becomes

$$R(\alpha,\gamma) = b + \beta\frac{\alpha}{1-\beta}[\gamma \int_R^{\bar{y}_P}(y_P - R(\alpha,\gamma))\,dG(y_P) + (1-\gamma)\int_R^{\bar{y}_F}(y_F - R(\alpha,\gamma))\,dG(y_F)]$$

Rearranging yields

$$R(\alpha,\gamma) = b + \beta\frac{\alpha}{1-\beta}[\gamma \int_R^{\bar{y}_P}(y_P - R(\alpha,\gamma))\,dG(y_P) + (1-\gamma)\int_R^{\bar{y}_F}(y_F - R(\alpha,\gamma))\,dG(y_F)]$$

reservation earnings are set to equal the flow value of unemployment and the expected surplus associated with job offers. Note that for all values that R can take, the expected surplus from a full-time job exceeds the expected surplus from a part-time job. Rearranging yields

$$R(\alpha,\gamma) = b + \beta\frac{\alpha}{1-\beta}[\int_R^{\bar{y}_F}(y_F - R(\alpha,\gamma))\,dG(y_F) - \gamma(\int_R^{\bar{y}_F}(y_F - R(\alpha,\gamma))\,dG(y_F) -$$
$$\int_R^{\bar{y}_P}(y_P - R(\alpha,\gamma))\,dG(y_P))]$$

Substituting $y_F = w$ and $y_P = \theta w$, and using the fact that $G(y_F) = F(w)$ and $w_{R,P} = \frac{R}{\theta}$, I rewrite the expression that implicitly defines reservation earnings as:

(A.1)
$$R(\alpha,\gamma) = b + \beta\frac{\alpha}{1-\beta}[\int_R^{\bar{w}}(w - R(\alpha,\gamma))\,dF(w) - \gamma(\int_R^{\bar{w}}(w - R(\alpha,\gamma))\,dF(w) -$$
$$\int_{\frac{R}{\theta}}^{\bar{w}}(\theta w - R(\alpha,\gamma))\,dF(w))]$$

Differentiating both sides of equation (A.1) with respect to $\alpha$ and applying the Leibniz

rule yields

$$\frac{\partial R(\alpha,\gamma)}{\partial \alpha} = \frac{\beta}{1-\beta} \left[\int_R^{\bar{w}} (w - R(\alpha,\gamma))\, dF(w) - \gamma(\int^{\bar{w}} (w - R(\alpha,\gamma))\, dF(w) - \int^{\bar{w}} (\theta w - R(\alpha,\gamma))\, dF(w))\right]$$

$$+ \frac{\beta\alpha}{1-\beta} \left[- \frac{\partial R(\alpha,\gamma)}{\partial \alpha}(1 - F(R)) + \gamma\frac{\partial R(\alpha,\gamma)}{\partial \alpha}(1 - F(R)) - \gamma\frac{\partial R(\alpha,\gamma)}{\partial \alpha}(1 - F(\frac{R}{\theta}))\right]$$

Rearranging, I get to the following expression

$$\frac{\partial R(\alpha,\gamma)}{\partial \gamma} = \frac{\frac{\beta}{1-\beta}\left[\int_R^{\bar{w}} (w - R(\alpha,\gamma))\, dF(w) - \gamma(\int_R^{\bar{w}} (w - R(\alpha,\gamma))\, dF(w) - \int_R^{\bar{w}} (\theta w - R(\alpha,\gamma))\, dF(w))\right]}{1 + \frac{\beta\alpha}{1-\beta}\left[(1 - F(R))(1 - \gamma) + \gamma(1 - F(\frac{R}{\theta}))\right]}$$

Both the numerator and the denominator of the right-hand side are positive. Hence, $\frac{\partial R(\alpha,\gamma)}{\partial \gamma} > 0$, i.e. reservation earnings are increasing the perceived probability of receiving a job offer. □

## Proof of Proposition 2

PROOF. Differentiating both sides of equation (A.1) with respect to $\gamma$ and applying the Leibniz rule yields

$$\frac{\partial R(\alpha,\gamma)}{\partial \gamma} = \frac{\beta\alpha}{1-\beta}\left[-(1 - F(R))\frac{\partial R(\alpha,\gamma)}{\partial \gamma} - (\int_R^{\bar{w}} (w - R(\alpha,\gamma))\, dF(w) - \int^{\bar{w}} (\theta w - R(\alpha,\gamma))\, dF(w)) \right.$$

$$\left. - \gamma(-\frac{\partial R(\alpha,\gamma)}{\partial \gamma}(1 - F(R)) + \frac{\partial R(\alpha,\gamma)}{\partial \gamma}(1 - F(\frac{R}{\theta})))\right]$$

which yields the following expression

$$\frac{\partial R(\alpha,\gamma)}{\partial \gamma} = -\frac{(\int_R^{\bar{w}} (w - R(\alpha,\gamma))\, dF(w) - \int_R^{\bar{w}} (\theta w - R(\alpha,\gamma))\, dF(w))}{[1 + \frac{\beta\alpha}{1-\beta}((1 - F(R))(1 - \gamma) + \gamma(1 - F(\frac{R}{\theta})))]}$$

Because the expected surplus from a full-time always exceeds that of a part-time job, the numerator is positive. The denominator is also positive. It follows that $\frac{\partial R(\alpha,\gamma)}{\partial \gamma} < 0$, i.e. reservation earnings are decreasing in a workers' beliefs of receiving a part-time relative to a full-time job offer. □

# B   Additional Figures and Tables

TABLE A.1. Summary Statistics of Demographics, Performance and Family Background

| | Female | | | Male | | | |
|---|---|---|---|---|---|---|---|
| | Mean | SD | Obs | Mean | SD | Obs | P-value |
| **Individual characteristics** | | | | | | | |
| Age at enrollment | 24.3 | 4.0 | 182792 | 24.5 | 4.1 | 133678 | 0.00 |
| GPA during Master | 27.8 | 1.5 | 182792 | 27.4 | 1.7 | 133678 | 0.00 |
| Final grade during Master | 108.6 | 5.6 | 182792 | 107.4 | 6.3 | 133678 | 0.00 |
| Time to completion of Master (years) | 2.5 | 0.6 | 182792 | 2.6 | 0.6 | 133678 | 0.00 |
| Bachelor grade | 101.3 | 7.4 | 162091 | 99.1 | 8.2 | 116258 | 0.00 |
| High school: general track (%) | 84.1 | 36.6 | 182792 | 71.3 | 45.2 | 133678 | 0.00 |
| science (%) | 40.2 | 49.0 | 182792 | 56.6 | 49.6 | 133678 | 0.00 |
| humanities (%) | 21.2 | 40.9 | 182792 | 10.4 | 30.6 | 133678 | 0.00 |
| foreign language (%) | 10.5 | 30.6 | 182792 | 1.9 | 13.8 | 133678 | 0.00 |
| social sciences (%) | 10.3 | 30.4 | 182792 | 1.4 | 11.6 | 133678 | 0.00 |
| arts (%) | 1.9 | 13.6 | 182792 | 1.0 | 9.8 | 133678 | 0.00 |
| High school: technical track (%) | 12.8 | 33.5 | 182792 | 25.0 | 43.3 | 133678 | 0.00 |
| High school: vocational track (%) | 1.2 | 10.9 | 182792 | 1.7 | 12.9 | 133678 | 0.00 |
| High school grade | 83.6 | 11.6 | 178593 | 80.8 | 12.1 | 130134 | 0.00 |
| **Field of study** | | | | | | | |
| Science, chemistry, biology (%) | 13.3 | 34.0 | 182792 | 13.1 | 33.7 | 133678 | 0.06 |
| Engineering (%) | 8.2 | 27.5 | 182792 | 27.0 | 44.4 | 133678 | 0.00 |
| Humanities (%) | 24.7 | 43.1 | 182792 | 10.4 | 30.5 | 133678 | 0.00 |
| Political and social sciences (%) | 11.6 | 32.0 | 182792 | 7.9 | 26.9 | 133678 | 0.00 |
| Economics and statistics (%) | 18.5 | 38.9 | 182792 | 24.3 | 42.9 | 133678 | 0.00 |
| Psychology (%) | 11.7 | 32.1 | 182792 | 3.1 | 17.4 | 133678 | 0.00 |
| Healthcare (%) | 4.0 | 19.7 | 182792 | 2.1 | 14.4 | 133678 | 0.00 |
| Architecture (%) | 3.9 | 19.5 | 182792 | 4.9 | 21.6 | 133678 | 0.00 |
| Agriculture (%) | 1.9 | 13.8 | 182792 | 2.9 | 16.9 | 133678 | 0.00 |
| **Family background** | | | | | | | |
| Mother: university education (%) | 18.9 | 39.2 | 167637 | 22.0 | 41.4 | 119745 | 0.00 |
| Father: university education (%) | 20.0 | 40.0 | 167637 | 24.0 | 42.7 | 119745 | 0.00 |
| Mother: high-school education (%) | 50.3 | 50.0 | 167637 | 50.9 | 50.0 | 119745 | 0.00 |
| Father: high-school education (%) | 46.0 | 49.8 | 167637 | 47.2 | 49.9 | 119745 | 0.00 |
| Mother is employed (%) | 71.0 | 45.4 | 163753 | 73.2 | 44.3 | 116921 | 0.00 |
| Father is employed (%) | 99.3 | 8.0 | 162735 | 99.4 | 7.5 | 117051 | 0.00 |
| Mother: low SES (%) | 59.5 | 49.1 | 163753 | 55.8 | 49.7 | 116921 | 0.00 |
| Mother: medium SES (%) | 30.2 | 45.9 | 163753 | 32.6 | 46.9 | 116921 | 0.00 |
| Mother: high SES (%) | 10.4 | 30.5 | 163753 | 11.6 | 32.0 | 116921 | 0.00 |
| Father: low SES (%) | 45.6 | 49.8 | 162735 | 40.5 | 49.1 | 117051 | 0.00 |
| Father: medium SES (%) | 23.1 | 42.1 | 162735 | 24.2 | 42.8 | 117051 | 0.00 |
| Father: high SES (%) | 31.3 | 46.4 | 162735 | 35.3 | 47.8 | 117051 | 0.00 |

Notes. The table compares mean characteristics between female and male students in the sample. Variables in this panel were collected in the administrative data and in the institutional survey.

TABLE A.2. Summary Statistics of Initial Job Characteristics, Search Behavior and Civil Status

| | Female | | | Male | | | |
|---|---|---|---|---|---|---|---|
| | Mean | SD | Obs | Mean | SD | Obs | P-value |
| Respond to follow-up survey (%) | 73.7 | 44.0 | 182792 | 73.2 | 44.3 | 133678 | 0.00 |
| Married/cohabiting with partner (%) | 16.1 | 36.8 | 134506 | 9.5 | 29.3 | 97709 | 0.00 |
| Has children (%) | 3.7 | 18.8 | 134514 | 2.2 | 14.6 | 97724 | 0.00 |
| **LM Participation at time of survey** | | | | | | | |
| Employed (%) | 53.9 | 49.8 | 134681 | 61.8 | 48.6 | 97823 | 0.00 |
| Internship (%) | 13.4 | 34.1 | 134411 | 10.0 | 30.0 | 97615 | 0.00 |
| Searching for a job (%) | 20.5 | 40.3 | 134558 | 15.1 | 35.9 | 97707 | 0.00 |
| Graduate education (%) | 5.9 | 23.5 | 134414 | 4.0 | 19.6 | 97613 | 0.00 |
| In a PhD program (%) | 5.7 | 23.2 | 134476 | 8.1 | 27.3 | 97663 | 0.00 |
| **Job characteristics (if employed)** | | | | | | | |
| Net monthly earnings (€) | 1077.8 | 499.3 | 69659 | 1324.5 | 509.6 | 57494 | 0.00 |
| Weekly hours worked | 32.9 | 13.2 | 72378 | 38.6 | 10.9 | 60297 | 0.00 |
| Full-time job (%) | 69.3 | 46.1 | 69659 | 86.2 | 34.5 | 57494 | 0.00 |
| Hourly wage | 8.9 | 6.4 | 69659 | 8.9 | 5.7 | 57494 | 0.67 |
| High earnings occupation (%) | 36.6 | 48.2 | 71062 | 61.3 | 48.7 | 59558 | 0.00 |
| High fulltime occupation (%) | 51.1 | 50.0 | 71062 | 74.3 | 43.7 | 59558 | 0.00 |
| High earnings industry (%) | 34.1 | 47.4 | 71264 | 47.8 | 50.0 | 59715 | 0.00 |
| High fulltime industry (%) | 38.0 | 48.5 | 71264 | 61.7 | 48.6 | 59715 | 0.00 |
| Permanent contract (%) | 22.8 | 42.0 | 72280 | 29.3 | 45.5 | 60237 | 0.00 |
| Fixed-term contract (%) | 50.9 | 50.0 | 72280 | 49.1 | 50.0 | 60237 | 0.00 |
| Self-employment | 16.5 | 37.2 | 72280 | 16.1 | 36.8 | 60237 | 0.04 |
| No contract (%) | 6.7 | 25.0 | 72280 | 3.1 | 17.2 | 60237 | 0.00 |
| Private sector (%) | 76.7 | 42.3 | 72420 | 86.2 | 34.5 | 60348 | 0.00 |
| Public sector (%) | 16.5 | 37.1 | 72420 | 11.0 | 31.3 | 60348 | 0.00 |
| No profit (%) | 6.8 | 25.2 | 72420 | 2.8 | 16.4 | 60348 | 0.00 |
| **Mismatch** | | | | | | | |
| Use skills acquired during Master (%) | 42.1 | 49.4 | 72446 | 47.6 | 49.9 | 60352 | 0.00 |
| Master necessary for current job (%) | 47.6 | 49.9 | 72437 | 54.8 | 49.8 | 60319 | 0.00 |
| Training in the Master good for job (%) | 51.3 | 50.0 | 72413 | 56.6 | 49.6 | 60312 | 0.00 |
| **Job search process** | | | | | | | |
| Keep job found during studies (%) | 25.8 | 43.8 | 72539 | 21.6 | 41.2 | 60415 | 0.00 |
| Current job: first job after grad. (%) | 80.2 | 39.8 | 72568 | 81.1 | 39.2 | 60435 | 0.00 |
| Numbers of jobs from grad. | 1.3 | 0.6 | 72788 | 1.2 | 0.6 | 60685 | 0.00 |
| Start of job search:months from grad. | 0.4 | 1.9 | 71524 | 0.4 | 1.7 | 59255 | 0.00 |
| Accepted job offer :months from grad. | 2.8 | 3.8 | 72284 | 2.5 | 3.5 | 60291 | 0.00 |
| On the job search (%) | 20.9 | 40.7 | 134674 | 19.9 | 39.9 | 97818 | 0.00 |

Notes. The table compares mean characteristics between female and male students in the sample. Variables in this panel were collected in the follow-up survey conducted one year after graduation.

FIGURE A.1. Heatmaps of other indicators of gender culture

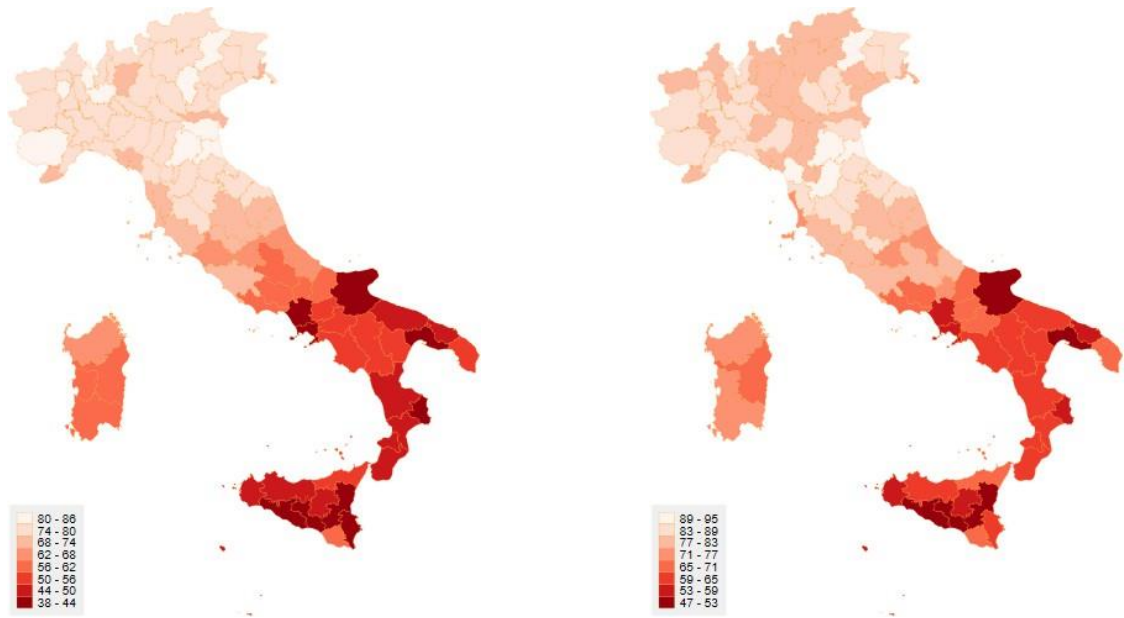

(a) Female LFP (%) - Age: 25-34          (b) Ratio of female to male LFP (%) - Age: 25-34

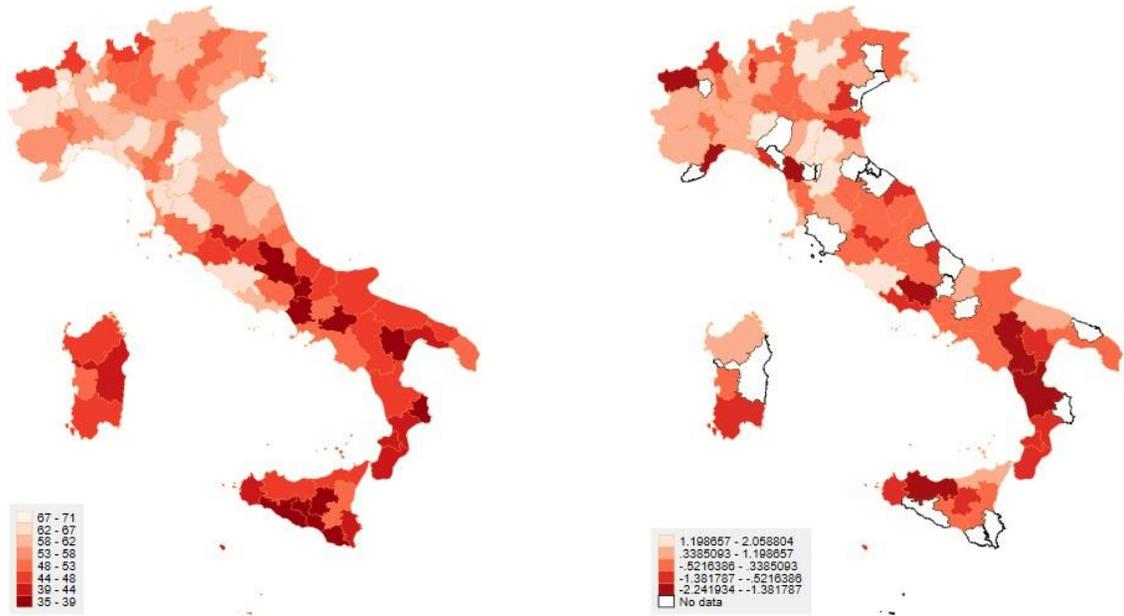

(c) Percentage of firms without preference for male workers

(d) Standardised index of individual gender attitudes

Notes. Panel (a) and (b) present the FLFP and the ratio FLFP/MLFP of young women (25-34) in Italy. These are constructed as averages from 2004-2007. Panel (c) presents the percentage of firms without hiring preferences for male workers in 2003. Panel (d) presents a standardised index of gender culture, based on individual answers to questions related to gender attitudes in the World Value Survey (2000). Provinces white represent those with missing values. All of these measured are defined at the province level.

TABLE A.3.  Summary statistics of measures of gender culture in the sample, by gender

| | Female | | Male | | |
|---|---|---|---|---|---|
| | Mean | SD | Mean | SD | P-value |
| Female labor force participation (age: 15-64) | 49.3 | 11.2 | 50.2 | 11.1 | 0.00 |
| Female/Male labor force participation (age: 15-64) | 66.3 | 11.9 | 67.2 | 11.8 | 0.00 |
| Female labor force participation (age: 25-34) | 64.6 | 15.3 | 65.8 | 15.1 | 0.00 |
| Female/Male labor force participation (age: 25-34) | 73.9 | 13.1 | 74.9 | 12.9 | 0.00 |
| Male labor force participation (age: 15-64) | 73.7 | 4.6 | 74.0 | 4.5 | 0.00 |
| Male labor force participation (age: 25-34) | 86.5 | 6.5 | 87.0 | 6.4 | 0.00 |
| % of female graduates in full-time job | 56.0 | 9.6 | 56.8 | 9.5 | 0.00 |
| % of female/male graduates in full-time job | 71.6 | 6.7 | 72.1 | 6.6 | 0.00 |
| % of firms without hiring pref. for male workers | 34.5 | 7.8 | 35.1 | 7.8 | 0.00 |

Notes: The Table presents summary statistics f the measures of gender culture 1-5 presented in Section 3, by gender. The sample of female and male students include, respectively, 182,792 and 133,678 students. Students are assigned to provinces based on their residence province prior to enrollment in the Master.

TABLE A.4. Mobility patterns by gender in the sample

| | Female | | Male | | |
|---|---|---|---|---|---|
| | Mean | SD | Mean | SD | P-value |
| Moved to another province for Master (%) | 58.9 | 49.2 | 55.4 | 49.7 | 0.00 |
| Moved to another region for Master (%) | 31.3 | 46.4 | 29.1 | 45.4 | 0.00 |
| Bachelor and Master in same univ. (%) | 71.5 | 45.1 | 75.7 | 42.9 | 0.00 |
| **Gender culture in province of university** | | | | | |
| Female labor force participation (age: 15-64) | 52.9 | 10.5 | 53.7 | 10.2 | 0.00 |
| Female/Male labor force participation (age: 15-64) | 70.2 | 11.3 | 71.0 | 11.0 | 0.00 |
| Female labor force participation (age: 25-34) | 69.1 | 14.2 | 70.2 | 13.7 | 0.00 |
| Female/Male labor force participation (age: 25-34) | 78.2 | 11.9 | 79.0 | 11.5 | 0.00 |
| Male labor force participation (age: 15-64) | 74.9 | 4.0 | 75.1 | 3.8 | 0.00 |
| Male labor force participation (age: 25-34) | 87.7 | 6.1 | 88.2 | 5.9 | 0.00 |
| % of female graduates in full-time | 58.8 | 8.8 | 59.7 | 8.6 | 0.00 |
| % of female/male graduates in full-time | 73.3 | 6.0 | 73.9 | 5.8 | 0.00 |
| % of firms without hiring pref. for male workers | 58.5 | 8.2 | 59.1 | 7.8 | 0.00 |

Notes. The table provides summary statistics regarding students' mobility for their studies. Besides mobility rates by gender, it provides information on the local gender culture in the province of studies of students. The sample of female and male students include, respectively, 182,792 and 133,678 students.

| | Female | | Male | | |
|---|---|---|---|---|---|
| | Mean | SD | Mean | SD | P-value |
| Work in province of studies (%) | 45.2 | 49.8 | 43.7 | 49.6 | 0.0 |
| Work in region of studies (%) | 68.4 | 46.5 | 65.2 | 47.6 | 0.0 |
| Work abroad (%) | 5.0 | 21.8 | 5.3 | 22.4 | 0.0 |
| Work outside province of origin (%) | 44.0 | 49.6 | 51.4 | 50.0 | 0.0 |
| **Gender culture in province of work** | | | | | |
| Female labor force participationn (age:15-64) | 54.6 | 9.8 | 55.6 | 9.2 | 0.0 |
| Female/Male labor force participationn (age:15-64) | 71.7 | 10.4 | 72.8 | 9.8 | 0.0 |
| Female labor force participationn (age:25-34) | 71.7 | 13.2 | 73.0 | 12.3 | 0.0 |
| Female/Male labor force participationn (age:25-34) | 79.7 | 11.0 | 80.8 | 10.3 | 0.0 |
| Male labor force participationn (age:15-64) | 75.7 | 3.8 | 76.0 | 3.6 | 0.0 |
| Male labor force participationn (age:25-34) | 89.2 | 5.7 | 89.8 | 5.4 | 0.0 |
| % of female graduates in full-time | 60.6 | 8.5 | 61.6 | 8.1 | 0.0 |
| % of female/male graduates in full-time | 74.6 | 5.9 | 75.2 | 5.6 | 0.0 |
| % of firms without hiring pref. for male workers | 58.5 | 8.3 | 59.3 | 8.1 | 0.0 |

Notes. The table provides summary statistics regarding students' mobility to local labor markets in their first job after graduation. Besides mobility rates by gender, it provides information on the local gender culture in the province of work of students. The sample includes female and male students who are employed at the moment of the survey, corresponding to 69,659 and 57,494 students.

FIGURE A.2. Students from below-median FLFP provinces within degrees

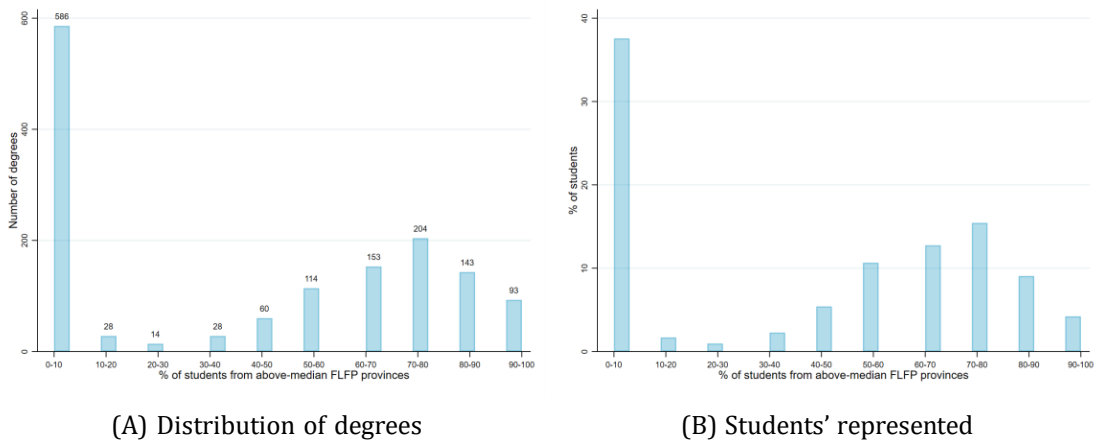

(A) Distribution of degrees        (B) Students' represented

Notes. Panel (A) represents degrees by the % of students from above-median FLFP areas in 2016, categorize in brackets 0-10, ..., 90-100. One unit corresponds to a degree (N=1,572). Panel (B) shows the percentage of students represented by degrees that fall within each of these brackets.

TABLE A.6. Mobility patterns by gender culture in province of origin (Female sample)

| | Q1 FLFP (N=48,896) | | Q4 FLFP (N=44,103) | | |
| --- | --- | --- | --- | --- | --- |
| | Mean | SD | Mean | SD | P-value |
| Moved to another province for Master (%) | 57.8 | 49.4 | 63.0 | 48.3 | 0.00 |
| Moved to another region for Master (%) | 37.2 | 48.3 | 27.1 | 44.5 | 0.00 |
| Work in different province than birth (%) | 54.8 | 49.8 | 37.8 | 48.5 | 0.00 |
| **Types of mobility** (only for movers) | | | | | |
| FLFP (age: 15-64) in prov. of university | 49.7 | 11.7 | 60.0 | 4.4 | 0.00 |
| Size of university | 33797.4 | 16373.0 | 36606.5 | 18288.8 | 0.00 |
| Nb. of students in the degree | 80.3 | 61.8 | 80.9 | 59.4 | 0.21 |
| % of female students in the degree | 69.0 | 18.6 | 65.6 | 18.5 | 0.00 |
| % of movers in the degree | 60.4 | 20.6 | 72.0 | 14.8 | 0.00 |
| % of movers (region) in the degree | 32.6 | 25.7 | 41.9 | 19.4 | 0.00 |
| % of peers from above-median FLFP prov | 27.0 | 31.3 | 67.1 | 17.4 | 0.00 |
| **Field of study** | | | | | |
| Science, chemistry, biology (%) | 13.8 | 34.5 | 12.6 | 33.2 | 0.00 |
| Engineering (%) | 8.3 | 27.5 | 6.2 | 24.2 | 0.00 |
| Humanities (%) | 25.7 | 43.7 | 27.1 | 44.5 | 0.00 |
| Political and social sciences (%) | 11.7 | 32.2 | 11.9 | 32.4 | 0.45 |
| Economics and statistics (%) | 14.2 | 34.9 | 17.0 | 37.6 | 0.00 |
| Psychology (%) | 15.6 | 36.3 | 12.3 | 32.9 | 0.00 |
| Healthcare (%) | 4.3 | 20.2 | 3.8 | 19.1 | 0.00 |
| Architecture (%) | 2.7 | 16.1 | 4.8 | 21.5 | 0.00 |
| Agriculture (%) | 1.4 | 11.8 | 2.2 | 14.6 | 0.00 |

Notes. The table provides summary statistics regarding students' mobility in the sample of female students, contrasting students from provinces in the first vs. fourth quartiles of FLFP. Besides mobility rates by gender, it provides information on the characteristics of mobility in the sample of movers (57.8% and 63% of the two samples).

TABLE A.7. Average characteristics of degrees

|  | Mean | SD | p50 | Min | Max |
|---|---|---|---|---|---|
| Size of degree | 47.0 | 43.0 | 34.0 | 4.0 | 410.0 |
| % of female students | 55.6 | 21.3 | 56.3 | 3.8 | 97.1 |
| % of movers | 55.3 | 23.2 | 56.3 | 0.0 | 100.0 |
| % of movers (region) | 28.5 | 23.1 | 25.0 | 0.0 | 91.7 |
| % of students from above-median FLFP provs. | 40.8 | 36.3 | 50.0 | 0.0 | 100.0 |
| % of students with BSc at same univ. | 72.9 | 22.6 | 78.4 | 0.0 | 100.0 |

The table presents average characteristics of degrees related to students' composition. The unit of observation is a degree. There are 1,572 degrees in the sample.

TABLE A.8. Gender differences in the extensive margin of labor supply

|  | (1) Has contract in LM | (2) Employed today | (3) Internship today | (4) Has worked |
|---|---|---|---|---|
| Female | 0.000 | -0.014*** | 0.015*** | -0.002 |
|  | (0.000) | (0.003) | (0.002) | (0.003) |
| Degree FEs | ✓ | ✓ | ✓ | ✓ |
| Cohort FEs | ✓ | ✓ | ✓ | ✓ |
| Observations | 232,026 | 232,504 | 232,026 | 232,504 |
| R-squared | 0.114 | 0.128 | 0.078 | 0.104 |

Notes: The table reports coefficients from regressions of graduates' labor market participation on a female dummy, after including degree and cohort fixed effects. The dependent variables are as follows. Column 1: indicator for whether a student is participating in the labor market with either a standard contract or an internship. Column 2: indicator for whether a student is employed with a standard contract one year after graduation. Column 3: indicator for whether a student is employed with an internship contract. Column 4: indicator for whether a student has been employed at any point during the first year after graduation. Standard errors are clustered at degree level.

TABLE A.9. The gender earnings gap at labor market entry, with controls for job types

|  | (1) Log(monthly earnings) | (2) Log(weekly hours) | (3) Pr(fulltime) | (4) Log(wage) |
|---|---|---|---|---|
| Female | -0.087*** | -0.057*** | -0.032*** | -0.030*** |
|  | (0.003) | (0.003) | (0.002) | (0.003) |
| GPA | ✓ | ✓ | ✓ | ✓ |
| Job characteristics | ✓ | ✓ | ✓ | ✓ |
| Degree FEs | ✓ | ✓ | ✓ | ✓ |
| Cohort FEs | ✓ | ✓ | ✓ | ✓ |
| Observations | 127,153 | 127,153 | 127,153 | 127,153 |
| R-squared | 0.407 | 0.382 | 0.381 | 0.181 |

Notes: The table reports coefficients from regressions of graduates' labor market outcomes on a female dummy, after including degree and cohort fixed effects and controlling for covariates (GPA, prov. of work FEs, occupation FEs (20 classes), industry FEs (21 classes)). The sample consists of female and male students who are employed one year post graduation. Standard errors are clustered at the degree level.

TABLE A.10. The gender earnings gap excluding individuals with children or married

|  | (1) Log(monthly earnings) | (2) Log(weekly hours) | (3) Pr(fulltime) | (4) Log(wage) |
|---|---|---|---|---|
| Female | -0.104*** | -0.079*** | -0.045*** | -0.030*** |
|  | (0.004) | (0.004) | (0.003) | (0.003) |
| Degree FEs | ✓ | ✓ | ✓ | ✓ |
| Cohort FEs | ✓ | ✓ | ✓ | ✓ |
| Observations | 106,360 | 106,360 | 106,360 | 106,360 |
| Nb. of degrees | 1,570 | 1,570 | 1,570 | 1,570 |
| R-squared | 0.313 | 0.269 | 0.309 | 0.093 |

Notes: The table reports coefficients from regressions of graduates' labor market outcomes on a female dummy, after including degree and cohort fixed effects. The sample consists of female and male students who are employed one year after graduation, excluding individuals with children or those who are married or cohabiting with a partner. Standard errors are clustered at the degree (master x university) level.

TABLE A.11. Selection of female movers by FLFP in place of birth

|  | (1) Low FLFP | (2) High FLFP | (3) P-value | (4) Observations |
|---|---|---|---|---|
| **Individual characteristics** | | | | |
| Age at enrollment | 23.84 | 23.83 | 0.84 | 16,496 |
| Bachelor grade | 101.01 | 101.84 | 0.00 | 14,736 |
| High school type: general track (%) | 88.53 | 81.31 | 0.00 | 16,496 |
| science (%) | 47.50 | 41.50 | 0.00 | 16,496 |
| humanities (%) | 23.63 | 15.16 | 0.00 | 16,496 |
| foreign language (%) | 9.10 | 13.13 | 0.00 | 16,496 |
| social sciences (%) | 7.17 | 9.64 | 0.00 | 16,496 |
| arts (%) | 1.12 | 1.88 | 0.00 | 16,496 |
| High school type: technical track (%) | 9.87 | 15.81 | 0.00 | 16,496 |
| High school type: vocational track (%) | 0.90 | 0.80 | 0.54 | 16,496 |
| **Family background** | | | | |
| Years of educucation mother | 12.22 | 12.48 | 0.00 | 14,958 |
| Years of educucation father | 12.13 | 12.09 | 0.63 | 14,958 |
| Mother: university degree (%) | 19.03 | 19.67 | 0.51 | 15,185 |
| Father: university degree (%) | 19.82 | 19.58 | 0.82 | 15,185 |
| Mother: high-school degree (%) | 48.74 | 50.27 | 0.14 | 14,958 |
| Father: high-school degree (%) | 45.60 | 44.07 | 0.18 | 14,958 |
| Mother is in the LF (%) | 62.24 | 81.84 | 0.00 | 14,866 |
| Father is in the LF (%) | 99.36 | 99.38 | 0.93 | 14,766 |
| Mother: low SES (%) | 37.61 | 45.79 | 0.00 | 10,503 |
| Mother: medium SES (%) | 48.83 | 39.84 | 0.00 | 10,503 |
| Mother: high SES (%) | 13.55 | 14.37 | 0.38 | 10,503 |
| Father: low SES (%) | 45.28 | 45.21 | 0.95 | 14,671 |
| Father: medium SES (%) | 26.39 | 21.11 | 0.00 | 14,671 |
| Father: high SES (%) | 28.33 | 33.68 | 0.00 | 14,671 |

Notes: This table examines the selection of female movers into degree programs based on their province of birth, categorized by the FLFP (top versus bottom quartile). The table compares the ability, educational histories and socio-economic background of female movers by place of birth. For each pre-determined characteristic, a separate regression is estimated following the empirical model 1. Predicted values for each group are presented in Columns 1 and 2, while Column 3 reports the p-value from a significance test on $\alpha$.

| | (1) | (2) | (3) | (4) | (5) | (6) |
|---|---|---|---|---|---|---|
| | Log(weekly hours) | | | Pr(fulltime) | | |
| Q4 vs. Q1 FLFP | 0.039*** | 0.039*** | 0.038*** | 0.011* | 0.008 | 0.08 |
| | (0.008) | (0.008) | (0.009) | (0.006) | (0.006) | (0.006) |
| Province of job FEs | | ✓ | ✓ | | ✓ | ✓ |
| GPA | | | ✓ | | | ✓ |
| Mother's occupation | | | ✓ | | | ✓ |
| Father's occupation | | | ✓ | | | ✓ |
| Degree FEs | ✓ | ✓ | ✓ | ✓ | ✓ | ✓ |
| Cohort FEs | ✓ | ✓ | ✓ | ✓ | ✓ | ✓ |
| N | 15,597 | 15,595 | 14,014 | 15,597 | 15,595 | 14,014 |

Notes: The table reports coefficients from separate regressions of men's labor market outcomes on a dummy variable indicating whether the student originates from a province with FLFP in the highest vs. lowest quartile. All regressions include controls for degree and cohort fixed effects. The sample consists of male movers, defined as men working in a different province from their birth province, who are employed one year post-graduation. Variations in sample sizes across columns arise from missing parental background data for some students. Standard errors are clustered at the degree level.

TABLE A.13. Balancing tests for cohort composition - Female students

**Panel A. Educational history and ability**

| | (1) Age | (2) HS science | (3) HS humanities | (4) HS foregin languages | (5) HS social sciences | (6) HS vocational | (7) BSc grade | (8) Bsc grade > p50 |
|---|---|---|---|---|---|---|---|---|
| (Mean) | (24.3) | (0.40) | (0.21) | (0.11) | (0.10) | (0.01) | (101.3) | (0.50) |
| $\hat{\delta}^{FP}$ | -0.030 (0.092) | -0.003 (0.006) | 0.002 (0.005) | 0.001 (0.004) | 0.003 (0.004) | 0.002 (0.001) | 0.078 (0.124) | -0.001 (0.008) |
| $\hat{\delta}^{MP}$ | -0.110 (0.095) | -0.009* (0.005) | -0.000 (0.005) | 0.002 (0.003) | -0.004 (0.004) | 0.001 (0.001) | -0.009 (0.090) | -0.001 (0.006) |
| Degree FEs | ✓ | ✓ | ✓ | ✓ | ✓ | ✓ | ✓ | ✓ |
| Cohort FEs | ✓ | ✓ | ✓ | ✓ | ✓ | ✓ | ✓ | ✓ |
| Observations | 182,792 | 182,792 | 182,792 | 182,792 | 182,792 | 182,792 | 162,091 | 162,091 |
| R-squared | 0.168 | 0.139 | 0.107 | 0.125 | 0.128 | 0.020 | 0.204 | 0.149 |

**Panel B. Parental background**

| | (1) Mother: univ. | (2) Father: univ. | (3) Mother: HS | (4) Father: HS | (5) Mother: low SES | (6) Mother: high SES | (7) Father: low SES | (8) Father: high SES |
|---|---|---|---|---|---|---|---|---|
| (Mean) | (0.19) | (0.20) | (0.48) | (0.44) | (0.43) | (0.15) | (0.45) | (0.32) |
| $\hat{\delta}^{FP}$ | -0.002 (0.006) | -0.009 (0.006) | 0.007 (0.007) | -0.005 (0.007) | -0.009 (0.008) | -0.004 (0.006) | 0.007 (0.007) | -0.007 (0.007) |
| $\hat{\delta}^{MP}$ | -0.004 (0.004) | -0.006 (0.004) | -0.001 (0.005) | 0.006 (0.005) | 0.007 (0.006) | -0.002 (0.005) | 0.008 (0.006) | 0.001 (0.005) |
| Degree FEs | ✓ | ✓ | ✓ | ✓ | ✓ | ✓ | ✓ | ✓ |
| Cohort FEs | ✓ | ✓ | ✓ | ✓ | ✓ | ✓ | ✓ | ✓ |
| Observations | 167,637 | 167,637 | 167,637 | 167,637 | 116,206 | 116,206 | 161,674 | 161,674 |
| R-squared | 0.043 | 0.042 | 0.019 | 0.015 | 0.039 | 0.026 | 0.036 | 0.038 |

Notes: The table reports OLS estimates of $\hat{\delta}^{FP}$ and $\hat{\delta}^{MP}$ from the baseline model (equation 2). Each column corresponds to a different regression, where the dependent variables are pre-determined covariate of a student, related to educational history and ability (Panel A), and parental background (Panel B). Regressions include cohort and degree fixed effects. The sample consists of all female students in the sample (N=182,792). Variations in sample sizes across columns arise from missing information on some of the covariates. All regressors are standardised. Standard errors are clustered at degree level.

TABLE A.14. Balancing tests for cohort composition - Male students

**Panel A. Educational history and ability**

| | (1) Age | (2) HS science | (3) HS humanities | (4) HS foreign languages | (5) HS social sciences | (6) HS vocational | (7) BSc grade | (8) Bsc grade > p50 |
|---|---|---|---|---|---|---|---|---|
| (Mean) | (24.5) | (0.56) | (0.10) | (0.02) | (0.01) | (0.02) | (99.1) | (0.50) |
| $\hat{\delta}^{FP}$ | 0.045 | 0.010 | -0.006 | 0.000 | -0.001 | 0.000 | 0.239* | 0.017** |
| | (0.059) | (0.006) | (0.004) | (0.002) | (0.001) | (0.002) | (0.129) | (0.008) |
| $\hat{\delta}^{MP}$ | -0.070 | -0.003 | 0.004 | -0.002 | 0.003 | -0.000 | -0.136 | -0.005 |
| | (0.074) | (0.007) | (0.005) | (0.002) | (0.002) | (0.002) | (0.124) | (0.008) |
| Degree FEs | ✓ | ✓ | ✓ | ✓ | ✓ | ✓ | ✓ | ✓ |
| Cohort FEs | ✓ | ✓ | ✓ | ✓ | ✓ | ✓ | ✓ | ✓ |
| Observations | 133,678 | 133,678 | 133,678 | 133,678 | 133,678 | 133,678 | 116,256 | 116,256 |
| R-squared | 0.218 | 0.106 | 0.121 | 0.070 | 0.055 | 0.044 | 0.234 | 0.173 |

**Panel B. Parental background**

| | (1) Mother: univ. | (2) Father: univ. | (3) Mother: HS | (4) Father: HS | (5) Mother: med SES | (6) Mother: high SES | (7) Father: med SES | (8) Father: high SES |
|---|---|---|---|---|---|---|---|---|
| (Mean) | (0.22) | (0.24) | (0.49) | (0.45) | (0.40) | (0.16) | (0.40) | (0.36) |
| $\hat{\delta}^{FP}$ | -0.000 | -0.012** | -0.012* | 0.005 | -0.001 | -0.000 | 0.009 | -0.007 |
| | (0.006) | (0.006) | (0.007) | (0.007) | (0.008) | (0.006) | (0.007) | (0.007) |
| $\hat{\delta}^{MP}$ | -0.005 | -0.001 | 0.000 | 0.004 | -0.004 | -0.004 | -0.015** | 0.014* |
| | (0.006) | (0.006) | (0.007) | (0.007) | (0.008) | (0.006) | (0.007) | (0.007) |
| Degree FEs | ✓ | ✓ | ✓ | ✓ | ✓ | ✓ | ✓ | ✓ |
| Cohort FEs | ✓ | ✓ | ✓ | ✓ | ✓ | ✓ | ✓ | ✓ |
| Observations | 119,743 | 119,743 | 119,743 | 119,743 | 85,590 | 85,590 | 116,389 | 116,389 |
| R-squared | 0.037 | 0.036 | 0.020 | 0.019 | 0.038 | 0.030 | 0.033 | 0.041 |

Notes: The table reports OLS estimates of $\hat{\delta}^{FP}$ and $\hat{\delta}^{MP}$ from the baseline model (equation 2). Each column corresponds to a different regression, where the dependent variables are pre-determined covariate of a student, related to educational history and ability (Panel A), and parental background (Panel B). Regressions include cohort and degree fixed effects. The sample consists of all male students in the sample (N=133,678). Variations in sample sizes across columns arise from missing information on some of the covariates. All regressors are standardised. Standard errors are clustered at degree level.

FIGURE A.3. Year-to-Year Variation in Students' Geographical Origins

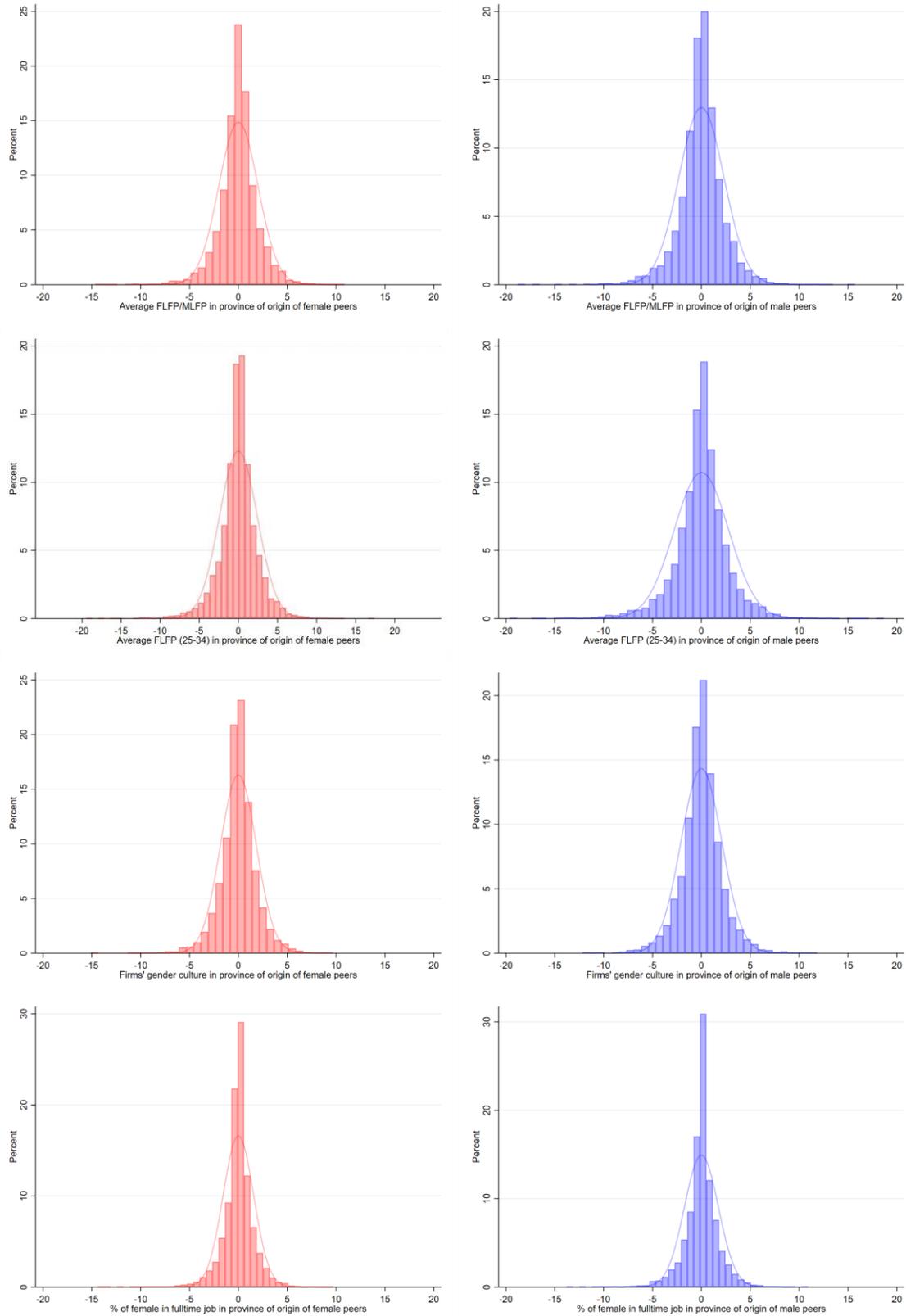

Notes: The figure plots the distribution of residuals from a OLS regression of the average characteristic in the province of origin of female (red) or male students (blue) on cohort and degree fixed effects. One observation corresponds to a degree-cohort pair. Histograms are presented by bins of 0.75. The normal distribution is plotted for comparison.

TABLE A.15. Raw and Residual Variation of Additional Measures of Peers' Gender Culture

|  | Mean | SD | Min | Max |
|---|---|---|---|---|
| **A: Avg FLFP (25-34) in province of origin of female peers** | | | | |
| Raw cohort variable | 65.00 | 11.93 | 39.85 | 85.00 |
| Residuals: net of degree and cohort fixed effects | 0.00 | 2.43 | -19.43 | 16.92 |
| **B: Avg FLFP (25-34) in province of origin of male peers** | | | | |
| Raw cohort variable | 65.09 | 12.02 | 39.33 | 85 |
| Residuals: net of degree and cohort fixed effects | 0.00 | 2.81 | -24.04 | 18.37 |
| **C: Avg FLFP/MLFP in province of origin of female peers** | | | | |
| Raw cohort variable | 66.76 | 8.88 | 43.62 | 85.36 |
| Residuals: net of degree and cohort fixed effects | 0.00 | 2.01 | -14.64 | 10.59 |
| **D: Avg FLFP/MLFP in province of origin of male peers** | | | | |
| Raw cohort variable | 66.81 | 8.99 | 43.02 | 85.29 |
| Residuals: net of degree and cohort fixed effects | 0.00 | 2.31 | -18.80 | 15.40 |
| **E: % of fulltime female graduates in prov. of female peers** | | | | |
| Raw cohort variable | 55.93 | 7.53 | 40.11 | 68.93 |
| Residuals: net of degree and cohort fixed effects | 0.00 | 1.56 | -14.37 | 9.04 |
| **F: % of fulltime female graduates in prov. of male peers** | | | | |
| Raw cohort variable | 55.95 | 7.59 | 40.11 | 68.93 |
| Residuals: net of degree and cohort fixed effects | 0.00 | 1.74 | -13.79 | 10.48 |
| **G: Firms gender culture in province of origin of female peers** | | | | |
| Raw cohort variable | 53.88 | 5.88 | 37.00 | 71.00 |
| Residuals: net of degree and cohort fixed effects | 0.00 | 1.83 | -15.09 | 9.31 |
| **H: Firms gender culture in province of origin of male peers** | | | | |
| Raw cohort variable | 54.09 | 6.04 | 36.00 | 68.00 |
| Residuals: net of degree and cohort fixed effects | 0.00 | 2.09 | -12.18 | 11.08 |

Notes: The table reports descriptive statistics for the main measures of gender culture in the province of origin of female (Panel A) and male (Panel B) peers, before and after removing degree and cohort fixed effects. The unit of observation is a degree-cohort pair, leading to a total of 7,160 observations.

FIGURE A.4. Time series of exposure to peers by deciles of program's size

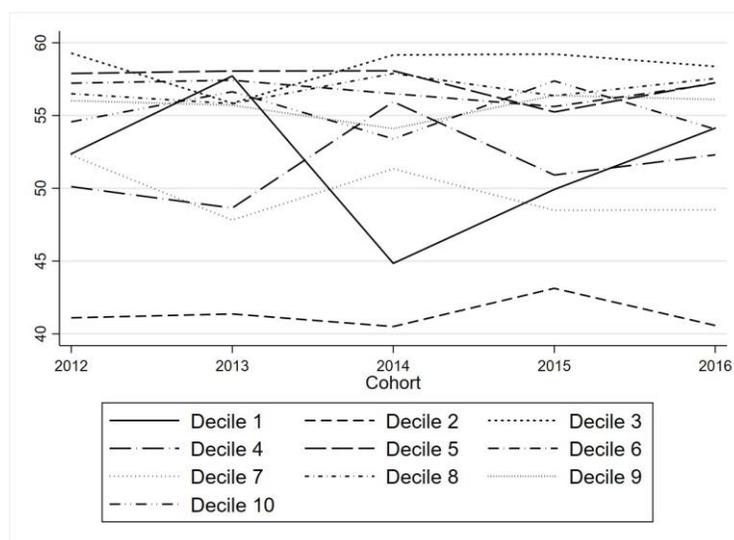

Notes: This figure plots the evolution in time series of the average FLFP in the province of origin of female peers within 10 randomly picked degrees. All programs were divided into deciles based on the average size across all years. One program was randomly chosen within each decile.

TABLE A.16. Raw and Residual Variation of Peers' Gender Culture by Quintiles of Degree Size

| | Mean | SD | Min | Max |
|---|---|---|---|---|
| **A: Avg FLFP in province of origin of female peers** | | | | |
| **Quintile 1** (<21 students) | | | | |
| Raw cohort variable | 48.68 | 8.87 | 32.09 | 66.18 |
| Residuals: net of degree and cohort FEs | -0.00 | 2.37 | -13.83 | 9.05 |
| **Quintile 2** (21-31 students) | | | | |
| Raw cohort variable | 50.79 | 7.80 | 31.14 | 66.18 |
| Residuals: net of degree and cohort FEs | 0.00 | 2.01 | -9.68 | 8.21 |
| **Quintile 3** (32-42 students) | | | | |
| Raw cohort variable | 50.00 | 8.45 | 29.89 | 66.18 |
| Residuals: net of degree and cohort FEs | 0.00 | 1.91 | -12.82 | 8.40 |
| **Quintile 4** (43-70 students) | | | | |
| Raw cohort variable | 49.73 | 8.38 | 31.02 | 65.99 |
| Residuals: net of degree and cohort FEs | 0.00 | 1.54 | -9.77 | 10.78 |
| **Quintile 5** (70-413 students) | | | | |
| Raw cohort variable | 49.72 | 8.35 | 30.06 | 63.19 |
| Residuals: net of degree and cohort FEs | 0.00 | 1.23 | -10.72 | 4.77 |
| | | | | |
| **B: Avg FLFP in province of origin of male peers** | | | | |
| **Quintile 1** (<21 students) | | | | |
| Raw cohort variable | 48.76 | 9.09 | 29.87 | 66.18 |
| Residuals: net of degree and cohort FEs | -0.00 | 2.37 | -11.36 | 11.33 |
| **Quintile 2** (21-31 students) | | | | |
| Raw cohort variable | 50.76 | 7.86 | 32.09 | 65.10 |
| Residuals: net of degree and cohort FEs | 0.00 | 2.24 | -13.17 | 9.65 |
| **Quintile 3** (32-42 students) | | | | |
| Raw cohort variable | 50.18 | 8.54 | 29.49 | 65.33 |
| Residuals: net of degree and cohort FEs | 0.00 | 2.11 | -11.47 | 9.11 |
| **Quintile 4** (43-70 students) | | | | |
| Raw cohort variable | 49.78 | 8.40 | 29.77 | 63.71 |
| Residuals: net of degree and cohort FEs | 0.00 | 1.91 | -9.50 | 14.47 |
| **Quintile 5** (71-413 students) | | | | |
| Raw cohort variable | 49.85 | 8.36 | 29.48 | 66.37 |
| Residuals: net of degree and cohort FEs | 0.00 | 1.55 | -9.61 | 11.76 |

Notes: The table reports descriptive statistics for the average FLFP in the province of origin for of female (Panel A) and male (Panel B) peers. These statistics are provided for groups of degrees categorized by quintiles based on the size of the degree programs. Quintiles are determined using the average size of programs calculated over five cohorts. The number of students in each program is indicated in parentheses next to the corresponding quintile. For instance, degrees in the first quintile include programs with fewer than 21 students.

TABLE A.17. Estimates of Peer Effects on Job Characteristics - Female sample

|  | (1) Permanent | (2) No contract | (3) Self-employment | (4) Public | (5) No-profit |
|---|---|---|---|---|---|
| $\delta^{FP}$ | -0.002 (0.010) | -0.007 (0.005) | -0.003 (0.008) | 0.001 (0.010) | 0.007 (0.005) |
| $\delta^{MP}$ | -0.008 (0.007) | -0.004 (0.004) | -0.012* (0.007) | 0.001 (0.008) | 0.003 (0.005) |
| Degree FE | ✓ | ✓ | ✓ | ✓ | ✓ |
| Cohort FE | ✓ | ✓ | ✓ | ✓ | ✓ |
| Observations | 69,417 | 69,417 | 69,417 | 69,556 | 69,556 |
| R-squared | 0.137 | 0.098 | 0.144 | 0.203 | 0.153 |

Notes: OLS estimates of regressions of types of contract and sector one year after graduation on: the average FLFP in the provinces of origin of female and male peers and the FLFP in the own province of origin. All the dependent variables are indicator variables. All the estimates are done on the sample of women who are employed one year after graduation and with non-missing information on the dependent variables. Standard errors clustered at degree level. All regressors are standardised.

TABLE A.18. Estimates of Peer Effects on Female Earnings - Controls for Degree Trends

|  | (1) Log(monthly earnings) | (2) Log(weekly hours) | (3) Pr(fulltime) | (4) Log(hourly wage) |
|---|---|---|---|---|
| $\hat{\delta}^{FP}$ | 0.045*** (0.014) | 0.039*** (0.014) | 0.029*** (0.010) | 0.007 (0.014) |
| $\hat{\delta}^{MP}$ | -0.002 (0.011) | -0.009 (0.011) | -0.005 (0.008) | 0.006 (0.011) |
| Degree FE | ✓ | ✓ | ✓ | ✓ |
| Degree trends | ✓ | ✓ | ✓ | ✓ |
| Cohort FE | ✓ | ✓ | ✓ | ✓ |
| Observations | 69,645 | 69,645 | 69,645 | 69,645 |
| R-squared | 0.308 | 0.266 | 0.299 | 0.124 |

Notes: OLS estimates of a regression of women's earnings and labor supply one year after graduation on: the average FLFP in the provinces of origin of female and male peers and the FLFP in the own province of origin. Regressions include cohort and degree fixed effects, as well as degree-specific linear time trends. All the estimates are done on the sample of women who are employed one year after graduation and with non-missing information on the dependent variables. Standard errors clustered at degree level. All regressors are standardised.

TABLE A.19. Estimates of Peer Effects on Female Earnings - Controls for Region Trends

| | (1) Log(monthly earnings) | (2) Log(weekly hours) | (3) Pr(fulltime) | (4) Log(hourly wage) |
|---|---|---|---|---|
| $\hat{\delta}^{FP}$ | 0.036*** | 0.030** | 0.019** | 0.005 |
| | (0.013) | (0.012) | (0.009) | (0.012) |
| $\hat{\delta}^{MP}$ | -0.001 | -0.001 | -0.002 | -0.001 |
| | (0.010) | (0.009) | (0.007) | (0.009) |
| Degree FE | ✓ | ✓ | ✓ | ✓ |
| Region trends | ✓ | ✓ | ✓ | ✓ |
| Cohort FE | ✓ | ✓ | ✓ | ✓ |
| Observations | 69,645 | 69,645 | 69,645 | 69,645 |
| R-squared | 0.288 | 0.246 | 0.280 | 0.100 |

Notes: OLS estimates of a regression of women's earnings and labor supply one year after graduation on: the average FLFP in the provinces of origin of female and male peers and the FLFP in the own province of origin. Regressions include cohort and degree fixed effects, as well as region (of studies)-specific linear time trends. All the estimates are done on the sample of women who are employed one year after graduation and with non-missing information on the dependent variables. Standard errors clustered at degree level. All regressors are standardised.

TABLE A.20. Estimates of Peer Effects on Female Earnings Excl. Degrees with Trends in Size

|  | (1) Log(monthly earnings) | (2) Log(weekly hours) | (3) Pr(fulltime) | (4) Log(hourly wage) |
|---|---|---|---|---|
| $\hat{\delta}^{FP}$ | 0.052*** | 0.029* | 0.030*** | 0.021 |
|  | (0.015) | (0.015) | (0.011) | (0.014) |
| $\hat{\delta}^{MP}$ | 0.003 | 0.000 | -0.002 | 0.002 |
|  | (0.012) | (0.011) | (0.008) | (0.010) |
| Degree FE | ✓ | ✓ | ✓ | ✓ |
| Cohort FE | ✓ | ✓ | ✓ | ✓ |
| Observations | 47,246 | 47,246 | 47,246 | 47,246 |
| R-squared | 0.286 | 0.250 | 0.278 | 0.095 |

Notes: OLS estimates of a regression of women's earnings and labor supply one year after graduation on: the average FLFP in the provinces of origin of female and male peers and the FLFP in the own province of origin. Regressions include cohort and degree fixed effects. The sample excludes degrees that experience trends in size over time.The estimates are done on the sample of women, studying in these degrees, who are employed one year after graduation and with non-missing information on the dependent variables. Standard errors clustered at degree level. All regressors are standardised.

TABLE A.21. Validity of the Empirical Strategy - Estimates of Peer Effects on Female Earnings

| | (1) Benchmark | (2) Δ size<=p75 | (3) Δ size<=p50 | (4) Δ avg grades<=p75 | (5) Δ avg grades<=p25 | (6) Δ sd grades<=p75 | (7) Δ sd grades<=p25 |
|---|---|---|---|---|---|---|---|
| $\hat{\delta}^{FP}$ | 0.037*** | 0.049*** | 0.053*** | 0.037** | 0.041* | 0.036** | 0.054** |
| | (0.013) | (0.013) | (0.015) | (0.015) | (0.025) | (0.014) | (0.022) |
| $\hat{\delta}^{MP}$ | -0.000 | -0.002 | -0.001 | -0.003 | -0.002 | 0.001 | -0.004 |
| | (0.010) | (0.011) | (0.013) | (0.011) | (0.014) | (0.012) | (0.014) |
| | | | | | | | |
| Degree FE | ✓ | ✓ | ✓ | ✓ | ✓ | ✓ | ✓ |
| Cohort FE | ✓ | ✓ | ✓ | ✓ | ✓ | ✓ | ✓ |
| Nb. of degrees | 1,572 | 1,163 | 770 | 1,170 | 390 | 1,171 | 389 |
| Observations | 69,645 | 58,363 | 42,518 | 59,040 | 25,564 | 60,613 | 28,741 |
| R-squared | 0.287 | 0.294 | 0.309 | 0.280 | 0.251 | 0.278 | 0.284 |

Notes: The table reports estimates of the baseline specification 2. The dependent variable is log(monthly earnings) in all columns. Regressions include cohort and degree fixed effects. In each column, a different sample of degrees is used. Column (1) and (2) refer to the sample of degrees that experience cross-cohort changes in size below the 75th and 50th percentiles (see definitions in Section 6.1). Column (3) and (4) refer to the sample of degrees that experience cross-cohort changes in the average ability of students below the 75th or 25th percentile. Column (5) and (6) refer to the sample of degrees that experience cross-cohort changes in the standard deviation of ability of students below the 75th or 25th percentile. The estimates are done on the sample of women, studying in these degrees, who are employed one year after graduation and with non-missing information on the dependent variables. Standard errors clustered at degree level. All regressors are standardised.

TABLE A.22. Other Checks - Estimates of peer effects on female earnings and labor supply

| | (1) Log(monthly earnings) | (2) Log(weekly hours) | (3) Pr(fulltime) | (4) Log(hourly wage) |
|---|---|---|---|---|
| $\hat{\delta}^{FP}$ | 0.037*** | 0.033*** | 0.018* | 0.003 |
| | (0.013) | (0.012) | (0.009) | (0.012) |
| $\hat{\delta}^{MP}$ | -0.001 | -0.000 | -0.003 | -0.002 |
| | (0.010) | (0.009) | (0.007) | (0.010) |
| Province of origin FEs | ✓ | ✓ | ✓ | ✓ |
| Degree FE | ✓ | ✓ | ✓ | ✓ |
| Cohort FE | ✓ | ✓ | ✓ | ✓ |
| Observations | 69,645 | 69,645 | 69,645 | 69,645 |
| R-squared | 0.290 | 0.248 | 0.282 | 0.102 |

Notes: this table reports OLS estimates from specification 2. Relative to the baseline specification, this includes controls for province of origin fixed effects, instead of the FLFP in the province of origin of the student. Regressions include cohort and degree fixed effects. All the estimates are done on the sample of women who are employed one year after graduation and with non-missing information on these variables. Standard errors clustered at degree level. All regressors are standardised.

TABLE A.23. Sensitivity to Sample Restrictions - Estimates of Peer Effects on Female Earnings

| | (1) | (2) | (3) | (4) | (5) | (6) | (7) | (8) |
|---|---|---|---|---|---|---|---|---|
| | | | | Degree size | | | Students with Bsc in same uni | |
| | Benchmark | >p10 | <=p90 | <=mean | >mean | >p90 | <=p90 | <=p25 |
| $\hat{\delta}^{FP}$ | 0.037*** | 0.032** | 0.039*** | 0.050*** | 0.017 | 0.017 | 0.037*** | 0.048** |
| | (0.013) | (0.013) | (0.014) | (0.016) | (0.020) | (0.029) | (0.013) | (0.019) |
| $\hat{\delta}^{MP}$ | -0.000 | -0.000 | -0.004 | -0.014 | 0.015 | 0.011 | 0.000 | 0.008 |
| | (0.010) | (0.010) | (0.012) | (0.012) | (0.016) | (0.017) | (0.010) | (0.013) |
| Degree FE | ✓ | ✓ | ✓ | ✓ | ✓ | ✓ | ✓ | ✓ |
| Cohort FE | ✓ | ✓ | ✓ | ✓ | ✓ | ✓ | ✓ | ✓ |
| Nb. of degrees | 1,572 | 1,403 | 1,399 | 1,037 | 519 | 157 | 1,404 | 391 |
| Observations | 69,645 | 68,409 | 46,721 | 22,804 | 46,841 | 22,924 | 65,453 | 21,886 |
| R-squared | 0.287 | 0.287 | 0.264 | 0.254 | 0.300 | 0.332 | 0.284 | 0.254 |

Notes: The table reports estimates of the baseline specification 2. The dependent variable is log(monthly earnings) in all columns. Regressions include cohort and degree fixed effects. In each column, a different sample of degrees is used. Column (1) and (2) refer to the sample of degrees that experience cross-cohort changes in size below the 75th and 50th percentiles (see definitions in Section 6.1). Column (3) and (4) refer to the sample of degrees that experience cross-cohort changes in the average ability of students below the 75th or 25th percentile. Column (5) and (6) refer to the sample of degrees that experience cross-cohort changes in the standard deviation of ability of students below the 75th or 25th percentile. The estimates are done on the sample of women, studying in these degrees, who are employed one year after graduation and with non-missing information on the dependent variables. Standard errors clustered at degree level. All regressors are standardised.

TABLE A.24. Sensitivity to Measures of Gender Culture - Estimates of Peer Effects on Female Earnings

| | (1) | (2) | (3) | (4) | (5) | (6) | (7) |
|---|---|---|---|---|---|---|---|
| | | | Measures of gender culture in province of origin of peers | | | | |
| | FLFP | FLFP (young) | FLFP/MLFP | FLFP/MLFP (young) | % of female grad. fulltime | female/male grad. fulltime | Firms' culture |
| $\hat{\delta}^{FP}$ | 0.037*** | 0.041*** | 0.035*** | 0.036*** | 0.040*** | 0.041*** | 0.016** |
| | (0.013) | (0.014) | (0.012) | (0.013) | (0.012) | (0.011) | (0.008) |
| $\hat{\delta}^{MP}$ | -0.000 | 0.003 | 0.000 | 0.002 | 0.010 | 0.009 | 0.004 |
| | (0.010) | (0.011) | (0.010) | (0.010) | (0.010) | (0.009) | (0.006) |
| Degree FE | ✓ | ✓ | ✓ | ✓ | ✓ | ✓ | ✓ |
| Cohort FE | ✓ | ✓ | ✓ | ✓ | ✓ | ✓ | ✓ |
| Nb. of degrees | 1,572 | 1,572 | 1,572 | 1,572 | 1,572 | 1,572 | 1,572 |
| Observations | 69,645 | 69,645 | 69,645 | 69,645 | 69,645 | 69,645 | 69,645 |
| R-squared | 0.287 | 0.287 | 0.287 | 0.287 | 0.288 | 0.288 | 0.287 |

The table presents estimates from the baseline specification 2, using alternative measures of peers' gender culture. The dependent variable in all columns is log(monthly earnings). Each column uses a different proxy for peers' gender culture, as specified in the column labels. Regressions include cohort and degree fixed effects. The estimates are done on the sample of women, who are employed one year after graduation and with non-missing information on the dependent variables. Standard errors clustered at degree level. All regressors are standardised.

TABLE A.25. Robustness checks - Estimates of Peer Effects on Female Earnings Controlling for Other Peers' Characteristics

| | Dependent variable: log(monthly earnings) | | | | | | | | |
|---|---|---|---|---|---|---|---|---|---|
| | (1) | (2) | (3) | (4) | (5) | (6) | (7) | (8) | (9) |
| $\hat{\delta}^{FP}$ | 0.041*** | 0.041*** | 0.040*** | 0.039*** | 0.039*** | 0.036*** | 0.040*** | 0.037*** | 0.037*** |
| | (0.012) | (0.012) | (0.012) | (0.012) | (0.012) | (0.012) | (0.012) | (0.013) | (0.013) |
| $\hat{\delta}^{MP}$ | 0.002 | 0.003 | 0.003 | 0.003 | 0.003 | 0.002 | -0.000 | -0.000 | -0.000 |
| | (0.010) | (0.010) | (0.010) | (0.010) | (0.010) | (0.010) | (0.010) | (0.010) | (0.010) |
| Share of peers with work. mother | ✓ | | | | | | | | |
| Share of peers with high-SES mother | | ✓ | | | | | | | |
| Share of peers with high-SES father | | | ✓ | | | | | | |
| Share of peers with college educ mother | | | | ✓ | | | | | |
| Share of peers with college educ father | | | | | ✓ | | | | |
| Share of high-ability peers | | | | | | ✓ | | | |
| Share of peers from general track | | | | | | | ✓ | | |
| Degree size | | | | | | | | ✓ | |
| Share of female peers | | | | | | | | | ✓ |
| Degree FE | ✓ | ✓ | ✓ | ✓ | ✓ | ✓ | ✓ | ✓ | ✓ |
| Cohort FE | ✓ | ✓ | ✓ | ✓ | ✓ | ✓ | ✓ | ✓ | ✓ |
| Nb. of degrees | 1,548 | 1,548 | 1,548 | 1,549 | 1,549 | 1,546 | 1,556 | 1,556 | 1,556 |
| Observations | 62,857 | 62,857 | 62,451 | 64,242 | 64,242 | 62,098 | 69,553 | 69,645 | 69,645 |
| R-squared | 0.293 | 0.293 | 0.293 | 0.291 | 0.291 | 0.292 | 0.288 | 0.287 | 0.288 |

Notes: The table presents estimates from the baseline specification 2 on log(monthly earnings). Each column represents a different regression, with an added control for an alternative peer characteristic, disaggregated by gender. For instance, in Column 1, I include controls for the proportion of female and male peers with working mothers. All regressions account for cohort and degree fixed effects and are conducted on the sample of women employed one year post-graduation, with non-missing data on the relevant variables. Variation in sample size across columns results from missing values in certain covariates. Standard errors are clustered at the degree level, and all regressors are standardized.

|  | (1)<br>GPA | (2)<br>Final grade | (3)<br>Time to<br>completion | (4)<br>Pr(delayed<br>grad.) |
|---|---|---|---|---|
| (Mean) | (27.8) | (108.6) | (2.5) | (0.35) |
| $\delta^{FP}$ | 0.047 | 0.071 | -0.004 | -0.007 |
|  | (0.029) | (0.102) | (0.010) | (0.008) |
| $\delta^{MP}$ | 0.039 | 0.066 | -0.006 | -0.005 |
|  | (0.024) | (0.085) | (0.008) | (0.007) |
| Degree FE | ✓ | ✓ | ✓ | ✓ |
| Cohort FE | ✓ | ✓ | ✓ | ✓ |
| Observations | 182,792 | 182,792 | 182,792 | 182,792 |
| R-squared | 0.244 | 0.174 | 0.161 | 0.148 |

Notes: OLS estimates of a regression of indicators of academic performance on: the average FLFP in the provinces of origin of female and male peers and the FLFP in the own province of origin. Regressions include cohort and degree fixed effects. All the estimates are done on the full sample of women. All regressors are standardised. All regressors are standardised, while the dependent variables are not. The mean values of the dependent variables are provided in the table. Standard errors clustered at degree level.

TABLE A.27. Estimates of peer effects on geographic mobility - Female sample

| | (1) FLFP in prov. of work | (2) Prov work = univ. | (3) Reg work = univ. | (4) Prov of work ≠ birth |
|---|---|---|---|---|
| (Mean) | (54.6) | (0.45) | (0.68) | (0.44) |
| | | | | |
| $\delta^{FP}$ | 0.155 | 0.007 | 0.013 | 0.007 |
| | (0.151) | (0.011) | (0.011) | (0.010) |
| $\delta^{MP}$ | 0.126 | -0.010 | -0.005 | 0.009 |
| | (0.123) | (0.008) | (0.008) | (0.007) |
| | | | | |
| Degree FE | ✓ | ✓ | ✓ | ✓ |
| Cohort FE | ✓ | ✓ | ✓ | ✓ |
| Observations | 68,751 | 72,367 | 72,367 | 72,367 |
| R-squared | 0.586 | 0.156 | 0.152 | 0.181 |

Notes: OLS estimates of a regression of indicators of women's geographic mobility on: the average FLFP in the provinces of origin of female and male peers and the FLFP in the own province of origin. Regressions include cohort and degree fixed effects. All the estimates are done on the sample of women employed one year post-graduation, with non-missing data on the relevant variables. The dependent variables are as follows: FLFP in the province of employment (Column 1), an indicator of whether the province of employment is the same as that of the university attended (Column 2), an indicator of whether the region of employment matches the university's region (Column 3), and an indicator of whether the province of employment differs from the province of birth (Column 4). The means of these variables are reported in parentheses. Standard errors clustered at degree level. All regressors are standardised.

TABLE A.28. Robustness Checks - Estimates of Peer Effects on Female Earnings and Labor Supply Controlling for Share of Local Students

|  | (1) Log(monthly earnings) | (2) Log(weekly hours) | (3) Pr(fulltime) | (4) Log(hourly wage) |
|---|---|---|---|---|
| $\delta^{FP}$ | 0.045*** | 0.041*** | 0.022** | 0.003 |
|  | (0.013) | (0.012) | (0.010) | (0.013) |
| $\delta^{MP}$ | -0.002 | -0.002 | -0.001 | -0.000 |
|  | (0.010) | (0.010) | (0.007) | (0.010) |
| Share of female stayers | -0.011* | -0.010* | -0.004 | -0.000 |
|  | (0.006) | (0.006) | (0.004) | (0.005) |
| Share of male stayers | 0.003 | 0.006 | -0.001 | -0.003 |
|  | (0.005) | (0.005) | (0.004) | (0.004) |
| Degree FE | ✓ | ✓ | ✓ | ✓ |
| Cohort FE | ✓ | ✓ | ✓ | ✓ |
| Observations | 69,645 | 69,645 | 69,645 | 69,645 |
| R-squared | 0.288 | 0.246 | 0.280 | 0.100 |

Notes: OLS estimates of a regression of women's earnings and labor supply one year after graduation on: the average FLFP in the provinces of origin of female and male peers and the FLFP in the own province of origin, as well as the share of *local* female and male peers. A student is defined as *local* if she studies at university in her province of birth. Regressions include cohort and degree fixed effects. All the estimates are done on the sample of women employed one year post-graduation, with non-missing data on the relevant variables. Standard errors clustered at degree level. All regressors are standardised.

TABLE A.29.  Estimates of Peer Effects on Job-Search Preferences

|  | (1)<br>Index Pecuniary | (2)<br>Index Flexibility | (3)<br>Job's social utility |
|---|---|---|---|
| $\delta^{FP}$ | 0.003 | -0.027* | -0.012* |
|  | (0.009) | (0.015) | (0.007) |
| $\delta^{MP}$ | 0.001 | 0.006 | 0.001 |
|  | (0.007) | (0.011) | (0.005) |
| Degree FE | ✓ | ✓ | ✓ |
| Cohort FE | ✓ | ✓ | ✓ |
| Observations | 165,116 | 163,855 | 164,214 |
| R-squared | 0.089 | 0.043 | 0.093 |

Notes: OLS estimates of regressions of valuation of job attributes on: the average FLFP in the provinces of origin of female and male peers and the FLFP in the own province of origin. The dependent variables in Columns (1)-(3) measure the importance students place on different job characteristics. Answers come the question: "How much do you value attribute X in the job you are searching?" (scale 1-5). Specifically, Column (1) reflects preferences for pecuniary job attributes, i.e. as salary and career progression, based on a standardized index constructed from students' rankings on a 1-5 scale. The index in Column (2) is constructed by averaging students' rankings of job attributes related to flexibility (i.e. leisure time and hours flexibility). Both indexes in (1) and (2) have been standardised. The dependent variable in Column (3) is an indicator variable for whether a student gives maximum value to the social utility of a job. Regressions include cohort and degree fixed effects. The estimates are done on the sample of women who fill in the institutional pre-graduation survey (91.7%). Standard errors clustered at degree level. All regressors are standardised.

TABLE A.30. Main Characteristics of the Sample of Prospective Students - Original Survey

| | All | | | Low FLFP | | High FLFP | |
|---|---|---|---|---|---|---|---|
| | Mean | SD | N | Mean | SD | Mean | SD |
| **Background Characteristics** | | | | | | | |
| Age | 23.4 | 1.8 | 487 | 23.6 | 2.3 | 23.3 | 1.5 |
| Changed province for Master (%) | 88.8 | 31.6 | 490 | 100.0 | 0.0 | 82.6 | 37.9 |
| Changed region for Master (%) | 69.6 | 46.0 | 490 | 100.0 | 0.0 | 53.0 | 50.0 |
| FLFP in province of origin | 54.6 | 11.2 | 489 | 41.9 | 8.9 | 61.5 | 4.0 |
| Mother: university level (%) | 31.4 | 46.5 | 468 | 27.3 | 44.7 | 33.7 | 47.3 |
| Father: university level (%) | 28.0 | 44.9 | 465 | 32.7 | 47.1 | 25.3 | 43.6 |
| Mother: full-time at childbirth (%) | 49.9 | 50.1 | 465 | 45.7 | 50.0 | 52.2 | 50.0 |
| Mother: part-time at childbirth (%) | 30.1 | 45.9 | 465 | 23.2 | 42.3 | 33.9 | 47.4 |
| Mother: no work at childbirth (%) | 20.0 | 40.0 | 465 | 31.1 | 46.4 | 14.0 | 34.7 |
| **Field of study** | | | | | | | |
| Major: Economics (%) | 19.2 | 39.4 | 480 | 20.0 | 40.1 | 18.7 | 39.1 |
| Major: Humanities (%) | 45.2 | 49.8 | 480 | 40.0 | 49.1 | 48.1 | 50.0 |
| Major: Science (%) | 20.4 | 40.4 | 480 | 23.5 | 42.5 | 18.7 | 39.1 |
| Major: Social Sciences (%) | 15.2 | 35.9 | 480 | 16.5 | 37.2 | 14.5 | 35.3 |
| First year (%) | 65.1 | 47.7 | 490 | 61.8 | 48.7 | 66.9 | 47.1 |
| Second year (%) | 33.5 | 47.2 | 490 | 38.2 | 48.7 | 30.9 | 46.3 |
| Above second year (%) | 1.4 | 11.9 | 490 | 0.0 | 0.0 | 2.2 | 14.7 |
| **Civil Status and Fertility Expectations** | | | | | | | |
| Single (%) | 48.1 | 50.0 | 468 | 46.1 | 50.0 | 49.2 | 50.1 |
| Has a partner (%) | 46.4 | 49.9 | 468 | 48.5 | 50.1 | 45.2 | 49.9 |
| Cohabits with partner (%) | 5.6 | 22.9 | 468 | 5.5 | 22.8 | 5.6 | 23.1 |
| Partner in same program (%) | 3.4 | 18.2 | 468 | 2.4 | 15.4 | 4.0 | 19.5 |
| Intend to have children (%) | 54.0 | 49.9 | 470 | 52.7 | 50.1 | 54.8 | 49.9 |
| Maybe children (%) | 33.2 | 47.1 | 470 | 35.8 | 48.1 | 31.8 | 46.6 |
| Does not intend to have children (%) | 12.6 | 33.2 | 470 | 11.5 | 32.0 | 13.1 | 33.8 |
| Has children already (%) | 0.2 | 4.6 | 470 | 0.0 | 0.0 | 0.3 | 5.7 |
| Expected age at first child | 31.3 | 2.8 | 351 | 31.7 | 3.2 | 31.1 | 2.6 |
| **Intended Job Search** | | | | | | | |
| Intend to search for a job (%) | 79.7 | 40.3 | 488 | 80.8 | 39.5 | 79.1 | 40.7 |
| Intend to pursue further education (%) | 19.1 | 39.3 | 488 | 18.0 | 38.6 | 19.6 | 39.8 |
| Intend to keep job (%) | 1.2 | 11.0 | 488 | 1.2 | 10.8 | 1.3 | 11.2 |
| Job location: North (%) | 61.2 | 48.8 | 485 | 62.2 | 48.6 | 60.7 | 48.9 |
| Job location: Centre (%) | 15.7 | 36.4 | 485 | 14.0 | 34.8 | 16.6 | 37.3 |
| Job location: South (%) | 2.7 | 16.2 | 485 | 7.0 | 25.5 | 0.3 | 5.7 |
| Job location: Abroad (%) | 20.4 | 40.3 | 485 | 16.9 | 37.5 | 22.4 | 41.7 |

Notes: this table summarizes the main characteristics of the sample of prospective students that participated in the original survey. It reports the mean and standard deviation of variables related to students' background, fields of study, civil status, partner information, fertility expectations, and labor market intentions. These statistics are reported for the overall sample (490 students), as well as for the two subsamples of students from above-median (317 students) and below-median (173 students) FLFP provinces.

TABLE A.31. Baseline and Updated Beliefs on the Job Offer Distribution - Robustness Checks

| | Below-med FLFP | | Above-med FLFP | | |
|---|---|---|---|---|---|
| | Pred | SE | Pred | SE | P-value |
| **a. Baseline Beliefs (T=0)** | | | | | |
| $\alpha$: Expected arrival rate of job offers (%) | 32.30 | 1.80 | 35.05 | 1.26 | 0.23 |
| $\gamma_P$: Expected % of part-time job offers | 57.48 | 2.43 | 50.33 | 1.70 | 0.02 |
| Perceived uncertainty (1-5) | 2.80 | 0.13 | 2.93 | 0.09 | 0.44 |
| Prob. to accept part-time job offer | 67.26 | 2.17 | 59.66 | 1.50 | 0.01 |
| | | | | | |
| **b. Updated Beliefs (T=1)** | | | | | |
| $\alpha$: Expected arrival rate of job offers (%) | 32.82 | 2.47 | 32.43 | 1.93 | 0.91 |
| $\gamma_P$: Expected % of part-time job offers | 52.41 | 3.11 | 51.89 | 2.43 | 0.90 |
| Perceived uncertainty (1-5) | 2.63 | 0.15 | 2.74 | 0.12 | 0.57 |
| Prob. to accept part-time job offer | 62.03 | 2.94 | 63.94 | 2.28 | 0.63 |

Notes: This table presents predictions from a linear regression model, where the dependent variable is regressed on an indicator for whether the FLFP in the birth province is above or below the median, along with fixed effects for the field of study and controls for students' background characteristics (age, parents' education), job search intentions, and expected job location. Each row represents a different regression, with the dependent variable specified in Column 1. For each regression, the table reports the predicted dependent variable for women from provinces with low versus high FLFP, along with the standard errors. The last column provides the p-value for the difference between these two groups. In Panel (a), the sample consists of all first-year female Master's students without missing information on the covariates (291), and in Panel (b), it includes all second-year female Master's students without missing information on the covariates (148). Between 60% and 65% of the students are from provinces with above-median FLFP.

TABLE A.32. Beliefs on arrival rates of job offers and acceptance of part-time jobs

|  | (1) | (2) | (3) | (4) |
|---|---|---|---|---|
|  | Probability to accept part-time job offer | | | |
| Expected percentage of part-time offers (γ) | 0.327** (0.066) | 0.272** (0.056) |  |  |
| Expected arrival rate of job offers (α) |  |  | -0.148** (0.028) | -0.078 (0.035) |
| Field FEs |  | ✓ |  | ✓ |
| Observations | 463 | 463 | 464 | 464 |
| R-squared | 0.125 | 0.171 | 0.014 | 0.101 |

The table presents estimated coefficients from regressions of the elicited probability of accepting a part-time job offer on workers' expected probability of receiving a job offer (Columns 1-2) or the expected percentage of part-time offers (Columns 3-4). In Columns 2 and 4, I include controls for the field of study. The sample consists of all female students with non-missing values for these variables, drawn from both the first and second year of the program.

| | Below-med FLFP | | Above-med FLFP | | |
|---|---|---|---|---|---|
| | Pred | SE | Pred | SE | P-value |
| **a. Baseline Expectations (T=0)** | | | | | |
| Fertility: yes | 0.50 | 0.05 | 0.54 | 0.04 | 0.50 |
| Fertility: don't know | 0.38 | 0.05 | 0.35 | 0.03 | 0.61 |
| Fertility: no | 0.12 | 0.03 | 0.11 | 0.02 | 0.76 |
| Age of expected fertility | 31.58 | 0.31 | 30.88 | 0.23 | 0.07 |
| Labor supply at motherhood (Scenario 1) | | | | | |
|     Work full-time | 0.49 | 0.05 | 0.43 | 0.04 | 0.40 |
|     Work part-time | 0.49 | 0.05 | 0.55 | 0.04 | 0.38 |
|     No work | 0.03 | 0.02 | 0.03 | 0.02 | 0.90 |
| Labor supply at motherhood (Scenario 2) | | | | | |
|     Work full-time | 0.70 | 0.05 | 0.68 | 0.04 | 0.81 |
|     Work part-time | 0.28 | 0.05 | 0.31 | 0.04 | 0.58 |
|     No work | 0.03 | 0.02 | 0.01 | 0.01 | 0.21 |
| | | | | | |
| **b. Updated Expectations (T=1)** | | | | | |
| Fertility: yes | 0.57 | 0.07 | 0.61 | 0.05 | 0.63 |
| Fertility: don't know | 0.32 | 0.06 | 0.23 | 0.04 | 0.27 |
| Fertility: no | 0.10 | 0.04 | 0.15 | 0.04 | 0.41 |
| Age of expected fertility | 32.65 | 0.44 | 31.13 | 0.35 | 0.01 |
| Labor supply at motherhood (Scenario 1) | | | | | |
|     Work full-time | 0.67 | 0.06 | 0.41 | 0.06 | 0.00 |
|     Work part-time | 0.33 | 0.06 | 0.55 | 0.06 | 0.02 |
|     No work | 0.04 | 0.03 | | | |
| Labor supply at motherhood (Scenario 2) | | | | | |
|     Work full-time | 0.79 | 0.05 | 0.80 | 0.05 | 0.85 |
|     Work part-time | 0.21 | 0.05 | 0.20 | 0.05 | 0.85 |
|     No work | 0.00 | | 0.00 | | |

Notes: This table presents predictions from logistic regressions, where the dependent variable is regressed on an indicator for whether the FLFP in the birth province is above or below the median, along with fixed effects for the field of study. Each row represents a different regression, with the dependent variable specified in Column 1. For each regression, the table reports the predicted dependent variable for women from provinces with low versus high FLFP, along with the standard errors. The last column provides the p-value for the difference between these two groups. In Panel (a), the sample consists of all first-year female Master's students without missing information on the dependent variables, and in Panel (b), it includes all second-year female Master's students without missing information on the dependent variables. Between 60% and 65% of the students are from provinces with above-median FLFP.